\documentclass[11pt]{article}
 \usepackage{titlesec}
\usepackage{hyperref}

\titleclass{\subsubsubsection}{straight}[\subsection]

\usepackage[utf8]{inputenc}
\usepackage{amsmath,amsfonts,amssymb,stackengine,graphicx}
\title{appendix H2}
\author{sagnotti }
\date{June 2020}

\usepackage[utf8]{inputenc}
\newif\iffigs\figstrue

\usepackage[title]{appendix}
\usepackage{epsfig,latexsym}
\usepackage{hyperref}
\usepackage{amsmath}
\usepackage{verbatim}
\usepackage{color}
\usepackage{mathrsfs}  
\usepackage{slashed}
\usepackage{amssymb}

\DeclareMathAlphabet{\mathpzc}{OT1}{pzc}{m}{it}

 \csname
@addtoreset\endcsname{equation}{section}
  
\def\gz0{\gamma^{0}}

\def\sign{\rm sign}

\def\beq{\begin{equation}}
\def\eeq{\end{equation}}
\def\bea{\begin{eqnarray}}
\def\eea{\end{eqnarray}}
\def\ba{\begin{array}}
\def\ea{\end{array}}
\def\bec{\begin{center}}
\def\ec{\end{center}}
\def\ba{\begin{align}}
\def\ena{\end{align}}

\def\12{\frac{1}{2}}

\def\pr{\partial}

\newcounter{subsubsubsection}[subsubsection]
\renewcommand\thesubsubsubsection{\thesubsubsection.\arabic{subsubsubsection}}

\titleformat{\subsubsubsection}
  {\normalfont\normalsize\bfseries}{\thesubsubsubsection}{1em}{}
\titlespacing*{\subsubsubsection}
{0pt}{3.25ex plus 1ex minus .2ex}{1.5ex plus .2ex}

\makeatletter
\renewcommand\paragraph{\@startsection{paragraph}{5}{\z@}%
  {3.25ex \@plus1ex \@minus.2ex}%
  {-1em}%
  {\normalfont\normalsize\bfseries}}
\renewcommand\subparagraph{\@startsection{subparagraph}{6}{\parindent}%
  {3.25ex \@plus1ex \@minus .2ex}%
  {-1em}%
  {\normalfont\normalsize\bfseries}}
\def\toclevel@subsubsubsection{4}
\def\toclevel@paragraph{5}
\def\toclevel@paragraph{6}
\def\l@subsubsubsection{\@dottedtocline{4}{7em}{4em}}
\def\l@paragraph{\@dottedtocline{5}{10em}{5em}}
\def\l@subparagraph{\@dottedtocline{6}{14em}{6em}}
\makeatother

\begin{document}

\begin{flushright}
{\today}
\end{flushright}

\vspace{10pt}

\begin{center}


{\Large\sc On Warped String Vacuum Profiles and Cosmologies, II}\\
{\large\sc Non--Supersymmetric Strings}


\vspace{25pt}
{\sc J.~Mourad${}^{\; a}$  \ and \ A.~Sagnotti${}^{\; b}$\\[15pt]

${}^a$\sl\small APC, UMR 7164-CNRS \\
Universit\'e de Paris  \\
10 rue Alice Domon et L\'eonie Duquet \\75205 Paris Cedex 13 \ FRANCE
\\ e-mail: {\small \it
mourad@apc.univ-paris7.fr}\vspace{10
pt}

{${}^b$\sl\small
Scuola Normale Superiore and INFN\\
Piazza dei Cavalieri, 7\\ 56126 Pisa \ ITALY \\
e-mail: {\small \it sagnotti@sns.it}}\vspace{10pt}
}

\vspace{30pt} {\sc\large Abstract}\end{center}
\noindent
We investigate the effects of the leading tadpole potentials of 10D tachyon-free non--super\-symmetric strings in warped products of flat geometries of the type $M_{p+1} \times R \ \times \ T_{10-p-2}$ depending on a single coordinate. In the absence of fluxes and for $p<8$, there are two families of these vacua for the orientifold disk-level potential, both involving a finite internal interval. Their asymptotics are surprisingly captured by tadpole-free solutions, isotropic for one family and anisotropic at one end for the other. In contrast, for the heterotic torus-level potential there are four types of vacua. Their asymptotics are always tadpole-dependent and isotropic at one end lying at a finite distance, while at the other end, which can lie at a finite or infinite distance, they can be  tadpole--dependent isotropic or tadpole--free anisotropic.
We then elaborate on the general setup for including symmetric fluxes, and present the three families of exact solutions that emerge when the orientifold potential and a seven--form flux are both present. These solutions include a pair of boundaries, which are always separated by a finite distance. In the neighborhood of one, they all approach a common supersymmetric limit, while the asymptotics at the other boundary can be tadpole-free isotropic, tadpole-free anisotropic or again supersymmetric. We also discuss corresponding cosmologies, with emphasis on their climbing or descending behavior at the initial singularity. In some cases the toroidal dimensions can contract during the cosmological expansion.

\setcounter{page}{1}

\pagebreak

\newpage
\tableofcontents
\newpage
\baselineskip=20pt
\section{\sc  Introduction and Summary}\label{sec:intro}

\vskip 12pt

Above and beyond the clear interest of broken supersymmetry~\cite{SUSY} for Particle Physics, the very existence of a handful of non--supersymmetric ten-dimensional string models~\cite{strings}, free of tachyons and satisfying all known consistency rules, makes it imperative to explore their compactifications, paying properly attention to their stability properties. Supersymmetry is absent in two of these models, the heterotic $SO(16)\times SO(16)$ string of~\cite{so16so16} and the $U(32)$ orientifold~\cite{orientifolds} model of~\cite{susy95}, while in the third model, Sugimoto's $USp(32)$ orientifold~\cite{bsb}, it is non-linearly realized~\cite{dmnonlinear} (for recent reviews see~\cite{reviews}). In all three cases, the breaking of supersymmetry in the original ten--dimensional Minkowski space induces a back-reaction that manifests itself via the emergence, in the low--energy effective field theory, of a new contribution associated to a runaway ``tadpole potential''
\beq{}
\Delta\,{\cal S} \ = \ - \ T \int d^{10} x \, \sqrt{-g}\ e^{\gamma \phi}  \ . \label{tadpole_pot}
\eeq
String theory yields a sharp prediction for the strengths $T$ of these potentials and for the exponents $\gamma$, which reflect their origin in string perturbation theory. In the Einstein frame, $\gamma=\frac{5}{2}$ for the heterotic model of~\cite{so16so16}, which only involves closed strings, so that the leading contribution emerges from the torus amplitude, while $\gamma=\frac{3}{2}$ for the orientifold models of~\cite{susy95} and~\cite{bsb}, where the leading contribution emerges from the two disk-level amplitudes. A direct consequence of the potential~\eqref{tadpole_pot} is that ten--dimensional flat space is not an acceptable vacuum for these theories. Moreover, tadpole potentials of this type emerge, in lower dimensions, as a result of compactifications that break supersymmetry, and in particular in the widely explored Scherk--Schwarz reductions~\cite{scherkschwarz}. 

Ideally, one would like to address dynamical questions on the vacuum directly within String Theory, but the currently available tools make it imperative to rely on the low--energy effective field theory. Therefore, vacuum solutions for the Einstein equations coupled to collections of matter fields are the key ingredient to gather some information on the problem, with one proviso. Their indications are fully reliable and significant for String Theory only within regions where the string coupling $g_s=e^\phi$ and spacetime curvature invariants are bounded since, from the vantage point of String Theory, low--energy descriptions based on General Relativity are merely the leading contribution to a double series expansion in curvatures and $g_s$. 

In this paper, we explore the effects on the vacuum of the tadpole potential~\eqref{tadpole_pot} for the three non--supersymmetric non--tachyonic strings, and more generally we trace their dependence on $\gamma$. To this end, we focus on geometries that are warped products of two flat spaces, and are described by metric tensors of the form
\beq
ds^{\,2}\ = \ e^{2A(r)}\, \eta_{\mu\nu}\, dx^\mu\,dx^\nu \ + \ e^{2B(r)}\,dr^2\ + \ e^{2C(r)}\, \delta_{mn}\, d y^m\,d y^n \ , \label{metric_sym_intro}
\eeq
depending on a single coordinate $r$,
where $\mu=0,\ldots , p$ and the $y^m$, with $m=1,\ldots , 8-p$, are coordinates on an internal torus. We also allow for symmetric dilaton and form profiles of all types that are compatible with their symmetries. Backgrounds for supersymmetric strings with these isometries were explored in detail in~\cite{ms_vacuum_1}, to which we shall refer at times as I, and of which this paper is a sequel. 

The plan of the paper is as follows. In Section~\ref{sec:symmetries} we briefly set up our notation, reviewing the effective action and the symmetric field profiles that are allowed in the class of metrics~\eqref{metric_sym_intro}, we identify the convenient harmonic gauge and present the set of equations for $A(r)$, $B(r)$, $C(r)$, the dilaton profile $\phi(r)$, and the allowed form field strengths. The reader can find a more detailed discussion of all these steps in~I. In Section~\ref{sec:susybT} we construct, within the above framework, all static vacuum solutions in the presence of the tadpole potential in~\eqref{tadpole_pot}, but in the absence of form fluxes. We also highlight the surprising sub--dominance of the tadpole potential in many asymptotic regions and connect these solutions, which are anisotropic generalizations of the $p=8$ Dudas--Mourad vacuum of~\cite{dm_vacuum}, to the Kasner--like flux--free geometries of~\cite{ms_vacuum_1} that are also reviewed in Appendix~\ref{app:kasner}. The special role of the ``critical'' orientifold potential with $\gamma=\gamma_c=\frac{3}{2}$ manifests itself clearly in this analysis, and in particular in the new solutions with $p<8$. For $\gamma=\gamma_c$ we find two families of such solutions: both approach, at one end, a tadpole-free isotropic solution of Appendix~\ref{app:kasner}, while at the other end the asymptotics is governed by a tadpole-free isotropic or anisotropic solution, even if the string coupling diverges there.
For $\gamma < \gamma_c$ the backgrounds approach, at both ends of the internal interval, which is of finite length, the anisotropic tadpole-free solutions of I, and the string coupling diverges at least at one end. For $\gamma>\gamma_c$ there are four types of solutions. They all approach an isotropic tadpole--dependent limit at one end, which lies at a finite distance, while at the other end, which can lie at a finite or infinite distance, the limiting behavior can be either isotropic and tadpole--dependent or anisotropic and tadpole--free. The string coupling diverges again at least at one end. 
In Section~\ref{sec:susybT_c} we discuss corresponding cosmological models, with emphasis on their climbing or descending behavior.
In Section~\ref{sec:Tn0hn0} we describe the general setup to analyze more complicated vacua where a form flux and a tadpole potential are simultaneously present. In Section~\ref{sec:exat5b} we describe in detail the exact solutions in the presence of the orientifold tadpole potential with $\gamma=\frac{3}{2}$ and of a ``magnetic'' three--form flux, and elaborate on their limiting behaviors. All these solutions include a finite interval and approach a supersymmetric limit at one end, while at the other the limiting behavior can be zero--flux tadpole--free isotropic or anisotropic, or again supersymmetric. The anisotropic case allows a finite string coupling in the asymptotic region.
Section~\ref{sec:exat5b} also contains a discussion of the corresponding cosmological solutions, which can combine climbing and descending behaviors with contractions of some set of coordinates.
Section~\ref{sec:conclusions} contains our conclusions, and includes tables summarizing the main properties of the solutions, together with some comments on possible future developments along these lines. Appendix~\ref{app:deq} collects some properties of a Newtonian model that recurs in our analysis, and finally Appendix~\ref{app:AdSxS} recovers the $AdS \times S$ vacua of~\cite{AdStimesS} in the harmonic gauge.

\section{\sc Symmetric Profiles and Equations of Motion}\label{sec:symmetries}

In this section we describe our basic setup and the notation that we shall use for the solutions of interest. 
We use the conventions of~I, so that in the string frame the bosonic portions of the low--energy effective field theories of interest include the terms
 \bea
 {\cal S} &=&\frac{1}{2\,(\alpha')^\frac{D-2}{2}}\ \int d^{D}x\,\sqrt{-\,G} \Big\{ e^{-2\phi}\Big[\, R\, + \,4(\partial\phi)^2 \Big] \, - \, {T} \, e^{\,\gamma_S\,\phi}\, - \, \frac{e^{-2\,\beta_S\,\phi}}{2\,(p+2)!}\, {\cal H}_{p+2}^2 \Big\} \ . \label{eqs1}
 \eea
In principle, one should consider different values of $p$, with the exponents that are collected in Table~\ref{table:tab_1}. 
This prototype action thus involves, in general, two types of fields aside from gravity: the dilaton $\phi$ and a $(p+1)$--form gauge potential ${\cal B}_{p+1}$ of field strength ${\cal H}_{p+2}$. Here the ``tadpole term'' sized by $T$ can describe, in principle, a non--critical sphere--level potential if $\gamma_S=-2$ and $D \neq 10$, with $T \sim D - 10$, a disk--level orientifold term if $\gamma_S=-1$ and $D=10$, or a genus--one contribution if $\gamma_S=0$ and $D=10$. In this paper, we shall focus on critical non--supersymmetric strings in $D=10$, and on the effects induced by an exponential potential determined by the choice of $\gamma$ that dominates at weak string coupling, but here and there we shall set up the formalism for generic values of $D$.

 \begin{table}
 \begin{center}
\begin{tabular}{ ||c||c|c|c|| } 
 \hline\hline
  Model & $p$ & $\beta_S$ & $\gamma_S$ \\ [0.5ex] 
  \hline\hline
   $USp(32)$ & $(1,5)$ & $(0,0)$ & $-1$  \\ [0.5ex] 
  \hline
  $U(32)$ & $I,V$;$(-\,1,1,3,5,7)$ & $1,\,-1$;\,$(0,0,0,0,0)$ & $-1$ \\ [0.5ex] 
  \hline
 $SO(16) \times SO(16)$ heter. & $I,V$ & $1,\,-1$ & 0  \\ [0.5ex] 
 \hline\hline
\end{tabular}
 \end{center}
 \vskip 12pt 
 \caption{\small String--frame parameters for the tachyon--free ten--dimensional string models. Roman numerals refer to NS-NS branes, entries within parentheses refer to RR ones, and dashes signal couplings that are not present in the low--energy effective theory.}\vskip 12pt
 \label{table:tab_1}
 \end{table}

In the Einstein frame, with the metric $g$ related to $G$ according to
\beq
G_{MN} \ = \ e^\frac{4\phi}{D-2} \ g_{MN} \ , \label{string_vs_einstein}
\eeq
the action of eq.~\eqref{eqs1} becomes
\bea
{\cal S} &=& \frac{1}{2(\alpha')^\frac{D-2}{2}}\int d^{D}x\sqrt{-g}\left[R\ - \ \frac{4}{D-2}\ (\partial\phi)^2\ - \ T \, e^{\,\gamma\,\phi}\ - \ \frac{e^{-2\,\beta_p\,\phi}}{2\,(p+2)!}\, {\cal H}_{p+2}^2 
\right] \ , \label{eqs4}
\eea
with
\beq
\beta_p \,=\, \beta_S \ - \ \frac{D-2(p+2)}{D-2} \ , \quad
\gamma \,=\, \gamma_S \ + \ \frac{2 D}{D-2} \ .
 \label{alphaE}
\eeq

The values of these quantities for the ten--dimensional tachyon--free string models are collected in Table~\ref{table:tab_2}, and the corresponding field equations read
\bea
R_{MN} \,-\, \frac{1}{2}\ g_{MN}\, R &=& \!\!\frac{4}{D-2}\  \pr_M\phi\, \pr_N\phi\, + \, \frac{e^{\,-\,2\,\beta_p\,\phi} }{2(p+1)!}\,  \left({\cal H}_{p+2}^2\right)_{M N}  \nonumber\\
&-& \!\!\frac{1}{2}\,g_{MN}\Big[\frac{4\,(\pr\phi)^2}{D-2}+ \frac{e^{\,-\,2\,\beta_p\,\phi}}{2(p+2)!}\,{\cal H}_{p+2}^2\,+\, V(\phi)\Big] \ ,  \nonumber
\\
\frac{8}{D-2} \ \Box\phi &=& \,-\,\frac{\beta_p\, e^{\,-\,2\,\beta_p\,\phi}}{(p+2)!} \ {\cal H}_{p+2}^2 \, +\, V^\prime(\phi)  \ , \nonumber 
\\
d\Big(e^{\,-\,2\,\beta_p\,\phi}\ {}^{*}\,{\cal H}_{p+2}\Big) &=& 0  \ . \label{eqsbeta}
\eea
where
\beq
V(\phi) \ = \ T \, e^{\gamma\,\phi} \ .
\eeq
 In the 0'B string there is also a five--form field strength, which satisfies the first--order self--duality equation
 \beq{}
{\cal H}_5 \ = \ {*}\,{\cal H}_{5} \ ,
\eeq{}
whose contribution is not captured by the preceding actions. We shall return to this case shortly.
 \begin{table}
 \begin{center}
\begin{tabular}{ ||c||c|c|c|| } 
 \hline\hline
  Model & $p$ & $\beta_p$ & $\gamma$ \\ [0.5ex] 
  \hline\hline
   $USp(32)$ & (1,5)  & $\left( -\,\frac{1}{2},\frac{1}{2}\right)$ & $\frac{3}{2}$  \\ [0.5ex] 
  \hline
  $U(32)$ & $I,V$;(-1,1,3,5,7)  & $\frac{1}{2},-\,\frac{1}{2}$;\,$\left(-\,1,-\,\frac{1}{2},0,\frac{1}{2},1\right)$ & $\frac{3}{2}$ \\ [0.5ex] 
  \hline
 $SO(16) \times SO(16)$ heter. & $I,V$ &  $\frac{1}{2},-\,\frac{1}{2}$ & $\frac{5}{2}$  \\ [0.5ex] 
 \hline\hline
\end{tabular}
 \end{center}
 \vskip 12pt 
 \caption{\small Einstein--frame parameters for the tachyon--free ten--dimensional string models. Roman numerals refer to NS-NS branes, entries within parentheses refer to RR ones, and dashes signal couplings not present in the low--energy effective theory.}\vskip 12pt
\label{table:tab_2}
 \end{table}

As we anticipated in the Introduction, we focus on metrics of the form~\eqref{metric_sym_intro}, which involve three dynamical functions of a single variable $r$. 
Furthermore, as in~\cite{ms_vacuum_1}, we shall find it convenient to work in the ``harmonic gauge'', whereby
\beq
B \ = \ (p+1)A \ + \ (D-p-2) C \ , \label{harm_gauge}
\eeq
which will simplify the resulting equations.
Moreover, we shall also explore counterparts of these solutions that are obtained via an analytic continuation of $r$ and $x^0$ to imaginary values. These build anisotropic cosmologies and generalize previous results.

As explained in~\cite{ms_vacuum_1}, there are four types of symmetric tensor profiles compatible with the Bianchi identities and the equations of motion,
\bea
{\cal H}_{p+2} &=& H_{p+2}\ e^{\,2\,\beta_p\phi + B +(p+1)A-(D-p-2)C} \ dx^0 \wedge \ldots \wedge dx^p \wedge dr \ , \nonumber \\
{\cal H}_{p+1} &=& h_{p+1} \ dx^0 \wedge \ldots \wedge dx^p \ , \label{Hhforms1}
\eea
and
\bea
{\cal H}_{D-p-1} &=& \tilde{H}_{D-p-1}\ e^{\,2\,\beta_{D-p-3}\phi + B -(p+1)A+(D-p-2)C} \ dy^1 \wedge \ldots \wedge dy^{D-p-2} \wedge dr \ , \nonumber \\
{\cal H}_{D-p-2} &=& \tilde{h}_{D-p-2} \ dy^1 \wedge \ldots \wedge dy^{D-p-2} \ , \label{Hhforms2}
\eea
where the $H$, $\tilde{H}$, $h$ and $\tilde{h}$ are constants.

There are also special tensor profiles that are relevant for the type--0'B theory in ten dimensions. They demand a few additional comments, since the corresponding field strength is self--dual. To begin with, one can start from the solution of the self--duality condition, which reads
\beq
{\cal H}_{5} \ = \ \frac{H_{5}}{2 \sqrt{2}}\,\left(e^{B +4\,A\,-\,5\,C}\, {\epsilon}_{(4)}  \wedge dr  \,+\, \widetilde{\epsilon}_{(5)} \right) \ , \label{H5H}
\eeq
since $\beta_S=0$ in this case. In a similar fashion, a second type of profile,
\beq
{\cal H}_{5} \ = \ \frac{h_{5}}{2 \sqrt{2}}\left(\epsilon_{(5)} \ + \ e^{ - 5 A + B + 4 C} \ dr \wedge \widetilde{\epsilon}_{(4)} \right) \ , \label{H5h}
\eeq
is the counterpart of eq.~\eqref{H5H} for the $h$ field strengths discussed above. 

In the ``harmonic'' gauge~\eqref{harm_gauge},
the equations of motion for $A$, $C$ and $\phi$ deduced from eqs.~\eqref{eqsbeta} are
 \bea
 A'' \!\!\!&=& - \ \frac{T}{(D-2)} \ e^{2\,B\,+\,\gamma\,\phi} \label{EqA_red}\\
  &+& \!\!\frac{(D-p-3)}{2\,(D-2)} \ e^{2\,B\,+\,2\,\beta_p\,\phi\,-\,2(D-p-2)C } H_{p+2}^2 \,+\, \frac{(D-p-2)}{2\,(D-2)} \ e^{2\,B\,-\,2\,\beta_{p-1}\,\phi\,-\,2 (p+1) A} h_{p+1}^2  \nonumber \  ,  \\
C''\!\!\!&=& - \ \frac{T}{(D-2)} \ e^{2\,B\,+\,\gamma\,\phi} \label{EqC_red} \\
   &-& \frac{(p+1)}{2\,(D-2)}\ e^{2\,B\,+\,2\,\beta_p\,\phi\,-\,2(D-p-2)C }  H_{p+2}^2\,-\, \frac{p}{2\,(D-2)}\ e^{2\,B\,-\,2\,\beta_{p-1}\,\phi\,-\,2 (p+1) A} h_{p+1}^2 \ ,  \nonumber  \\
   \phi'' \!\!\!&=& \frac{T\,\gamma\,(D-2)}{8}\ e^{2\,B\,+\,\gamma\,\phi} \label{Eqphi_red} \\ &+& \frac{\beta_p\,(D-2)}{8}\ e^{2\,B\,+\,2\,\beta_p\,\phi\,-\,2(D-p-2)C } H_{p+2}^2 \,+\, \frac{\beta_{p-1}\,(D-2)}{8}\ e^{2\,B\,-\,2\,\beta_{p-1}\,\phi\,-\,2 (p+1) A} h_{p+1}^2\ . \nonumber
  \eea
Moreover the equation for $B$, which is usually called ``Hamiltonian constraint'', reads
 \begin{align}
&(p+1)A'[p\,A' \,+\, (D-p-2)C']\,+\, (D-p-2)C'[(D-p-3)C'+(p+1)A'] \nonumber \\
 &- \, \frac{4\,(\phi')^2}{D-2} \, + \, {T} \, e^{\, 2\,B\,+\,\gamma\,\phi} \nonumber \\
 & + \, \frac{1}{2}\, e^{\,2\,\beta_p\,\phi\,+\,2\,B\,-\,2\,(D-p-2)\,C} \ H_{p+2}^2 \,-\, \frac{1}{2}\, e^{\,-\,2\,\beta_{p-1}\,\phi\,-\,2(p+1)A\,+\,2\,B} \ h_{p+1}^2  \,= \, 0 \ . \label{EqB_red}
 \end{align}
  
Notice that this system has an interesting discrete symmetry: its equations are left invariant by the redefinitions
\bea
&& \left[A,C,\,p\right] \ \longleftrightarrow \ \left[C,A,\,D-p-3 \right] \ , \nonumber \\
&& \left[H_{p+2}^2,\beta_p;h_{p+1}^2,\beta_{p-1} \right] \ \longleftrightarrow \ \left[-h_{p+1}^2,- \beta_{p-1};- H_{p+2}^2,-\beta_p  \right] \label{sym_AC}
\ . \eea

Two special cases, related to the type--0'B string, must be treated separately, since they involve fluxes of the self--dual five--form field strength, for which we refer the reader to eqs.~\eqref{H5H} and \eqref{H5h}, and also to~I and \cite{int4d_vacuum}.
The complete equations of motion for the first case are
\beq
R_{MN}  \ = \  \frac{1}{24}\ \left({\cal H}_{5}^2\right)_{M N} \ + \ \frac{1}{2}\, \partial_M\,\phi \,\partial_N\,\phi \ + \ \frac{T}{8} \, e^{\gamma\,\phi}\, g_{MN}\ ,
\eeq
and their reduced form for the class of metrics of interest in the ``harmonic'' gauge $B=4 A + 5 C$ and for the symmetric $H_5$ profile of eq.~\eqref{H5H} reads
\bea
A'' &=&  \frac{H_5^2}{8}\ e^{8 A} \ - \ \frac{T}{8}\,e^{2B\,+\,\gamma\,\phi} \ , \nonumber \\
C'' &=&  - \ \frac{H_5^2}{8}\ e^{8 A} \ - \ \frac{T}{8}\,e^{2B\,+\,\gamma\,\phi} \ , \nonumber \\
\phi'' &=& \frac{3}{2}\, T \ e^{2B\,+\,\gamma\,\phi} \ . \label{eqABC_sdual}
\eea
The corresponding Hamiltonian constraint is
\beq
3\left(A'\right)^2 \ + \ 10\, A'\, C' \ + \ 5 \left(C'\right)^2 \ = \ \frac{1}{8}\, \left(\phi'\right)^2 \ - \ \frac{H_5^2}{16} \ e^{8 A}  \ - \ \frac{T}{4}\,e^{2B\,+\,\gamma\,\phi} \ . \label{ham_sdual}
\eeq

The counterpart of these results for the $h_{p+1}$--fluxes corresponds to $p=4$, and in this case
\bea
A'' &=& \frac{h_5^2}{8}\ e^{8 C} \ - \ \frac{T}{8}\,e^{2B\,+\,\gamma\,\phi} \ , \nonumber \\
C'' &=&  - \ \frac{h_5^2}{8}\ e^{8 C}\ - \ \frac{T}{8}\,e^{2B\,+\,\gamma\,\phi} \ , \nonumber \\
\phi'' &=& \frac{3}{2}\, T \ e^{2B\,+\,\gamma\,\phi} \ .  \label{eqABC_sdual_h}
\eea

while the Hamiltonian constraint becomes
\beq
5\left(A'\right)^2 \ + \ 10\, A'\, C' \ + \ 3 \left(C'\right)^2 \ = \ \frac{1}{8}\, \left(\phi'\right)^2 \ + \ \frac{h_5^2}{16} \ e^{8 C} \ - \ \frac{T}{4}\,e^{2B\,+\,\gamma\,\phi} \ . \label{ham_sdual_h}
\eeq

\section{\sc Vacuum Solutions without Form Fluxes} \label{sec:susybT}

We can now see how a tadpole potential affects the vacuum solutions of supersymmetric strings described in~I that do not involve form fluxes, which are also reviewed in Appendix~\ref{app:kasner}. The following discussion applies to all three non--tachyonic models in ten dimensions.

 Eqs.~\eqref{EqA_red}--\eqref{Eqphi_red} simplify considerably in the present setting and become, in ten dimensions,
\bea
A'' &=& - \ \frac{T}{8} \ e^{\, 2\,X}  \  , \nonumber\\
  C''& =&  - \ \frac{T}{8} \ e^{\, 2\,X} \ , \nonumber \\
\phi'' &=& {T\,\gamma}\ e^{\, 2\,X} \label{eqACphiF2}   \ ,
\eea
where
\beq
X \ = \ (p+1)A+(8-p)C + \frac{\gamma}{2}\,\phi \ . \label{X_beta}
\eeq
Note that the harmonic gauge condition translates into
\beq{}{}
X \ = \ B \ + \ \frac{\gamma}{2}\,\phi \ . \label{XB}
\eeq
The r.h.s.'s of the three equations in~\eqref{eqACphiF2} are all proportional, and consequently the new variable $X$ satisfies
\beq
X'' \ = \ \frac{T}{2} \big( \gamma^2 \,-\, \gamma_c^2 \big) \ e^{\, 2\,X} \ , \label{X_eq}
\eeq
where
\beq
\gamma_c \ = \ \frac{3}{2} \ ,
\eeq
the value that pertains the two ten--dimensional orientifolds. We thus come across the quantity
\beq{}{}
\Delta^2 \ = \  \frac{T}{2}\ \left|\gamma^2\,-\,\gamma_c^2\right| 
 \ , \label{Delta_def}
 \eeq
which will play an important role in ensuing analysis.
The introduction of $X$ reduces the Hamiltonian constraint of eq.~\eqref{EqB_red} to
 \beq
\left(X'\,-\, \frac{\gamma}{2}\ \phi'\right)^2 \ - \ (p+1) \left(A'\right)^2 \ - \ (8-p) \left(C'\right)^2  \,-\, \, \frac{(\phi')^2}{2} \, + \, {T} \, e^{\, 2\,X}\, = \, 0 \ ,\label{hamc_Tnohkprime}
 \eeq
but the four variables $X$, $A$, $C$ and $\phi$ are clearly not independent.

The form of the original equations~\eqref{eqACphiF2} now suggests to work with $X$ and with the additional combinations
\bea
V &=& A \ - \ C \ , \nonumber \\
W &=& \phi \ + \ 8\,\gamma \ A \ , \label{VW_def}
\eea
whose equations of motion are simply
\bea
V '' &=& 0 \ , \nonumber \\
W'' &=& 0 \ . \label{eqs_VW}
\eea
Notice, however, the linear relation
\beq
X \,+\,(8-p)V \,-\, \frac{\gamma}{2}\, W \ = \ - \  4\left( \gamma^2 \ - \ \gamma_c^2\right) A \ ,
\eeq
so that the three variables $(V,W,X)$ are not independent in the special case $\gamma=\gamma_c$ that is relevant for the orientifold models of~\cite{susy95,bsb}, which is to be treated separately. Let us now begin our analysis from the special case $\gamma=\gamma_c$.

\subsection{\sc Vacuum Solutions with $\gamma = \gamma_c=\frac{3}{2}$}

For $\gamma = \gamma_c=\frac{3}{2}$ eq.~\eqref{X_eq} reduces to
\beq
X'' \ = \ 0 \ ,
\eeq
so that
\beq
X \ = \ \beta\,r \ + \ \beta_0 \ , \label{XA2}
\eeq
where $\beta$ and $\beta_0$ are a pair of constants, and it is now again convenient to distinguish two cases.

\subsubsection{\sc The Special Case $\beta=0$}\label{sec:tnohnobeta}

The special case $\beta=0$ results in technically simpler solutions, 
\bea
A &=& - \ \frac{T_0\, r^2}{16} \ + \ A_1\, r \ + \ A_2 \ , \nonumber \\
C &=& - \ \frac{T_0 \,r^2}{16} \ + \ C_1\, r \ + \ C_2 \ , \label{sol_ACbeta0T}
\eea
where $-\infty < r < + \infty$, the $A_i$ and $C_i$ are constants and 
\beq{}{}
T_0 \ = \ T\,e^{2\,\beta_0} \ , \label{t0}
\eeq
while the condition $X=\beta_0$ determines
\beq
\phi \ = \   \frac{3}{4}\,T_0\, r^2 \ + \ \frac{4}{3}\,\left[\beta_0 \,-\, (p+1)(A_1\,r+A_2)\,-\,(8-p)(C_1\,r+C_2)\right] \ . \label{sol_beta0T}
\eeq
Finally, in view of eq.~\eqref{XB}, the harmonic gauge translates into
\beq
B \ = \ - \ \frac{3}{4} \ \phi \,+\,\beta_0 \ = \ - \ \frac{9}{16}\,T_0\, r^2 \ + \ (p+1)(A_1\,r+A_2)+\,(8-p)(C_1\,r+C_2) \ , \label{sol_Bbeta0T}
\eeq
and taking these results into account the Hamiltonian constraint reduces to
\beq
(8-p)\left(A_1\,-\,C_1\right)^2   \ = \ \frac{9\,T_0}{ (p+1)}\ . \label{ham_beta0T}
\eeq
It can be solved consistently within this family only for $p<8$, and letting
\beq
A_1 \ = \  a \ + \ \frac{b}{2} \ , \qquad C_1 \ = \ a \ - \ \frac{b}{2} \ , 
\eeq
it determines
\beq
b \ = \  \pm\,\sqrt{\frac{9\,T_0}{(p+1)(8-p)}} \ . \label{bT0}
\eeq
Using the preceding results this family of solutions can be cast in the form
\bea
ds^2 &=& e^{\,-\,\frac{9}{8}\,T_0\,r^2 \, + \, 18\,a\,r\,-\,(7-2\,p)b\,r\, - \,
\frac{3}{2}\,\phi_2\,+\, 2\,\beta_0}\, dr^2 \,+\, e^{\,-\,\frac{T_0\,r^2}{8} \, + \, 2\,a\,r} \left(e^{\,b\,r}\, dx^2
 + \ e^{\,-\,b\,r}\, d\vec{y}^{\,2} \right) \ , \nonumber \\
e^\phi &=& e^{\phi_2}\, e^{\,\frac{3}{4}\,T_0\,r^2 \, - \, 12\,{a\,r}\,+\,\frac{2}{3}\,(7-2p){b\,r}} \ ,  \label{sol_T_a}
\eea
and, as we have stated, they only exists for $p<8$. The special tadpole-free solutions found in Section 5 of~I with $A_1=C_1$ are recovered in the limit $T_0 \to 0$.

These solutions apparently depend on three arbitrary parameters, $\beta_0$, $a$ and $\phi_2$. However, in the presence of a non--vanishing tension $T$, one can eliminate the contributions proportional to $a$ by a translation of $r$, a redefinition of $\phi_2$ and independent rescalings of the $x$ and $y$ coordinates. One is then left with $\phi_2$ and $\beta_0$, which only appears in the combination $r e^{\beta_0}$, so that a rescaling of $r$ can eliminate $\beta_0$ altogether. All in all, these solutions can be presented in the form
\bea
ds^2 &=& e^{\,-\,\frac{9}{8}\,T \,r^2\,-\,(7-2\,p)b\,r\,-\,\frac{3}{2}\,\phi_2}\, dr^2 \,+\, e^{\,-\,\frac{T\,r^2}{8}} \left(e^{\,b\,r}\, dx^2
 + \ e^{\,-\,b\,r}\, d\vec{y}^{\,2} \right)  \ , \nonumber \\
e^\phi &=& e^{\phi_2}\, e^{\,\frac{3}{4}\,T \,r^2 \,+\,\frac{2}{3}\,(7-2p){b\,r}} \ , \label{ds2phibeta0}
\eea
where $b$ is now as in eq.~\eqref{bT0} with $T_0$ replaced by $T$, so that they depend on a single parameter, $\phi_2$. Alternatively, one can choose in eqs.~\eqref{ds2phibeta0} $\beta_0=\frac{3}{4}\,\phi_2$, after removing $a$, so that only the combination $T e^{\frac{3}{2}\,\phi_2}$ enters the preceding expressions. The system is then invariant under shifts of $\phi_2$ combined with corresponding multiplicative redefinitions of $T$, as expected from the original form of the Lagrangian~\eqref{alphaE}.
In conclusion, $\phi_2$ is the only essential parameter on which this class of solutions depends.

For large values of $r$, the terms depending on the tension $T$ clearly dominate and, letting
\beq{}{}
u \ = \  e^{\,\frac{3}{4}\,T \,r^2} \ 
\eeq
turns the asymptotic form of eqs.~\eqref{ds2phibeta0} into
\bea
ds^2 &\simeq&  e^{\,-\,\frac{u}{6}} \left(dx^2 \ + \ d\vec{y}^{\,2} \right) \ + \  e^{\,-\,\frac{3}{2}\,u} \ \frac{du^2}{3\,T\, u} \nonumber \\
&\simeq&  e^{\,-\,\frac{u}{6}} \left(dx^2 \ + \ d\vec{y}^{\,2} \right) \ + \  e^{\,-\,\frac{3}{2}\,u} \ \frac{du^2}{T}  \, \nonumber \\
e^\phi &=& e^{u} \ , \label{ds2phibeta0_2}
\eea
since $\frac{3}{2} \ u \ + \ log \,u \ + \log 3\ \simeq \frac{3}{2} \ u$
for large values of $u$. The end result is the asymptotic form of the nine--dimensional Dudas--Mourad vacuum of~\cite{dm_vacuum} at its strong--coupling end, and in terms of the proper length $\xi \sim e^{\,-\,\frac{3}{4}\,u}$, 
\bea{}{}
ds^2 &\sim& \xi^\frac{2}{9}\left( dx^2 \ + \ dy^2\right) \ + \ d \xi^2 \ , \nonumber \\
e^\phi &\sim& \xi^{\,-\,\frac{4}{3}} \ , \label{dmstrong}
\eea
where $\xi=0$ at the boundaries, and where the dependence on $T$ was eliminated by a further rescaling of $\xi$.
Note that this is also, surprisingly, the isotropic strong--coupling solution obtained in~I in the absence of tension, which is briefly reviewed in Appendix~\ref{app:kasner}.

In these anisotropic spacetimes, the length $L$ in the $r$--direction is \emph{finite} in the presence of a non--vanishing tension $T$, and is given by
\beq
L \ = \ \int_{-\infty}^\infty \, e^B \,dr \ = \ \frac{4}{3} \, \sqrt{\frac{\pi}{T\,e^{\,\frac{3}{2}\,\phi_2}}}\,e^\frac{\left(7-2p\right)^2}{(p+1)(8-p)} \ .
\eeq
At the same time, the effective $(p+1)$--dimensional Planck mass can be finite if the $y$'s describe an internal torus, and then
\bea{}{}
m_{\mathrm{Pl}(p+1)}^{p-1} &=& m_{\mathrm{Pl}(10)}^8\,V_T \ e^{\,-\,\frac{3}{4}\,\phi_2} \int_{-\infty}^\infty dr \ e^{\,-\, T r^2 \,+\, 2 br(p-4)} \nonumber \\
&=& m_{\mathrm{Pl}(10)}^8\,V_T \ e^{\,-\,\frac{3}{4}\,\phi_2} \, \sqrt{\frac{\pi}{T}}\ e^{\frac{9(p-4)^2}{(p+1)(8-p)}} \ .
\eea
Consequently
\beq
m_{\mathrm{Pl}(p+1)}^{p-1} \ = \ \frac{3}{4}\, m_{\mathrm{Pl}(10)}^8\,V_T\, L \, e^{\frac{(p-5)(5p-19)}{(p+1)(8-p)}} \ ,
\eeq
and the factor is of order one for $2 \leq p \leq 6$ and of order 10 or so for $p=1$ and $p=7$.
However, in the three non--tachyonic ten--dimensional models there is strong coupling at both ends, since $T>0$.

This simple example is quite instructive. The key issue is that a positive tension translates, in general, into a {convex} dilaton profile, and increasing $\phi$ is tantamount to increasing even more the effective tension, which is determined by $T\,e^{\gamma\,\phi}$, with positive values of $\gamma$ in all cases of interest for ten--dimensional strings with broken supersymmetry. The dilaton profile has a minimum value in the interval,
and one can choose $\phi_2$ to obtain a small string coupling in a wide region away from the ends, where the effects of the tension pile up and the string coupling diverges.

These solutions are vacua of non--supersymmetric strings that have a $(p+1)$--dimensional Poincar\'e symmetry. They only exist for $p<8$, since they require two independent sets of $x$ and $y$ coordinates. In particular, for $p=3$ one gets a vacuum with four--dimensional Poincar\'e symmetry that, when combined with an internal torus, has an effective four--dimensional gravity. The special form of the Hamiltonian constraint implies that these vacua cannot be isotropic and do not admit cosmological counterparts, which would require a continuation of $A_1$ and $C_1$ to imaginary values. 
When $T = 0$, one cannot remove the constant $a$ in eqs.~\eqref{sol_T_a}, and one recovers the two--parameter family of tadpole-free solutions reviewed in Appendix~\ref{app:kasner}. 

\subsubsection{\sc Solutions with $\beta \neq 0$} \label{sec:tnohbeta}

If $\beta \neq 0$, up to a reflection of the radial coordinate one can confine the attention to positive values of $\beta$. $X$ acquires the linear term in eq.~\eqref{XA2}, and now $\beta_0$ can be absorbed by a translation of $r$. $A$ and $C$ are determined from eq.~\eqref{eqACphiF2}, and read $(-\infty < r < \infty)$
\bea
A &=& - \ \frac{T}{32} \ \frac{e^{2\,\beta\,r}}{\beta^2}\ + \ A_1\, r \ + \ A_2 \ , \nonumber \\
C &=& - \ \frac{T}{32} \ \frac{e^{2\,\beta\,r}}{\beta^2} \ + \ C_1\, r \ + \ C_2 \ . \label{ACbeta0000}
\eea
The definition of $X$ in eq.~\eqref{X_beta} then determines
\beq
\phi \ = \ \frac{3}{8}\,T \ \frac{e^{2\,\beta\,r}}{\beta^2} \ + \ \phi_1\, r \ + \ \phi_2 \ , \label{beta0000}
\eeq
with
\bea
\phi_1 &=& \frac{4}{3} \left[\beta \ - \ (p+1)A_1 \ - \ (8-p) C_1 \right] \ , \nonumber \\
\phi_2 &=& - \ \frac{4}{3} \big[ (p+1)A_2 \ + \ (8-p) C_2 \big] \ . \label{phi_12}
\eea
and finally the harmonic gauge condition $F=0$ determines
\beq
B \ = \ - \ \frac{9}{32}\,T \ \frac{e^{2\,\beta\,r}}{\beta^2}\  \ + \ \beta \, r \ - \ \frac{3}{4} \left( \phi_1\,r \ + \ \phi_2\right) \ . \label{Bbeta0000}
\eeq
Note also that $A_2$ and $C_2$ can be eliminated from the metric by rescalings in the $x$ and $y$ directions, while their combination $\phi_2$ remains in $\phi$ and $B$.

These solutions are to be considered again on the whole real $r$ axis.
All $r$--dependent terms drop out of the Hamiltonian constraint \eqref{hamc_Tnohkprime}, which reduces to a quadratic relation among the three constants $A_1$, $C_1$ and $\beta$, or equivalently $A_1$, $C_1$ and $\phi_1$:
\beq
\frac{4\,p\,\phi_1^2}{8(p+1)} \ = \ \bigg[p\,A_1 \ + (8-p)C_1\bigg]^2 \ - \ \frac{8(8-p)}{p+1}\, C_1^2 \ . \label{ham_reduced}
\eeq
This expression, which should be used for $p \neq 0$, is independent of $T$, and coincides with a corresponding result that emerged in~\cite{ms_vacuum_1} for vacua of supersymmetric strings. It can be cast in the two equivalent forms
\beq
\frac{\phi_1^2}{2} \ = \ p(p+1)A_1^2 \ + 2(p+1)(8-p) A_1 C_1 + (8-p)(7-p) C_1^2 \ , \label{ham_00case}
\eeq
and 
\beq
\frac{\phi_1^2}{2} \ + \ (p+1) \, A_1^2 \ + \ (8-p) \, C_1^2 \ = \ \left[ (p+1)A_1 \ + \ (8-p)C_1\right]^2 \ , \label{ham_00case2}
\eeq
which can also be used for $p=0$. This last form is also discussed in Appendix~\ref{app:kasner}.

Eq.~\eqref{XB} determines $\beta$, which is assumed not to vanish for the present class of solutions, as
\beq{}{}
\beta \ = \  \frac{3}{4}\, \phi_1 \ + \ (p+1)A_1 \ + \ (8-p) C_1 \ , \label{betaACphi}
\eeq
and consequently $A_1=C_1=\phi_1=0$ is not an acceptable choice. 

These results reveal an important property of these ``critical'' vacua for $\gamma=\frac{3}{2}$ and $\beta \neq 0$:  \emph{given a solution of the $T=0$ case, and thus an angle $\theta$ in Appendix~\ref{app:kasner}, eq.~\eqref{betaACphi} determines a corresponding value of $\beta$, and eqs.~\eqref{ACbeta0000} and \eqref{beta0000} determine a solution of the system in the presence of the ``critical'' tadpole potential with $\gamma=\gamma_c$.} These ``critical'' solutions are thus dressings of those for $T=0$, and yet this modification has the crucial effect of leading to $r$--intervals of finite length, as we can now see in detail.

\subsubsubsection{\sc The Dudas--Mourad Isotropic Solution for $\gamma = \frac{3}{2}$}

Let us begin our analysis from the most symmetric case, $p=8$.
In this case the Hamiltonian constraint~\eqref{ham_reduced} gives
\beq{}{}
\phi_1 \ = \ \pm 12\, A_1 \ ,
\eeq
and only the upper sign gives a non--vanishing $\beta$ in eq.~\eqref{betaACphi}. In this fashion
\beq{}{}
\beta \ = \ \frac{3}{2}\,\phi_1 \ ,
\eeq
and after letting
\beq{}{}
u \ = \ \frac{T}{6\,\phi_1^2}\ e^{3\,\phi_1\,r} \ ,
\eeq
and rescaling the $x$ coordinates, the solution takes the form
\bea{}
ds^2 &=& e^{\,-\,\frac{u}{6}} \, {u}^\frac{1}{18}\, dx^2 \,+\, \frac{2}{3\,T\,u^\frac{3}{2}}\ e^{\,-\,\frac{3}{2}\left(u+\phi_0\right)}\, {du^2} \nonumber \\
e^\phi &=& e^{\,u\,+\,{\phi}_0}\, {u}^\frac{1}{3} \ , \label{dmsol}
\eea
where $u \geq 0$ and
\beq{}{}
e^{\phi_0} \ = \ e^{\phi_2} \, \left(\frac{6\,\phi_1^2}{T}\right)^{\,\frac{1}{3}}  \ .
\eeq
One can now explore the behavior of this solution in the neighborhoods of the two boundaries, starting from the one at $u=0$. In this case, the additional substitution
\beq
\xi \ \sim \ u^\frac{1}{4} \ , 
\eeq
shows that this limiting behavior is dominated, as $\xi \to 0$,  by
\bea{}
ds^2 &\sim& \xi^\frac{2}{9} \, dx^2 \,+\, {d\xi^2} \ , \nonumber \\
e^\phi &\sim& \xi^{\frac{4}{3}} \ . \label{dmsol2_21}
\eea
This is again the isotropic \emph{tensionless} solution of~I reviewed in Appendix A, with a string coupling that vanishes at $\xi=0$.

On the other hand, for large values of $u$, where the powers have negligible effects compared to the exponential terms, the background in eqs.~\eqref{dmsol} approaches
\bea{}
ds^2 &=& e^{\,-\,\frac{u}{6}} \, dx^2 \,+\, \frac{1}{T}\ e^{\,-\,\frac{3}{2}\, u} {du^2} \ , \nonumber \\
e^\phi &=& e^{\,u} \ . \label{dmsol2}
\eea
Letting now $e^{\,-\,\frac{3}{4}\, u} \sim \xi$, or if you will in terms of the distance from the other boundary, the background takes the form
\bea{}
ds^2 &\sim& \xi^\frac{2}{9} \, dx^2 \,+\, {d\xi^2} \ , \nonumber \\
e^\phi &\sim& \xi^{-\frac{4}{3}} \ . \label{dmsol2_2}
\eea
This is again the isotropic \emph{tensionless} solution of~I reviewed in Appendix~\ref{app:kasner}, which emerges, surprisingly, in a region of strong coupling.
Remarkably, the tadpole ought to dominate at this end of the interval, but has somehow negligible effects even there for $\gamma=\gamma_c$.

Summarizing, we have described a one--parameter family of solutions depending on $\phi_0$,
as in~\cite{dm_vacuum}, which is indeed the Dudas--Mourad vacuum in a different parametrization. Its key property in that the internal space is an interval of \emph{finite length}
\beq{}{}
\ell \ = \ \sqrt{\frac{2}{3 T}}\,e^{\,-\,\frac{3}{4}\,\phi_0}\, \int_0^\infty  e^{\,-\,\frac{3}{4}\,u}\,u^{\,-\,\frac{3}{4}}\, du \ = \ \frac{2}{\sqrt{T}}\ \left(3\,e^{\,\phi_0}\right)^{\,-\,\frac{3}{4}}\, \Gamma\left(\frac{1}{4}\right) \ ,
\eeq
which decreases for increasing values of $\phi_0$. The corresponding nine--dimensional Planck mass is finite, and is given by
\bea
m_{\mathrm{Pl(9)}}^7 &=&  m_{\mathrm{Pl(10)}}^8 \, \sqrt{\frac{2}{3\,T}}\,e^{\,-\,\frac{3}{4}\,\phi_0}\,\int_0^\infty e^{\,-\,\frac{4}{3}\,u}\,u^{\,-\,\frac{5}{9}}\, du \ = \ m_{\mathrm{Pl(10)}}^8 \, \frac{3^{\frac{3}{4}}\,\ell}{\sqrt{6}\,\Gamma\left(\frac{1}{4}\right)}\ \left(\frac{3}{4}\right)^\frac{4}{9} \, \Gamma\left(\frac{4}{9}\right) \nonumber \\
&\simeq& 0.09 \ m_{\mathrm{Pl(10)}}^8 \, \ell \ .
\eea

\subsubsubsection{\sc The Anisotropic $p<8$ Cases}

As in~I and in Appendix~\ref{app:kasner}, it is now convenient to define
\beq{}{}
\alpha_A \ = \ \frac{A_1}{\mu} \ , \qquad \alpha_C \ = \ \frac{C_1}{\mu} , \qquad \alpha_\phi \ = \ \frac{\phi_1}{\mu} \ ,
\eeq
which are determined in terms on angle $\theta$ in eqs.~\eqref{param_theta}, and then
\beq{}{}
\beta \ = \ \mu \left( 1 \ + \ \sin\theta \right) \ .
\eeq
Note that the point $\theta=\pi$ must be left out in the current treatment, since $\beta$ vanishes there.
Consequently the solutions can be parametrized as
\bea
A &=& - \ \frac{T}{32\,\mu^2} \ \frac{e^{2\left(1 \ + \ \sin\theta\right)\mu\,r}}{\left(1 \ + \ \sin\theta\right)^2}\ + \ \alpha_A(\theta)\, \mu\, r \ + \ A_2 \ , \nonumber \\
B &=& - \ \frac{9}{32}\,\frac{T}{\mu^2} \ \frac{e^{2\left(1 \ + \ \sin\theta\right)\mu\,r}}{\left(1 \ + \ \sin\theta\right)^2} \  \ + \ \mu\,r \ - \ \frac{3}{4} \, \phi_2 \ , \nonumber \\
C &=& - \ \frac{T}{32\,\mu^2} \ \frac{e^{2\left(1 \ + \ \sin\theta\right)\mu\,r}}{\left(1 \ + \ \sin\theta\right)^2} \ + \ \alpha_C(\theta)\, \mu\, r \ + \ C_2 \ , \label{ACbeta0000p} \nonumber \\
\phi &=& \frac{3}{8}\,\frac{T}{\mu^2} \ \frac{e^{2\left(1 \ + \ \sin\theta\right)\mu\,r}}{\left(1 \ + \ \sin\theta\right)^2} \ + \ \frac{4}{3}\,\sin\theta \, \mu\, r \ + \ \phi_2 \ , \label{Tanistrop}
\eea
where
\beq
\phi_2 \ = \  - \ \frac{4}{3} \big[ (p+1)A_2 \ + \ (8-p) C_2 \big] \ .
\eeq
As for $p=8$, these expressions suggest a convenient change of variable,
\beq{}{}
u \ = \ u_0 \ e^{2\left(1 \ + \ \sin\theta\right)\mu\,r} \ , \label{ur}
\eeq
where
\beq{}{}
u_0 \ = \ \frac{3}{8}\,\frac{T}{\mu^2 \left(1 \ + \ \sin\theta\right)^2 }\ . \label{u0T}
\eeq
After rescaling the $x$ and $y$ coordinates, letting also
\beq{}{}
e^{\phi_0} \ = \ e^{\phi_2} \ u_0^{\,-\,\frac{2\,\sin\theta}{1 \ + \ \sin\theta}} \ ,
\eeq
the result can be finally cast in the form
\bea{}
ds^2 &=& e^{\,-\,\frac{u}{6}} \ {u}^\frac{\alpha_A(\theta)}{1 \ + \ \sin\theta}\, dx^2 \ + \ \frac{2}{3\,T} \, e^{\,-\,\frac{3}{2}\left(u+\phi_0\right)} {u}^{\frac{1}{1 \ + \ \sin\theta}\,-\,2} \, du^2 \ +\ e^{\,-\,\frac{u}{6}} \ {u}^\frac{\alpha_C(\theta)}{1 \ + \ \sin\theta}\, d \vec{y}^{\,2} \ , \nonumber \\
e^\phi &=& e^{u\,+\,\phi_0} u^\frac{2\,\sin\theta}{3 \left(1 \ + \ \sin\theta\right)} \ , \label{metricu}
\eea
where $0 \leq u < \infty$ and $\alpha_A\left(\theta\right)$, and $\alpha_C\left(\theta\right)$ can be found in eqs.~\eqref{param_theta}. Note that $\mu$ has disappeared, and these vacua have a two--dimensional moduli space: they are characterized by  $\phi_0$ and $\theta \neq \pi$. 

For small values of $u$ one can ignore the exponential terms and eqs.~\eqref{metricu}, and 
letting 
\beq
\xi \sim u^\frac{1}{2\left(1 \ + \ \sin\theta \right)} \ ,
\eeq
the solutions take the form
\bea{}
ds^2 &\sim & \xi^{2 \alpha_A(\theta)} \, dx^2 \,+\, \xi^{2 \alpha_C(\theta)} \, d\vec{y}^2 \,+\, {d\xi^2} \ , \nonumber \\
e^\phi &\sim& \xi^{\frac{4}{3}\, \sin \theta} \ . \label{dmsol2_2s}
\eea
Once more, this limiting behavior is captured by the \emph{anisotropic tensionless} solutions of~\cite{ms_vacuum_1} that are reviewed in Appendix~\ref{app:kasner}, independently of the limiting behavior of the string coupling, which can be zero, finite or infinite depending on the value of $\theta$. 

For large values of $u$, where the exponential terms in eqs.~\eqref{Tanistrop} dominate, the solutions approach the nine--dimensional Dudas--Mourad vacuum of~\cite{dm_vacuum} given in eq.~\eqref{dmsol2}, whose asymptotics is again dominated by the strong--coupling tensionless solution of~I reviewed in Appendix~\ref{app:kasner}.
To reiterate, the behavior of the background near the two boundaries is captured by tensionless solutions, independently of the limiting behavior of the string coupling. At one end the limiting form of the metric is generally anisotropic while at the other end it is isotropic, and the string coupling diverges at least at one end.
\begin{figure}[ht]
\centering
\begin{tabular}{cc}
\includegraphics[width=60mm]{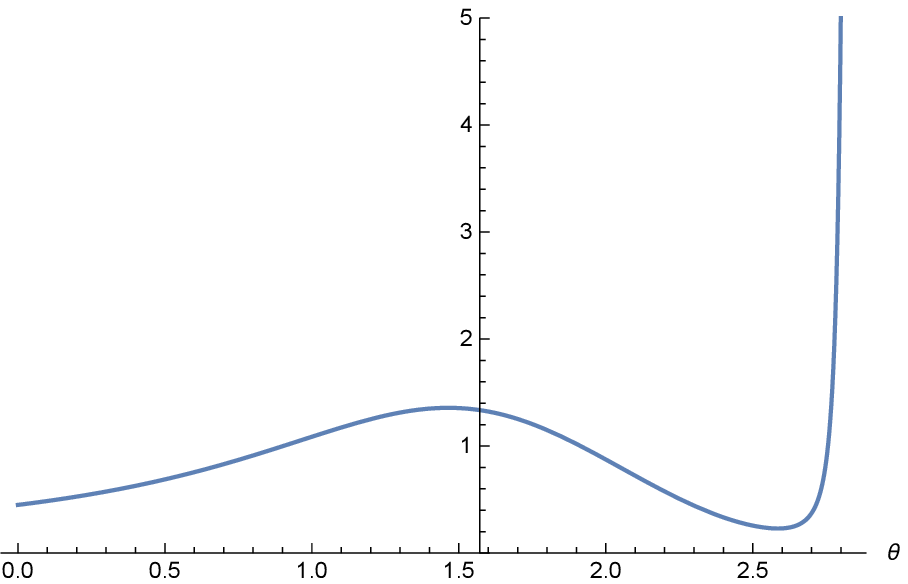} &
\includegraphics[width=60mm]{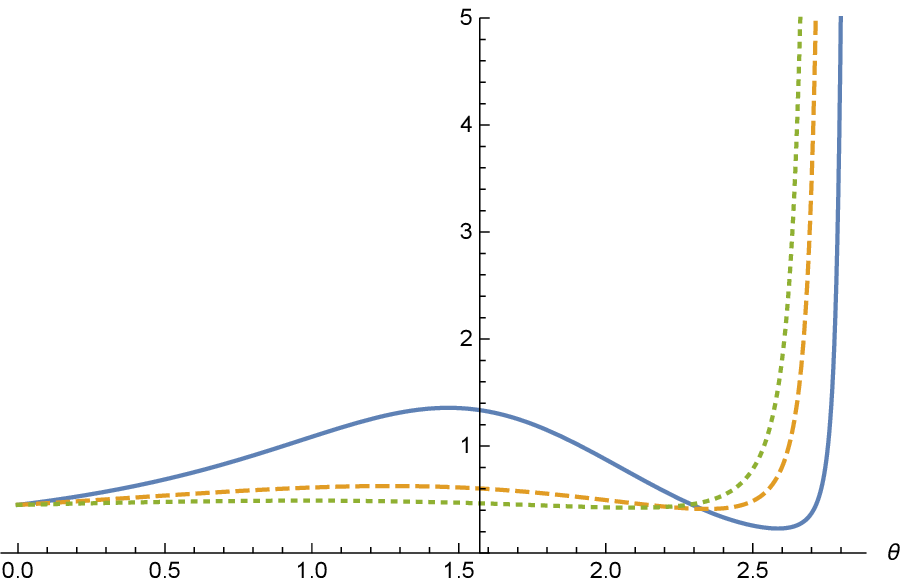} \\
\end{tabular}
\caption{\small Left panel: the length of the interval as a function of the angle $\theta$ that parametrizes the ellipse in eq.~\eqref{ellipse}. Right panel: the ratios of the effective $(p+1)$--dimensional Planck mass to the product of the length $L$ and the volume of the internal torus for $p=1,4,7$ (solid, dashed, dotted) as functions of $\theta$. The $p=1$ case has a bump but the other curves fall essentially on top of one another. The string coupling is finite at the origin in the left portion of these curves and is infinite in the right portion.}
\label{fig:figTp}
\end{figure}

The singularities at $u=0$ and $u= + \infty$ are separated by a distance
\beq{}{}
L \ = \ \sqrt{\frac{2}{3\,T}}\ e^{\,-\,\frac{3}{4}\,\phi_0}\left( \frac{4}{3}\right)^{\frac{1}{2\left((1 \ + \ \sin\theta\right)}} \ \Gamma\left[\frac{1}{2\left(1 \ + \ \sin\theta\right)} \right] \ ,
\eeq
which is finite in the allowed region, where $(1 \ + \ \sin\theta) > 0$,
and if the $y$ coordinates describe a torus of parametric volume $V_T$, the reduced Planck mass, which is determined by
\bea
m_{\mathrm{Pl}(p+1)}^{p-1} &=& m_{\mathrm{Pl}(p+1)}^8 V_T \, \sqrt{\frac{2}{3\,T}}\ e^{\,-\,\frac{3}{4}\,\phi_0}\,\int du \ e^{\,-\,\frac{4}{3}\,u} \ u^{\frac{1\,-\,\alpha_A}{\left(1+ \frac{3}{4}\,\alpha_\phi\right)} \, -\, 1} \nonumber \\
&=& m_{\mathrm{Pl}(p+1)}^8 \, V_T \sqrt{\frac{2}{3T}}\,e^{\,-\,\frac{3}{4}\,\phi_0}\, \left(\frac{3}{4}\right)^\frac{1-\alpha_A}{1+ \frac{3}{4}\,\alpha_\phi}\ \Gamma\left[\frac{1-\alpha_A}{1+\frac{3}{4}\,\alpha_\phi}\right] \nonumber \\
&=& m_{\mathrm{Pl}(p+1)}^8 \, V_T \, L \left(\frac{3}{4}\right)^\frac{1+2\left(1-\alpha_A\right)}{2\left(1+ \frac{3}{4}\,\alpha_\phi\right)} \frac{\Gamma\left[\frac{1-\alpha_A}{1+\frac{3}{4}\,\alpha_\phi}\right]}{\Gamma\left[\frac{1}{2\left(1+ \frac{3}{4}\,\alpha_\phi\right)} \right] } \ ,
\eea
is always finite in the region $\sin\theta> -1$, where they apply. Some examples are displayed in fig.~\ref{fig:figTp}. These solutions are vastly different from those obtained for $\beta=0$ in Section~\ref{sec:tnohnobeta}.

\subsection{\sc Vacuum Solutions for $\gamma \neq \frac{3}{2}$}
In this case one can work with the three variables $(V,W,X)$, defined in eqs.~\eqref{X_beta} and \eqref{VW_def}, which are now independent and determine $(A,C,\phi)$ according to
\bea
A &=&   - \ \frac{\Big[ X \ + \ (8-p)V \ - \  \frac{\gamma}{2}\, W \Big]}{4 \left(\gamma^2 \ - \ \gamma_c^2\right)}  \ , \nonumber \\
C &=&   - \ \frac{\Big[ X \ + \ \left[8-p\ + \ 4 \left(\gamma^2 \ - \ \gamma_c^2\right)\right]V \ - \  \frac{\gamma}{2}\, W \Big]  }{4 \left(\gamma^2 \ - \ \gamma_c^2\right)}   \ , \nonumber \\
\phi &=&  \frac{\Big[ 2\,\gamma\,X \ + \ 2\,\gamma\,(8-p)\,V \ - \  \gamma_c^2\, W \Big] }{\left(\gamma^2 \ - \ \gamma_c^2\right)}   \ .
\eea
One can write the solutions of eqs.~\eqref{eqs_VW} for $V$ and $W$ in the convenient form
\bea
V &=& \frac{v_1}{2}\, r \ + \ \frac{v_2}{2} \ , \nonumber \\
W &=& \left[ \left(1\ - \ \frac{\gamma^2}{\gamma_c^2} \right) \left(\phi_1\,r\,+\,\phi_2\right)  \ + \ \frac{\gamma}{\gamma_c^2} (8-p) \left(v_1\,r\,+\,v_2\right) \right] \ ,
\eea
where $v_1$, $v_2$, $\phi_1$ and $\phi_2$ are integration constants. Consequently
\bea
A &=&  \frac{1}{9} \left\{ - \ \frac{\gamma_c^2\, X}{\left(\gamma^2\,-\,\gamma_c^2\right)} \ + \ \frac{1}{2}\,\Big[(8-p)v_1 \,-\, {\gamma}\, \phi_1\Big]r \right\}\ + \ a_2 ,\nonumber \\
C&=& \frac{1}{9} \left\{- \ \frac{\gamma_c^2\,X}{\left(\gamma^2\,-\,\gamma_c^2\right)} \ -\ \frac{1}{2}\,\Big[(p+1)v_1\,+\,{\gamma}\, \phi_1\Big] r \right\} \ + \ c_2 \ , \nonumber \\
\phi &=& 2\, \frac{\gamma \,X}{\left(\gamma^2\,-\,\gamma_c^2\right)} \ + \ \phi_1\, r \,+\, \phi_2  \ , \label{ACphiX}
\eea
where $a_2$ and $c_2$ are a pair of constants determined by $v_2$ and $\phi_2$, whose detailed form does not play a role, since they can be removed from the metric by rescaling the $x$ and $y$ coordinates, and $B$ is determined as
\beq{}{}{}{}
B \ = \ X \ - \ \frac{\gamma}{2}\, \phi\ = \ - \ \frac{\gamma_c^2\, X}{\gamma^2 \,-\,\gamma_c^2} \ - \ \frac{\gamma}{2} \left( \phi_1\, r \,+\, \phi_2\right)\ .
\eeq

Finally, $X$ is determined by the Hamiltonian constraint, which reads
\beq
 (X')^2 \ = \ \epsilon\,\Delta^2\ e^{2\,X} \ + \ E \ , \label{ham_less}
\eeq

where
\bea
\Delta^2 &=& \frac{T}{2}\ \left|\gamma^2\,-\,\gamma_c^2\right| \ \ , \qquad
\epsilon \ = \  {\mathrm{sign}}\left(\gamma^2\,-\,\gamma_c^2\right) \ , \nonumber \\
E &=& \frac{1}{4} \bigg\{ \left(\frac{\gamma^2}{\gamma_c^2} \ - \ 1 \right)^2 \,\gamma_c^2\,\phi_1^2 \,- \,   \left(\frac{\gamma^2}{\gamma_c^2} \ - \ 1 \right) (p+1)\left(1 -\frac{p}{8}\right)\ v_1^2  \bigg\} \ , \label{dynamical_T}
\eea
and the two case $\gamma< \gamma_c$ and $\gamma > \gamma_c$ are to be discussed separately.

\subsubsection{\sc Vacuum Solutions for $\gamma < \frac{3}{2}$} \label{sec:gammalessgc}

Although this range of values is not realized in ten-dimensional String Theory, it is simple and instructive to include it in our analysis.
For $0 < \gamma < \gamma_c=\frac{3}{2}$, referring to Appendix~\ref{app:deq} one can see that the problem reduces to the one--dimensional dynamics of a Newtonian particle subject to a positive exponential potential, so that its energy $E$ must be positive. Up to a translation of the radial variable $r$, we have
\beq
X \ = \ -  \log\left[\Delta\,\rho \, \cosh \left(\frac{r}{\rho}\right) \right]\ , \qquad (- \infty < r < + \infty) \ ,
\eeq
where
\beq
\rho = \frac{1}{\sqrt{E}} \ .
\eeq
Letting now
\beq
\lambda \ = \ \frac{1}{9\left(1 \ -\ \frac{\gamma^2}{\gamma_c^2} \right)}\ ,  \label{lambda_sub}
\eeq
the allowed values of $v_1$ and $\phi_1$ can be parametrized via an angle $\eta$, as
\bea
\phi_1 \ \equiv \ \frac{\widetilde{\phi}_1}{\rho} &=& - \ \frac{18 \, \lambda}{\rho\,\gamma_c} \ \cos \eta \ , \nonumber \\
v_1  \ \equiv \ \frac{\widetilde{v}_1}{\rho} &=& \frac{12}{\rho} \, \sqrt{\frac{2\,\lambda}{(p+1)(8-p)}}\, \sin \eta  \ , \label{trigo_phichi_less}
\eea
and the solution reads
\bea
ds^2 &=& e^{\,-\, \frac{\gamma\,\phi_1\,r}{9}}\ \frac{e^{\,\frac{(8-p)v_1\,r}{9}}\,dx^2 \ + \ e^{\,\frac{-(p+1)v_1\,r}{9}}\,dy^2}{\left[\Delta\,\rho\,\cosh\left(\frac{r}{\rho}\right)\right]^{2\,\lambda}}  \ +\ \frac{e^{\,-\,\gamma\left(\phi_1\,r\,+\,\phi_2\right)}\ dr^2}{\left[\Delta\,\rho\,\cosh\left(\frac{r}{\rho}\right)\right]^{18\,\lambda}} \ , \nonumber \\
e^{\phi} &=&  \left[\Delta\,\rho\,\cosh\left(\frac{r}{\rho}\right)\right]^{\frac{18\,\gamma}{\gamma_c^2}\,\lambda} \ e^{\phi_1\,r\,+\,\phi_2} \ . \label{solgammalessgc}
\eea
Notice that $\eta$ parametrizes the anisotropy of this family of solutions, which become $9+1$ dimensional when it  vanishes. We shall return to this case in Section~\ref{sec:dm_9}.

In fact, contrary to what eqs.~\eqref{solgammalessgc} may suggest, finite values of $\rho$ can be removed also from these solutions, rescaling $r$ and the $x$ and $y$ coordinates and redefining $\phi_2$, while taking into account the definition of $\lambda$ in eq.~\eqref{lambda_sub}. This leads to
\bea
ds^2 &=& e^{\,-\, \frac{\gamma\,\widetilde{\phi}_1\,r}{9}}\ \frac{e^{\,\frac{(8-p)\widetilde{v}_1\,r}{9}}\,dx^2 \ + \ e^{\,\frac{-(p+1)\widetilde{v}_1\,r}{9}}\,dy^2}{\left[\Delta\,\cosh(r)\right]^{2\,\lambda}}  \ +\ \frac{e^{\,-\,\gamma\left(\widetilde{\phi}_1\,r\,+\,\widetilde{\phi}_2\right)}\ dr^2}{\left[\Delta\,\cosh(r)\right]^{18\,\lambda}} \ , \nonumber \\
e^{\phi} &=&  \left[\Delta\,\cosh(r)\right]^{\frac{18\,\gamma}{\gamma_c^2}\,\lambda} \ e^{\widetilde{\phi}_1\,r\,+\,\widetilde{\phi}_2} \ , \label{solgammalessgc2}
\eea
where $\widetilde{v}_1$ and $\widetilde{\phi}_1$ are defined in eqs.~\eqref{trigo_phichi_less} and depend on $\eta$. This family of solutions thus depends on the two parameters $\eta$ and $\widetilde{\phi}_2$. On the other hand, the $ \rho \to \infty$ limit in eqs.~\eqref{solgammalessgc} is singular.

As $r \to \pm \infty$, the exponential potential becomes negligible in eq.~\eqref{ham_less}, and
\beq
X \ \sim \ \mp \ r \ .
\eeq
Taking into account the definitions of $\lambda$ and $\widetilde{\phi}_1$ in eqs.~\eqref{trigo_phichi_less} and \eqref{lambda_sub}, one can then conclude that, as $r \to \pm \infty$,
\beq
e^B \, dr \ \sim \ e^{\,-\,9\,\lambda\left[\pm\,1 \ - \ \frac{\gamma}{\gamma_c}\, \cos \eta \, \right]r}\, dr \ , \label{B_spatial}
\eeq
so that the interval has a finite length for any choice of $\eta$. At the same time,
\beq
\phi \ \sim \ \frac{18\,\lambda\, r}{\gamma_c} \left( \pm\,\frac{\gamma}{\gamma_c}\ - \ \cos \eta \right) \ . \label{phi_spatial}
\eeq
so that the string coupling  diverges only at one end and vanishes at the other if
\beq
\left|\cos\,\eta\right| \ >  \frac{\gamma}{\gamma_c} \ ,
\eeq
while it diverges at both ends otherwise. Moreover, the reduced Planck mass is finite, if the internal space is a torus, since
\beq{}{}
1 \,-\,\epsilon\,\frac{\gamma}{\gamma_c}\,\cos\eta \ +\ \frac{1}{24} \frac{\epsilon}{\sqrt{\lambda}}\,\sin\eta \, \sqrt{\frac{8(8-p)}{p+1}}
\eeq
is positive for $0<p<8$.

In terms of the proper lengths, which behave as
\beq{}{}{}{}
\xi \ \sim \ e^{\,-\,9\,\lambda\left[\pm\,1 \ - \ \frac{\gamma}{\gamma_c}\, \cos \eta \, \right]r} 
\eeq
close to the two ends of the interval, so that they vanish there in all cases, the background approaches
\bea{}{}
ds^2 &\sim& \xi^{2\alpha_A^\pm}\, dx^2 \ + \ \xi^{2\alpha_C^\pm}\, d\vec{y}^2 \ + \ d \xi^2 \, \nonumber \\
e^\phi &\sim& \xi^{\alpha_\phi^\pm} \ ,
\eea
where the exponents are
\bea{}{}
2\,\alpha_A^\pm &=& \frac{2}{9}\ \frac{\left[\gamma_c \ \mp \ {\gamma}\, \cos \eta \ \pm \ \frac{2}{3}\,\gamma_c\,\sqrt{\frac{2\left(8-p\right)}{\lambda\left(p+1\right)}}\ \sin\eta\right]}{\left(\gamma_c \ \mp \ {\gamma}\, \cos \eta \, \right)}\ , \nonumber \\
2\,\alpha_C^\pm &=& \frac{2}{9}\ \frac{\left[\gamma_c \ \mp \ {\gamma}\, \cos \eta \ \mp \ \frac{2}{3}\,\gamma_c\,\sqrt{\frac{2\left(p+1\right)}{\lambda\left(8-p\right)}}\ \sin\eta\right]}{\left(\gamma_c \ \mp \ {\gamma}\, \cos \eta \, \right)}\ , \nonumber \\
\alpha_\phi^\pm &=& - \  \frac{2}{\gamma_c}\  \frac{\left(\gamma \ \mp \ \gamma_c\,\cos\eta\right)}{\left(\gamma_c \ \mp \ {\gamma}\, \cos \eta \, \right)} \ . \label{alphagammaless}
\eea
Taking into account that $\gamma_c=\frac{3}{2}$, one thus finds that the two conditions
\bea{}{}
&&(p+1)\alpha_A^\pm \ + \ (8-p)\alpha_C^\pm \ = \ 1 \ , \nonumber \\
&&(p+1)\alpha_A^\pm{}^2 \ + \ (8-p)\alpha_C^\pm{}^2 \ + \ \frac{1}{2}\,\alpha_\phi^\pm{}^2 \ = \ 1 \label{cond_T0}
\eea
hold, so that the two asymptotic regions are again described by Kasner--like flux--free backgrounds for the $T=0$ case, with parameters $\theta^\pm$ that depend on $\gamma$ and $\eta$ as
\beq{}{}{}{}
\sin\theta^\pm \ = \ - \  \frac{\left(\gamma \ \mp \ \gamma_c\,\cos\eta\right)}{\left(\gamma_c \ \mp \ {\gamma}\, \cos \eta \, \right)} \ .
\eeq
One can verify that, in all cases, the tadpole potential $V(\phi) \sim \xi^{\gamma\,\alpha_\phi^\pm}$ is sub-dominant as $\xi \to 0$ with respect to the scalar kinetic term.

In brief, for all values of $\gamma < \gamma_c$ there are two asymptotic regions, where the limiting behavior is dominated again by particular flux--free solutions of the $T=0$ system. The behavior at one end determines $\eta$, and thus also the behavior at the other end. 

\subsubsection{\sc Vacuum Solutions for $\gamma > \frac{3}{2}$}

For $\gamma>\gamma_c$, a range that is directly relevant for the $SO(16)\times SO(16)$ string, the potential is an inverted exponential and the sign of the energy $E$ in eqs.~\eqref{ham_less} is arbitrary, so that one must distinguish three cases.
\begin{itemize}

\item { If the energy ${{E=0}}$}, a value which obtains if
    \beq
    v_1 \ = \ \pm \ \gamma_c\,\phi_1\, \sqrt{\frac{8}{(p+1)(8-p)} \left(
    \frac{\gamma^2}{\gamma_c^2} \ - \ 1\right)} \ , \label{v1phi1}
    \eeq
 one can see from Appendix~\ref{app:deq} that, up to a translation of the radial variable $r$, $X$ is given by
\beq
X \ = \ - \ \log\big(\Delta \,r \big) \ ,  \qquad ( 0 < r < \infty ) \ .
\eeq
The solutions then read
\bea
ds^2 &=& e^{\,-\, \frac{\gamma\,\phi_1\,r}{9}}\ \left(\Delta\,r\right)^{2|\lambda|}\,\left(e^{\,\frac{(8-p)v_1\,r}{9}}\,dx^2 \ + \ e^{\,\frac{-(p+1)v_1\,r}{9}}\,dy^2\right) \nonumber \\ &+& \left(\Delta\,r\right)^{18|\lambda|}\, e^{\,-\,\gamma\left(\phi_1\,r\,+\,\phi_2\right)}\ dr^2 \ , \nonumber \\
e^{\phi} &=&  \left(\Delta\,r\right)^{\,-\,\frac{18\,\gamma}{\gamma_c^2}\,|\lambda|} \ e^{\phi_1\,r\,+\,\phi_2} \ . \label{gammamoree0}
\eea

There are singularities at the two ends of the range of $r$, $r=0$ and $r=\infty$, which are separated by a finite (infinite) distance if $\phi_1 > 0$ ($\phi_1 \leq 0$). In the former case the reduced Planck mass is also  finite if the internal space is a torus, while in the latter case it is infinite. There is always strong coupling at one end ($r=0$), and there is weak coupling at the other end if $\phi_1 \leq 0$ and strong coupling if $\phi_1 > 0$. 
In the vicinity of the origin the background approaches
\bea 
ds^2 &=&  (\Delta\, r)^{2|\lambda|}\, (dx^2 \ +\,dy^2)   + \left(\Delta\,r\right)^{18|\lambda|}\,\ dr^2 \ , \nonumber \\
e^{\phi} &=&  \left(\Delta\,r\right)^{\,-\,\frac{18\,\gamma}{\gamma_c^2}\,|\lambda|} \ e^{\phi_2} \ , \label{E0r}
\eea
which is the isotropic $\phi_1=0$ solution.
In terms of the proper length $\xi=r^{9|\lambda|+1}$, it becomes
\bea 
ds^2&=&\xi^{\frac{2\,\gamma_c^2}{9\,\gamma^2}}(dx^2+dy^2)+d\xi^2 \ , \nonumber \\
e^\phi&=& \xi^{\,-\,\frac{2}{\gamma}} \ e^{\phi_2} \ . \label{phi10}
\eea
 {This is once more a Kasner--like behavior but, in contrast with what we saw in the preceding cases, it is not the tensionless behavior of eqs.~\eqref{spontaneous_r_isotropic}, which is only approached as $\gamma \to \gamma_c$, and thus for $\lambda \to - \infty$}.

On the other hand, as $r \to +\infty$ for $\phi_1 \neq 0$, the limiting form of the background is
\bea{}{}
ds^2 &\sim& \xi^{2\alpha_A} \ dx^2 \ + \  \xi^{2\alpha_C} \ d\vec{y}^2 \ + d \xi^2 \ , \nonumber \\
e^\phi &\sim& \xi^{\alpha_\phi} \ , \label{limE0}
\eea
where
\beq{}{}{}{}
\xi \ \sim \ e^{\,-\,\frac{\gamma}{2}\,\phi_1\,r} \ , \label{xirmore}
\eeq
so that $\xi=0$ at the right boundary when $\phi_1>0$ and $\xi=\infty$ if $\phi_1<0$, and
\bea{}{}
\alpha_A &=& \frac{1}{9}\left[ 1 \ \mp \  \sqrt{\frac{8(8-p)}{(p+1)}\left(1 \ - \ 
    \frac{\gamma_c^2}{\gamma^2} \right)}\right]  \ , \nonumber \\
\alpha_C &=&  \frac{1}{9}\left[ 1 \ \pm \ \sqrt{\frac{8(p+1)}{(8-p)}\left(1 \ - \ 
    \frac{\gamma_c^2}{\gamma^2} \right)}\right] \ , \nonumber \\
\alpha_\phi &=& - \ \frac{2}{\gamma} \ .
\eea
Taking into account that $\gamma_c=\frac{3}{2}$, one can verify that
these expressions are of the form~\eqref{param_theta}, so that the $r \to \infty$ limit is dominated by an anisotropic tensionless solution with
\beq{}{}
 \sin \theta \ = \ - \ \frac{\gamma_c}{\gamma} \ ,
\eeq
independently of whether the coupling is weak or strong there,
and the two values of $\cos\theta$ correspond to the two branches in eq.~\eqref{v1phi1}.
Finally, for $\phi_1=0$ the limiting behavior is captured by eqs.~\eqref{phi10}.
\item {If the energy ${{E > 0}}$}, as discussed in Appendix~\ref{app:deq} one can choose the solution
\beq
X \ = \ -  \log\left[\Delta\,\rho \, \sinh \left(\frac{r}{\rho}\right) \right] \ \ \ (r > 0) \ 
\eeq
and work in the region $0< r< \infty$. There are two branches of solutions of the Hamiltonian constraint in eqs.~\eqref{dynamical_T} parametrized by a real variable $\zeta$,
\bea
\phi_1 &=&  \pm\,\frac{18|\,\lambda|}{\rho\,\gamma_c} \ \cosh \zeta \ = \ \frac{1}{\rho}\, \tilde{\phi}_1\ , \nonumber\\
v_1 &=& \frac{12}{\rho} \ \sinh \zeta\ \sqrt{\frac{2\,|\lambda|}{(p+1)(8-p)}}\ = \ \frac{1}{\rho}\, \tilde{v}_1  \ , \label{hamiltonian_hyperbolic}
\eea
where $\lambda$ is defined in eq.~\eqref{lambda_sub},
and the background reads
\bea
ds^2 &=& e^{\,-\, \frac{\gamma\,\phi_1\,r}{9}}\ {\left[\Delta\,\rho\,\sinh\left(\frac{r}{\rho}\right)\right]^{2\,|\lambda|}} \ \left(e^{\,\frac{(8-p)v_1\,r}{9}}dx^2 \ + \ e^{\,\frac{-(p+1)v_1\,r}{9}}\,dy^2\right)
  \nonumber \\ &+& \left[\Delta\,\rho\,\sinh\left(\frac{r}{\rho}\right)\right]^{18\,|\lambda|} \ {e^{\,-\,\gamma\left(\phi_1\,r\,+\,\phi_2\right)}\ dr^2} \ , \nonumber \\
e^{\phi} &=& \ \frac{e^{\phi_1\,r\,+\,\phi_2}}{\left[\Delta\,\rho\,\sinh\left(\frac{r}{\rho}\right)\right]^{\frac{18\,\gamma}{\gamma_c^2}\,|\lambda|} } \ .
\eea
Notice that $\rho$ can be removed from these solutions, rescaling $r$ and the $x$ and $y$ coordinates, and redefining $\phi_2$. As a result, this is a two--parameter family of solutions,  depending on $\phi_2$  and $\zeta$.

There are two singularities, at $r=0$ and at $r=+\infty$. Near $r=0$ the background approaches the $E=0$ solution in eqs.~\eqref{E0r} or, in terms of the proper length, in eqs.~\eqref{phi10}. The string coupling diverges at $r=0$, where the dominant behavior is isotropic and sensitive to the tension $T$. 

At the other singularity, as $r \to \infty$, the background can be conveniently expressed in terms of the proper length
\beq{}{}
\xi \ \sim \ e^{\frac{9\,|\lambda|r}{\rho}\left(1 \mp \frac{\gamma}{\gamma_c}\,\cosh \zeta \right)} \ ,
\eeq
which diverges in the lower branch and tends to zero in the upper branch, 
and reads
\bea{}{}
ds^2 &\sim& \xi^{2\alpha_A} \ dx^2 \ + \  \xi^{2\alpha_C} \ d\vec{y}^2 \ + d \xi^2 \ , \nonumber \\
e^\phi &\sim& \xi^{\alpha_\phi} \ , \label{limEpos}
\eea
where
\bea{}{}
\alpha_A &=&\frac{1}{9} \left[1 \ + \ \frac{\frac{2}{3}\,\sinh\zeta\, \sqrt{\frac{2(8-p)}{|\lambda|(p+1)}}}{1 \ \mp \ \frac{\gamma}{\gamma_c}\, \cosh \zeta} \right] \ , \nonumber \\
\alpha_C &=& \frac{1}{9} \left[1 \ - \ \frac{\frac{2}{3}\,\sinh\zeta\, \sqrt{\frac{2(p+1)}{|\lambda|(8-p)}}}{1 \ \mp \ \frac{\gamma}{\gamma_c}\, \cosh \zeta} \right] \ , \nonumber \\
\alpha_\phi &=& \frac{4}{3}\,\frac{-\,\cosh \zeta \ \pm \ \frac{\gamma}{\gamma_c}}{\left(\frac{\gamma}{\gamma_c}\, \cosh \zeta \ \mp \ 1\right)} \label{alphaACphi_epos}
 \ .
\eea
In all cases, these coefficients satisfy eqs.~\eqref{cond_T0}, so that the limiting behavior at the right end is captured, once more, by the tensionless flux--free solutions of~I reviewed in Appendix~\ref{app:kasner}, with
\beq
\sin\theta \ = \  \frac{-\,\cosh \zeta \ \pm \ \frac{\gamma}{\gamma_c}}{\frac{\gamma}{\gamma_c}\, \cosh \zeta \ \mp \ 1}  \ , \qquad \cos\theta \ = \ \frac{1}{3\sqrt{\left|\lambda\right|}} \ \frac{\sinh\zeta}{1 \ \mp \ \frac{\gamma}{\gamma_c}\, \cosh\zeta} \ .
\eeq
This also happened for $\gamma< \gamma_c$, but in that case this type of behavior was also approached at the other end, while for $\gamma> \gamma_c$ the limiting behavior at the other end is not captured by tension--free solutions.

In the first branch (upper sign) $\xi \to 0$ as $r \to \infty$ and the length is finite. Moreover, $\alpha_\phi$ can have both signs. When $\alpha_\phi \geq0$ the solutions exhibit a novel feature: a finite length in the $r$ direction can be accompanied by a string coupling that is also finite as $r \to \infty$. This occurs when
\beq
\cosh \zeta \ \le \ \frac{\gamma}{\gamma_c} \ .
\eeq
The solutions in the second branch (lower sign) have an \emph{infinite length} in the $r$ direction, so that eqs.~\eqref{limEpos} apply for large values of $\xi$. In this case $\alpha_\phi<0$, so that the string coupling tends to zero at this end.

\item  {If the energy ${{E < 0}}$}, letting $E = \,-\, \frac{1}{\rho^2}$, one can parametrize $\phi_1$ and $v_1$ according to
\bea
\phi_1 &=&  \frac{18\,|\lambda|}{\rho\,\gamma_c} \ \sinh \zeta \ , \nonumber\\
v_1 &=& \pm \ \frac{12}{\rho} \ \cosh \zeta\ \sqrt{\frac{2\,|\lambda|}{(p+1)(8-p)}}  \ , \label{hamiltonian_hyperbolic12}
\eea
so that $v_1$ cannot vanish and the solutions in this family are all \emph{anisotropic}. In this case, using the results in Appendix~\ref{app:deq}, $X$ can be presented in the form
\beq
X \ = \  -  \log\left[\Delta\,\rho \, \sin \left(\frac{r}{\rho}\right) \right]\ ,
\eeq
so that $0 < r < {\pi\,\rho}$, and the metric and the string coupling read
\bea
ds^2 &=& e^{\,-\, \frac{\gamma\,\phi_1\,r}{9}}\ {\left[\Delta\,\rho\,\sin\left(\frac{r}{\rho}\right)\right]^{2\,|\lambda|}} \ \left(e^{\,\frac{(8-p)v_1\,r}{9}}dx^2 \ + \ e^{\,\frac{-(p+1)v_1\,r}{9}}\,dy^2\right)
  \nonumber \\ &+& \left[\Delta\,\rho\,\sin\left(\frac{r}{\rho}\right)\right]^{18\,|\lambda|} \ {e^{\,-\,\gamma\left(\phi_1\,r\,+\,\phi_2\right)}\ dr^2} \ , \nonumber \\
e^{\phi} &=& \ \frac{e^{\phi_1\,r\,+\,\phi_2}}{\left[\Delta\,\rho\,\sin\left(\frac{r}{\rho}\right)\right]^{\frac{18\,\gamma}{\gamma_c^2}\,|\lambda|} } \ .
\eea
Now the interval has a finite length and the effective Planck mass is finite, but the string coupling diverges at both ends. Once more, $\rho$ can be removed from the solutions, rescaling $r$ and the $x$ and $y$ coordinates, and redefining $\phi_2$. 
This background approaches the $E=0$ solution at both ends, so that the limiting behavior is isotropic, is not captured by tension--free solutions and the string coupling diverges.
\end{itemize}

\subsubsection{\sc Isotropic solutions for $\gamma \neq \frac{3}{2}$} \label{sec:dm_9}

This class of solutions comprises the most symmetric backgrounds, which replace ten--dimensional flat space in the presence of the tadpole potential~\eqref{tadpole_pot}.
They can be recovered from the general ones letting $v_1=0$, but their special nature deserves a few additional comments.

\begin{itemize}
\item { For $\mathbf{\gamma} \,\mathbf{<} \,\mathbf{\gamma}_\mathbf{c}$} there are the two options $\theta=0,\pi$ in eqs.~\eqref{trigo_phichi_less}, which are equivalent up to an overall reflection of the radial coordinate $r \in \left(- \infty, \infty\right)$, so that it will suffice to consider
\beq
\widetilde{\phi}_1 \ = \ \frac{18\,\lambda}{\gamma_c} \ .
\eeq
The background then becomes
\bea
ds^2 &=& e^{\,-\, \frac{2\,\gamma\,\lambda\,r}{\gamma_c}}\ \frac{dx^2 \ + \ dy^2}{\left[\Delta\,\cosh\left({r}\right)\right]^{2\,\lambda}}  \ +\ \frac{e^{\,-\,\gamma\left(\frac{18 \,\lambda}{\gamma_c}\,r\,+\,\widetilde{\phi}_2\right)}\ dr^2}{\left[\Delta\,\cosh\left({r}\right)\right]^{18\,\lambda}} \ , \nonumber \\
e^{\phi} &=&  \left[\Delta\,\cosh\left({r}\right)\right]^{\frac{18\,\gamma}{\gamma_c^2}\,\lambda} \ e^{\frac{18 \,\lambda}{\gamma_c}\,r\,+\,\widetilde{\phi}_2} \ .
\eea
This class of solutions is a special case of what we discussed in Section~\ref{sec:gammalessgc}. It describes compactifications on intervals of finite length, where the string coupling vanishes at the one end and diverges at the other. The limiting behavior at both ends is captured, as was the case for $\gamma=\gamma_c$, by the isotropic tensionless solution of~\cite{ms_vacuum_1} or Appendix~\ref{app:kasner},
\bea{}{}
ds^2 &\sim& \xi^\frac{2}{9} \left(dx^2 + d\vec{y}^{\,2} \right) \ + \ d\xi^2 \ ,\nonumber \\
e^\phi &\sim& \xi^{\pm\,\frac{4}{3}} \ .
\eea
$\xi$ is close to zero in both cases, but this value corresponds to $r=-\infty$ at one end and to $r=+\infty$ at the other.

\item {For $\mathbf{\gamma} \,\mathbf{>}\, \mathbf{\gamma}_\mathbf{c}$} there are two classes of solutions, since for $E<0$ there are no isotropic solutions.
\begin{itemize}
\item {If the energy ${E=0}$}, the solutions are obtained from eqs.~\eqref{gammamoree0} letting $v_1=0$, and read
\bea
ds^2 &=&  r^{2|\lambda|}\,\big(dx^2 \ + \ dy^2\big) \,+\, \frac{r^{18\,|\lambda|}}{\Delta^2}\, e^{\,-\,\gamma\,\widetilde{\phi}_2}\ dr^2 \ , \nonumber \\
e^{\phi} &=&  r^{\,-\,\frac{18\,\gamma}{\gamma_c^2}\,|\lambda|} \ e^{\widetilde{\phi}_2} \ ,
\eea
where $0 < r < \infty$.
They describe intervals of infinite length, and the string coupling diverges at one end. Letting
\beq
\xi \ = \ e^{\,-\,\frac{\gamma}{2}\,\widetilde{\phi}_2}\ \frac{\left( r\right)^{9\,|\lambda| +1}}{\left(9\,|\lambda| +1\right)} \ ,
\eeq
the preceding expressions become, for $0 < \xi < \infty$,
\bea
ds^2 &=&  \xi^{\frac{2 \,\gamma_c^2}{9\,\gamma^2}}\,\big(dx^2 \ + \ dy^2\big) \,+\, \frac{d\xi^2}{\Delta^2} \ , \nonumber \\
e^{\phi} &=&  \xi^{\,-\,\frac{2}{\gamma}} \ e^{\widetilde{\phi}_2} \ , \label{near0e0}
\eea
after absorbing some multiplicative constants in rescalings of the $(x,y)$ coordinates and in redefinitions of $\widetilde{\phi}_2$. Contrary to what happens for $\gamma \leq \gamma_c$, the limiting behavior of these solutions at both ends is not captured by tension--free solutions, and the Kasner--like exponents satisfy the constraints of Appendix~\ref{app:kasner} only as $\gamma \to \gamma_c$.

\item {If the energy ${{E > 0}}$}, referring to the preceding section, there are two branches of solutions with $v_1=0$, and thus $
    \zeta=0$, for which
\bea
\widetilde{\phi}_1 &=&  \pm\,\frac{18\,|\lambda|}{\gamma_c} \ . \label{hamiltonian_hyperbolic_iso}
\eea
The corresponding form of the background is
\bea
ds^2 &=& e^{\,\mp\, \frac{2\,\gamma\,|\lambda|\,r}{\gamma_c}}\ {\left[\Delta\,\sinh\left({r}\right)\right]^{2\,|\lambda|}} \ \left( dx^2 \ + \ dy^2\right)
  \nonumber \\ &+& \left[\Delta\,\sinh\left({r}\right)\right]^{18\,|\lambda|} \ e^{\,\mp\, \frac{18\,\gamma\,|\lambda|\,r}{\gamma_c}}\ e^{\,-\,\gamma\,\phi_2}\ dr^2 \ , \nonumber \\
e^{\phi} &=& \ \frac{e^{\,\phi_2}\ e^{\,\pm\, \frac{18\,|\lambda|\,r}{\gamma_c}}}{\left[\Delta\,\sinh\left({r}\right)\right]^{\frac{18\,\gamma}{\gamma_c^2}\,|\lambda|} } \ . \label{isotropicepos}
\eea
In both branches, the string coupling vanishes as $r \to \infty$ but diverges at $r=0$. In the first branch (upper sign) the length of the $r$-interval is finite, while in the second (lower sign) it is infinite.
Close to $r=0$ the limiting behavior is as in eqs.~\eqref{near0e0}, and the string coupling is unbounded. On the other hand, as $r\to \infty$, the limiting behavior is what we described in eqs.~\eqref{alphaACphi_epos}, so that it is captured, for the two branches, by the isotropic tensionless solutions
\bea
ds^2 &\sim&  \xi^{\frac{2}{9}}\,\big(dx^2 \ + \ dy^2\big) \,+\, {d\xi^2} \ , \nonumber \\
e^{\phi} &\sim&  \xi^{\,\pm\,\frac{4}{3}} \ e^{\phi_2} \ , \label{isotropicgammamoreEpos}
\eea
where in the first case $\xi \to 0$ and in the second $\xi \to +\infty$. The string thus vanishes in both asymptotic regions.
\end{itemize}
\end{itemize}

Summarizing, for $\gamma \leq \gamma_c$ the asymptotic behavior of these isotropic backgrounds at both ends of the internal interval is captured by the tension--free solutions of I, although the string coupling vanishes at one end and diverges at the other. The length of the interval is always finite, and the solutions depend on one real parameter, $\phi_2$. For $\gamma> \gamma_c$, there are three types of isotropic solutions. For $E>0$ there are indeed two families of solutions, identified by the branches in eqs.~\eqref{isotropicepos}, while the third type of solutions correspond to $E=0$. The length is only finite for one of the branches with $E>0$, while the string coupling vanishes at one end and diverges at the other.

\section{\sc Cosmological Solutions without Form Fluxes} \label{sec:susybT_c}

We can now turn to the cosmological solutions, which can be obtained from the preceding results via an analytic continuation. 

\subsection{\sc Cosmological Solutions for $\gamma = \frac{3}{2}$ and $\beta \neq 0$} \label{sec:criticalbetan0}

There are cosmological counterparts of the solutions with $\beta \neq 0$, which can be obtained continuing $r$ to $i\,\tau$, and $A_1$, $C_1$ and $\phi_1$, and thus also $\beta$, to imaginary values. The end result is just an overall change of sign for $T$, so that
\bea
A &=& \frac{T}{32} \ \frac{e^{2\,\beta\,\tau}}{\beta^2}\ + \ A_1\, \tau \ , \nonumber \\
B &=& \frac{9\,T}{32} \ \frac{e^{2\,\beta\,\tau}}{\beta^2}\ + \ B_1\, \tau \ , \nonumber \\
C &=& \frac{T}{32} \ \frac{e^{2\,\beta\,\tau}}{\beta^2} \ + \ C_1\, \tau \ , \nonumber \\
\phi &=& - \ \frac{3\,T}{8} \ \frac{e^{2\,\beta\,\tau}}{\beta^2} \ + \ \phi_1\, \tau \ + \ \phi_2 \ , \label{beta0000_2}
\eea
where the range of $\tau$ is the whole real axis. 

Let us begin to describe these cosmologies, assuming that $\beta>0$. In this case, the Big Bang lies a finite amount of cosmic time in the past, and the solutions approach quickly a universal isotropic behavior for large positive values of $\tau$. When the exponential terms dominate, the relation
\beq
dt \ = \ e^B \, d \tau
\eeq
integrates indeed to
\beq
2\,\beta\, t \ \simeq \ e^B \ ,
\eeq
since $d \,e^{B} \simeq 2 \,\beta\,e^B\, d\,\tau$, so that
\bea
ds^2 &\simeq & - \, dt^2 \ + t^\frac{2}{9} \left(dx^2 + d\vec{y}^2 \right) \ , \nonumber \\
e^\phi & \simeq & \ (2\, \beta \, t)^{-12} \ .
\eea
Proceeding as in the previous section, one can recast these results in a form similar to eq.~\eqref{metricu},
\bea{}
ds^2 &=&  - \ \frac{2}{3\,T} \, e^{\,\frac{3}{2}\left(u+\phi_0\right)} {u}^{\frac{1}{1+\frac{3}{4}\,\alpha_\phi}\,-\,2} \, du^2 \ + \ e^{\,\frac{u}{6}} \ {u}^\frac{\alpha_A}{1+\frac{3}{4}\,\alpha_\phi}\, d\vec{x}^{\,2} \ +\ e^{\,\frac{u}{6}} \ {u}^\frac{\alpha_C}{1+\frac{3}{4}\,\alpha_\phi}\, d \vec{y}^{\,2} \ , \nonumber \\
e^\phi &=& e^{\,-\left(u\,+\,\phi_0\right)} u^\frac{\alpha_\phi}{2 \left(1+\frac{3}{4}\,\alpha_\phi\right)} \ , \label{metricuc}
\eea
where
\beq{}{}
u \ = \ \frac{3 T}{8}\, \frac{e^{2\beta\tau}}{\beta^2}  \ . \label{utau}
\eeq
Here the $\alpha$'s correspond again to the points of the ellipse of eq.~\eqref{ellipse}, and are given in eqs.~\eqref{param_theta}. The value $\theta=\pi$, which corresponds to $\alpha_\phi=-\frac{4}{3}$, would describe an isotropic descending solution, but is excluded, consistently with~\cite{dks}. Close to the initial singularity at $u=0$, in terms of the cosmic time,
\bea{}{}
ds^2 &\simeq& - \ dt^2 \ + \  t^{2\,\alpha_A}\, d\vec{x}^{\,2} \ + \ t^{2\,\alpha_C}\, d\vec{y}^{\,2} \ , \nonumber \\
e^\phi &\simeq& e^{\widetilde{\phi}_0}\, t^{\alpha_\phi} \ ,
\eea
and one recovers the behavior of the tension--free cosmological solution of~\cite{ms_vacuum_1}.

The values of $\alpha_\phi$ within the range $-\,\frac{4}{3} < \alpha_\phi< 0$ correspond to the descending behavior, which is possible in the anisotropic case also for $\gamma=\gamma_c$, while the values in the range $0 < \alpha_\phi< \frac{4}{3}$ correspond to the climbing behavior. Moreover, $\alpha_\phi=0$ is a novelty of these solutions, since in that case the dilaton descends from a finite height.
\begin{figure}[ht]
\centering
\begin{tabular}{cc}
\includegraphics[width=60mm]{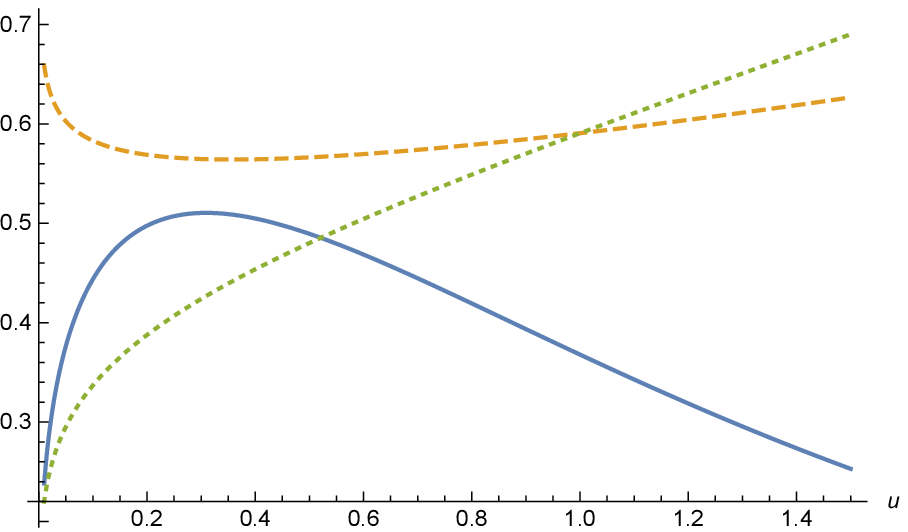} &
\includegraphics[width=60mm]{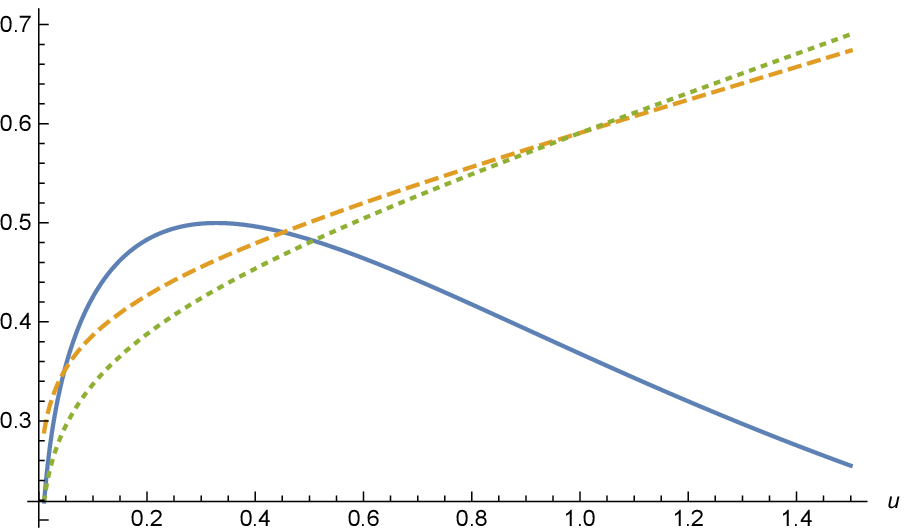} \\
\end{tabular}
\caption{\small $e^A$ (dashed), $e^C$ (dotted) and $e^\phi$ (solid) for a ``critical'' anisotropic \emph{climbing scalar} cosmology with $\alpha_A<0,\alpha_C>0,\alpha_\phi>0$, where the space--time $x$--coordinates undergo a bounce (left panel), and for an anisotropic \emph{climbing scalar} cosmology with $\alpha_A>0,\alpha_C>0,\alpha_\phi>0$, where all directions expand (right panel).}
\label{fig:fig01}
\end{figure}
\begin{figure}[ht]
\centering
\begin{tabular}{cc}
\includegraphics[width=60mm]{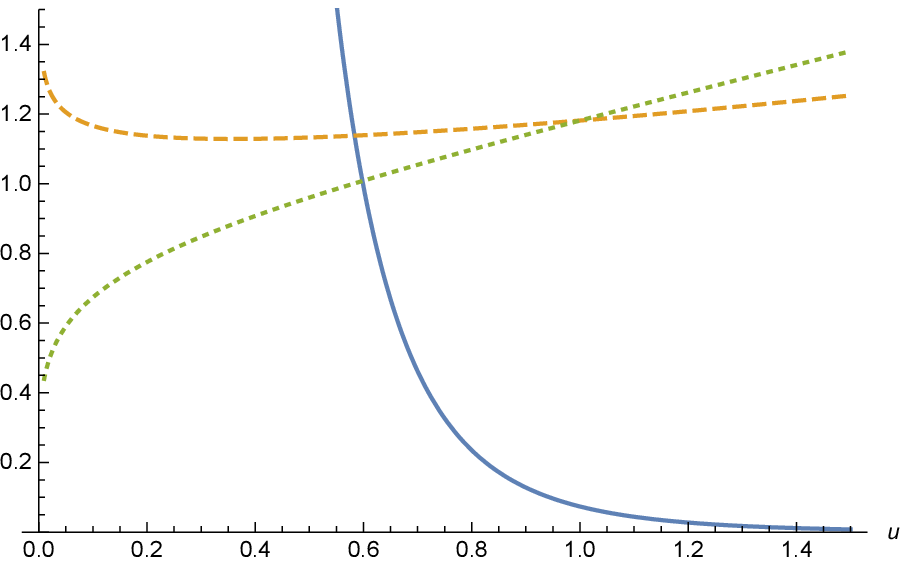} &
\includegraphics[width=60mm]{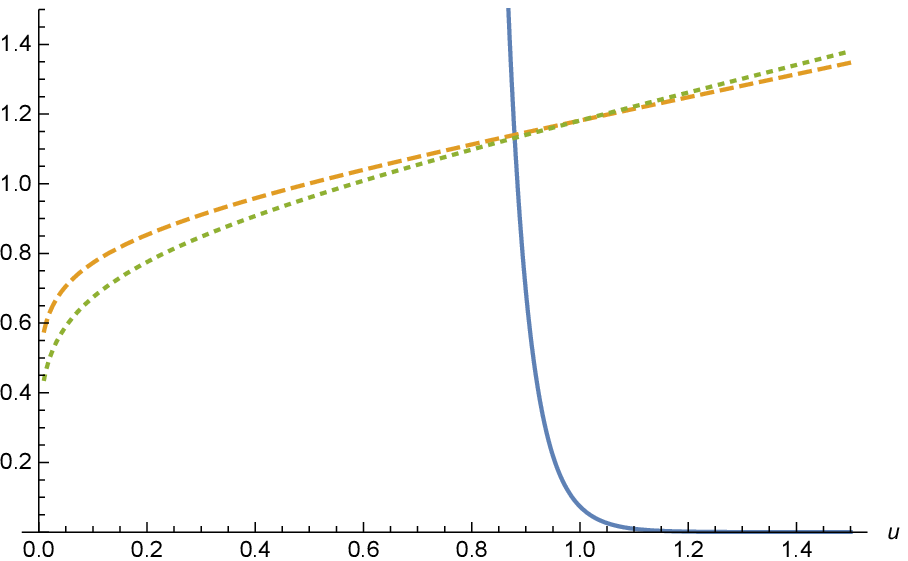} \\
\end{tabular}
\caption{\small $e^A$ (dashed), $e^C$ (dotted) and $e^\phi$ (solid) for a ``critical'' anisotropic \emph{descending scalar} cosmology with $\alpha_A<0,\alpha_C>0,\alpha_\phi<0$, where the space--time $x$--coordinates undergo a bounce (left panel), and for an anisotropic \emph{descending scalar} cosmology with $\alpha_A>0,\alpha_C>0,\alpha_\phi<0$, where all directions expand (right panel).}
\label{fig:fig02}
\end{figure}

Summarizing, for $\beta>0$ these cosmologies are generally anisotropic as the Universe emerges from the initial singularity at a finite cosmic time in the past, but always approach an isotropic expanding behavior for large values of the cosmic time. The two sets of coordinates can always expand, or one set can first contract to then expand, and the climbing and descending behaviors are both possible in the anisotropic case. On the other hand, if $\beta<0$ these cosmologies are generally isotropic and contracting as the Universe emerges from an initial singularity, but become generically anisotropic for large values of the cosmic time. These solutions could describe, in principle, dynamical compactifications of an internal torus if $\alpha_A>0$ and $\alpha_C<0$, but the Big Bang lies an infinite amount of cosmic time in the past.

\subsection{\sc Cosmological Solutions for $\gamma \neq \frac{3}{2}$}

There are also cosmological counterparts of these other families of solutions, with metrics
\beq
ds^2 \ = \ - \ e^{2B}\, d\tau^2 \ + \ e^{2A} \,d\vec{x}^{\,2} \ +\ e^{2C}\, d\vec{y}^{\,2} 
\eeq
that can be obtained letting $r \to i\,\tau$ and continuing $\phi_1$ and $v_1$ to imaginary values. In this fashion the relevant equations become
\bea
&& (\dot{X})^2 \ = \ \epsilon\,\Delta^2\ e^{2\,X} \ + \ E \ ,
\eea
where
\bea
\Delta^2 &=& \frac{T}{2}\ \left|\gamma^2\,-\,\gamma_c^2\right| \ \ , \nonumber \\
\epsilon &=& {\mathrm{sign}}\left(\gamma_c^2\,-\,\gamma^2\right) \ , \nonumber \\
E &=& \frac{1}{4} \bigg\{ \left(\frac{\gamma^2}{\gamma_c^2} \ - \ 1 \right)^2 \,\gamma_c^2\,\phi_1^2 \,- \,   \left(\frac{\gamma^2}{\gamma_c^2} \ - \ 1 \right)\frac{(p+1)(8-p)}{8}\ v_1^2  \bigg\} \ . \label{dynamical_T_c}
\eea
The corresponding relations between $X$ and the other quantities now read
\bea
A &=&  \frac{1}{9} \left\{ - \ \frac{\gamma_c^2\, X}{\left(\gamma^2\,-\,\gamma_c^2\right)} \ + \ \frac{1}{2}\,\Big[(8-p)v_1 \,-\, {\gamma}\, \phi_1\Big]\tau \right\} ,\nonumber \\
B&=& - \ \frac{\gamma_c^2\,X}{\left(\gamma^2\,-\,\gamma_c^2\right)} \ - \ \frac{\gamma}{2}\left(\phi_1\, \tau \,+\, \phi_2\right) \ , \nonumber \\
C&=& \frac{1}{9} \left\{- \ \frac{\gamma_c^2\,X}{\left(\gamma^2\,-\,\gamma_c^2\right)} \ -\ \frac{1}{2}\,\Big[(p+1)v_1\,+\,{\gamma}\, \phi_1\Big] \tau \right\} \ , \nonumber \\
\phi &=& 2\, \frac{\gamma \,X}{\left(\gamma^2\,-\,\gamma_c^2\right)} \ + \ \phi_1\, \tau \,+\, \phi_2  \ . \label{ACphiX_c}
\eea
{The sign of the exponential potential is thus reverted with respect to the spatial profiles, while the constant term maintains the same form}, after the independent variable and the old coefficients are all continued to imaginary values. Let us now investigate the behavior of these models.

\subsubsection{\sc Cosmological Solutions for $\gamma < \frac{3}{2}$}

For $\gamma<\gamma_c$ the Newtonian potential is a negative exponential while the third of eqs.~\eqref{dynamical_T_c} implies that $E \geq 0$, and one can capture the independent values of $\phi_1$ and $v_1$ via the trigonometric parametrization of eqs.~\eqref{trigo_phichi_less}. As before, we thus set $E = \frac{1}{\sqrt{\rho}}$. There are two classes of solutions where $X$ spans the whole real axis, and one can confine the attention to the choice where $X$ \emph{increases} as $\tau < 0$ also \emph{increases},
\beq
X \ = \ -  \log\left[\Delta\,\rho \, \sinh \left(\frac{-\tau}{\rho}\right) \right] \ \ \ (-\infty \,<\,\tau \,<\, 0) \ .
\eeq
Then, defining again
\beq
\lambda \ = \ \frac{1}{9\left(1 \ -\ \frac{\gamma^2}{\gamma_c^2} \right)}\ ,  \label{lambda_sub_2}
\eeq
as in eq.~\eqref{lambda_sub}, the solutions for the metric and the string coupling for $E>0$ read
\bea
ds^2 &=& - \ \frac{e^{\,-\,\gamma\left(\phi_1\,\tau\,+\,\phi_2\right)}\ d\tau^2}{\left[\Delta\,\rho\,\sinh\left(\,-\,\frac{\tau}{\rho}\right)\right]^{18\,\lambda}}\ + \ e^{\,-\, \frac{\gamma\,\phi_1\,\tau}{9}}\ \frac{e^{\,\frac{(8-p)v_1\,\tau}{9}}\,dx^2 \ + \ e^{\,\frac{-(p+1)v_1\,\tau}{9}}\,dy^2}{\left[\Delta\,\rho\,\sinh\left(\,-\,\frac{\tau}{\rho}\right)\right]^{2\,\lambda}}  \ , \nonumber \\
e^{\phi} &=&  \left[\Delta\,\rho\,\sinh\left(\,-\,\frac{\tau}{\rho}\right)\right]^{\frac{18\,\gamma\,\lambda}
{\gamma_c^2}} \ e^{\phi_1\,\tau\,+\,\phi_2} \ . \label{gammaless_cosmo}
\eea

Notice that, from eqs.~\eqref{gammaless_cosmo},
\beq
X_{\tau \to - \infty} \ \sim \ \frac{\tau}{\rho} \ , \qquad X_{\tau \to 0^-} \ \sim \ - \log |\emph{}\Delta \tau| \ ,
\eeq
and since $9\,\lambda >1$ the latter limit corresponds to large cosmic time $t \sim \left(-\tau\right)^{1-9\lambda}$, where the Universe approaches the isotropic expanding geometry
\beq
ds^2 \ \sim \  - \ \ d t^2 + \ t^{\,\frac{2\gamma_c^2}{9\gamma^2}} \left(dx^2 \ + \ dy^2 \right) \label{metric_lm}
\eeq
while the string coupling tends to zero as
\beq
e^\phi \ \sim  \ t^{-\frac{2}{\gamma}} \ . \label{phi_lm}
\eeq
This is actually the exact solution of the problem for $E=0$, in which case $\phi_1$ and $v_1$ must also vanish. Indeed, all these anisotropic cosmologies approach, for large cosmic time, this ten--dimensional counterpart of the isotropic Lucchin--Matarrese attractor~\cite{lm}.

On the  other hand, $\tau \to - \infty$ corresponds to the initial singularity, and the results are those in eqs.~\eqref{B_spatial} and~\eqref{phi_spatial}, up to the replacement of $|r|$ with $-\tau$, so that the limiting behaviors of metric and string coupling are 
\bea
ds^2 \!\!\!&\sim&-\ \left( \frac{2}{\Delta\,\rho}\right)^{18\lambda}\, e^{\,18\,\lambda\left[1 \ + \ \frac{\gamma}{\gamma_c}\, \cos \eta \right]\frac{\tau}{\rho}}\, e^{\,-\,\gamma\,\phi_2}\ d\tau^2  \\ &+& \!\!\!\left( \frac{2}{\Delta\,\rho}\right)^{2\,\lambda}\Biggl[ d\vec{x}^{\,2}\,e^{\frac{2\,\lambda\,\tau}{\rho}\bigl( 1 + \frac{\gamma}{\gamma_c}\, \cos \eta + \sqrt{\frac{8(8-p)}{9(p+1)\lambda}}\, \sin \eta \bigr)}\nonumber \\ &+&d\vec{y}^{\,2}\,e^{\frac{2\,\lambda\,\tau}{\rho}\bigl( 1 + \frac{\gamma}{\gamma_c} \cos \eta - \sqrt{\frac{8(p+1)}{9(8-p)\lambda}}\sin \eta \bigr)}\Biggr]\ , \nonumber \\
e^\phi &\sim& e^{\phi_2} \ \left(\frac{\Delta\,\rho}{2}\right)^\frac{18\,\gamma\,\lambda}{\gamma_c^2} e^{\,- \, \frac{18\,\lambda\, \tau}{\rho\,\gamma_c} \left( \frac{\gamma}{\gamma_c}\ + \ \cos \eta \right)} \ .
\eea

These results indicate that in these cosmologies:
\begin{itemize}
\item there is always a Big Bang singularity in the finite past, since the integral
$$\int_{-\infty}^{-\tau_0} d\tau\, e^B(\tau)$$
is finite for any value of $\eta$;
\item \emph{descending scalar} solutions, for which the string coupling diverges as $\tau \to - \infty$, exist for values of $\eta$ such that
    $$\frac{\gamma}{\gamma_c} \ + \ \cos \eta > \ 0 \ . $$
    The isotropic solutions with $\cos\eta=1$ are a special instance in this class;
    \item \emph{climbing scalar} solutions, which emerge from the initial singularity with vanishing string coupling, also exist in the complementary region
    $$ \frac{\gamma}{\gamma_c} \ + \ \cos \eta \ < \ 0 \ . $$
    The isotropic solutions with $\cos\eta=- 1$ are a special instance in this class.
 \item finite values of $\rho$ can be removed from these equations, up to rescalings of the $x$ and $y$ coordinates and up to a redefinition of $\phi_2$, for any value of $\gamma$. As a result, the solutions depend only on $\eta$ and $\phi_2$.
    \end{itemize}

\begin{figure}[ht]
\centering
\begin{tabular}{cc}
\includegraphics[width=60mm]{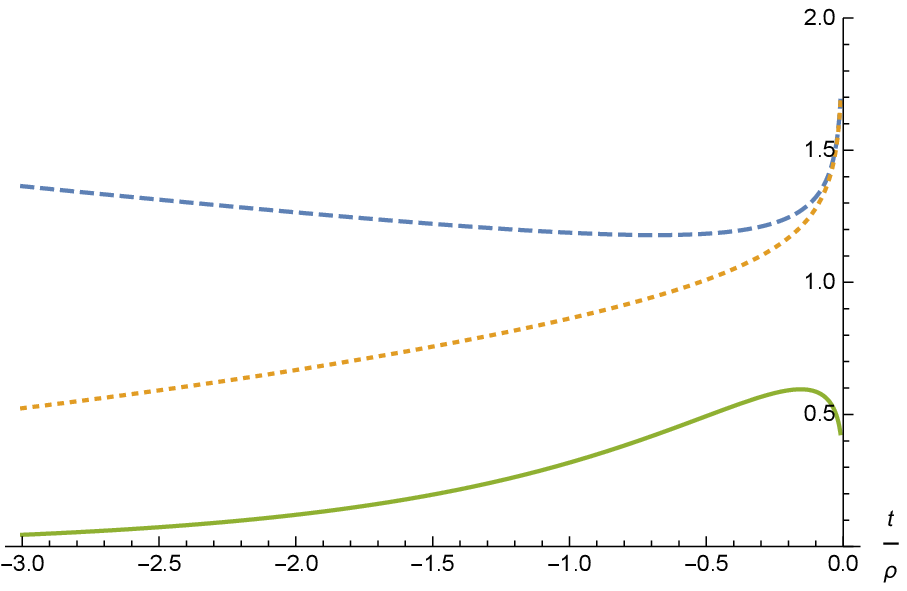} &
\includegraphics[width=60mm]{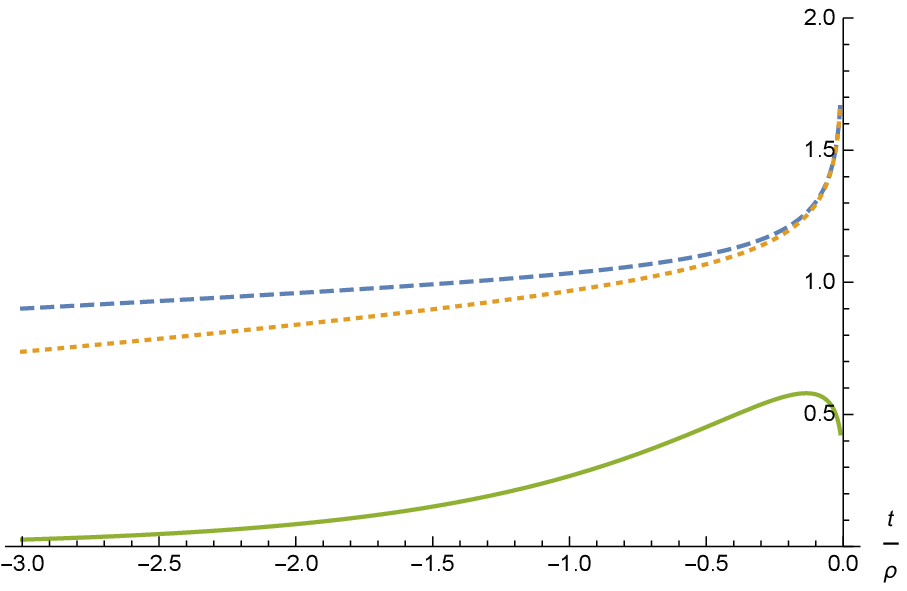} \\
\end{tabular}
\caption{\small $e^A$ (dashed), $e^C$ (dash--dotted) and $e^\phi$ (solid) for an anisotropic \emph{climbing scalar} cosmology where the space--time $x$--coordinates undergo a bounce $(\gamma=0.2 \gamma_c,\eta=-\frac{5\pi}{6})$ (left panel), and for an anisotropic \emph{climbing scalar} cosmology where all directions expand $(\gamma=0.2 \gamma_c,\eta=-\frac{5.8\pi}{6})$ (right panel).}
\label{fig:fig1}
\end{figure}

\begin{figure}[ht]
\centering
\begin{tabular}{cc}
\includegraphics[width=60mm]{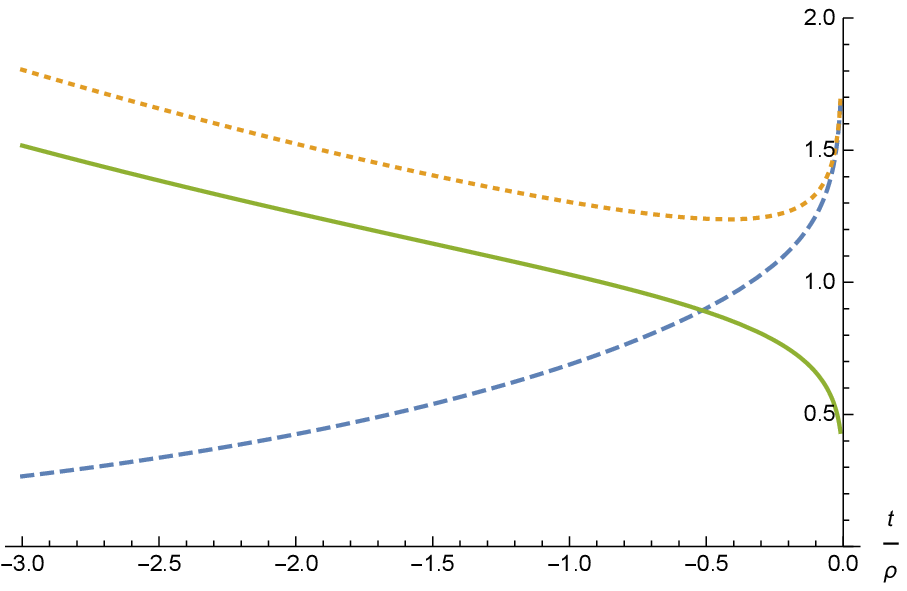} &
\includegraphics[width=60mm]{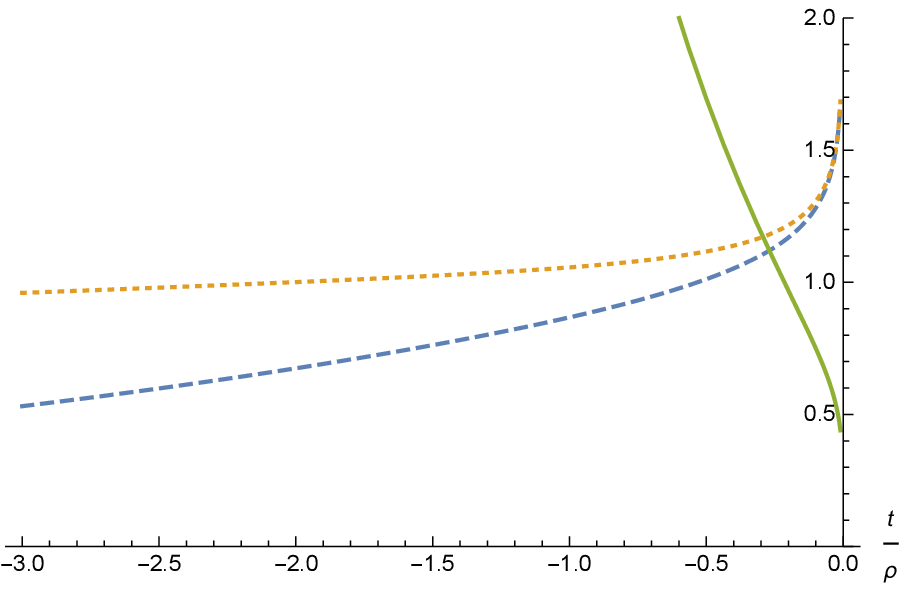} \\
\end{tabular}
\caption{\small $e^A$ (dashed), $e^C$ (dash--dotted) and $e^\phi$ (solid) for an anisotropic \emph{descending scalar} cosmology where the space--time $x$--coordinates undergo a bounce $(\gamma=0.2 \gamma_c,\eta=\frac{\pi}{2})$ (left panel), and for an anisotropic \emph{descending scalar} cosmology where all directions expand $(\gamma=0.2 \gamma_c,\eta=\frac{\pi}{10})$ (right panel).}
\label{fig:fig2}
\end{figure}

Up to the interchange of the $x$ and $y$ coordinates, the independent cases are $p=0,1,2,3$, and they all allow for $A$ and $C$ both increasing as cosmic time increases (a Big Bang singularity), or one increasing and one initially decreasing (a bounce for one group of coordinates and a Big Bang singularity for the rest). The plots in figs.~\ref{fig:fig1} and  \ref{fig:fig2} illustrate these types of behavior, which are manifest in the special cases $\eta=0,\pm \frac{\pi}{2}$, for both climbing and descending scalars.

\subsubsection{\sc Cosmological Solutions for $\gamma > \frac{3}{2}$}

\begin{figure}[ht]
\centering
\begin{tabular}{cc}
\includegraphics[width=60mm]{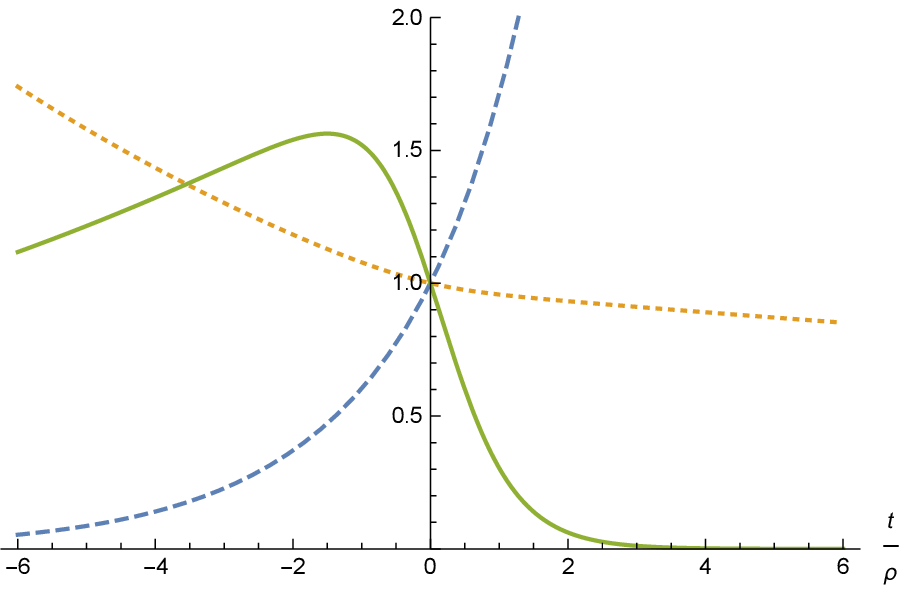} &
\includegraphics[width=60mm]{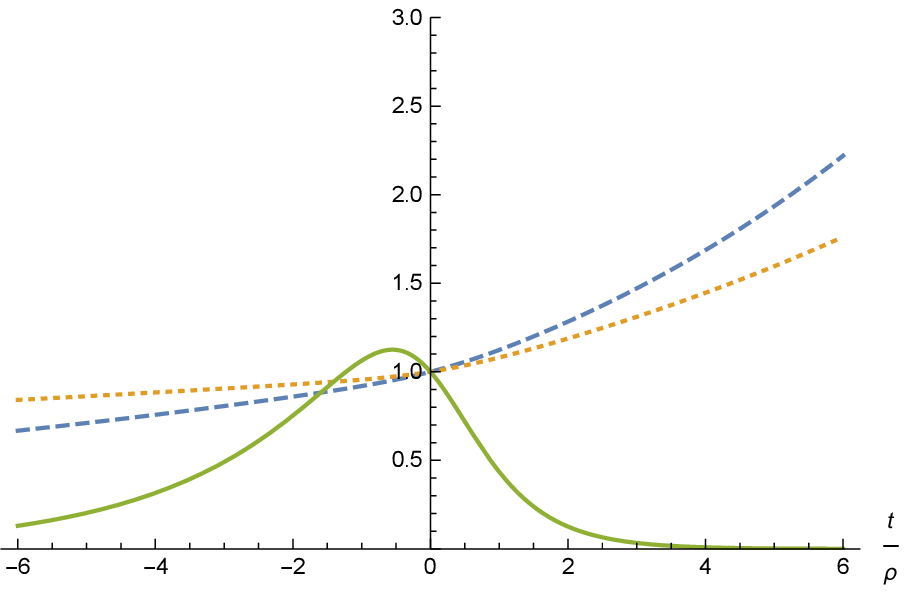} \\
\end{tabular}
\caption{\small $e^A$ (dashed), $e^C$ (dash--dotted) and $e^\phi$ (solid) for an anisotropic four--dimensional \emph{climbing scalar} cosmology where space--time expands while the $y$ internal space contracts $(\gamma=2 \gamma_c,\zeta=1.2)$ (left panel), and for an anisotropic \emph{climbing scalar} cosmology where all directions expand $(\gamma=2 \gamma_c,\zeta=0.1)$ (right panel).}
\label{fig:fig3}
\end{figure}

\begin{figure}[ht]
\centering
\begin{tabular}{cc}
\includegraphics[width=60mm]{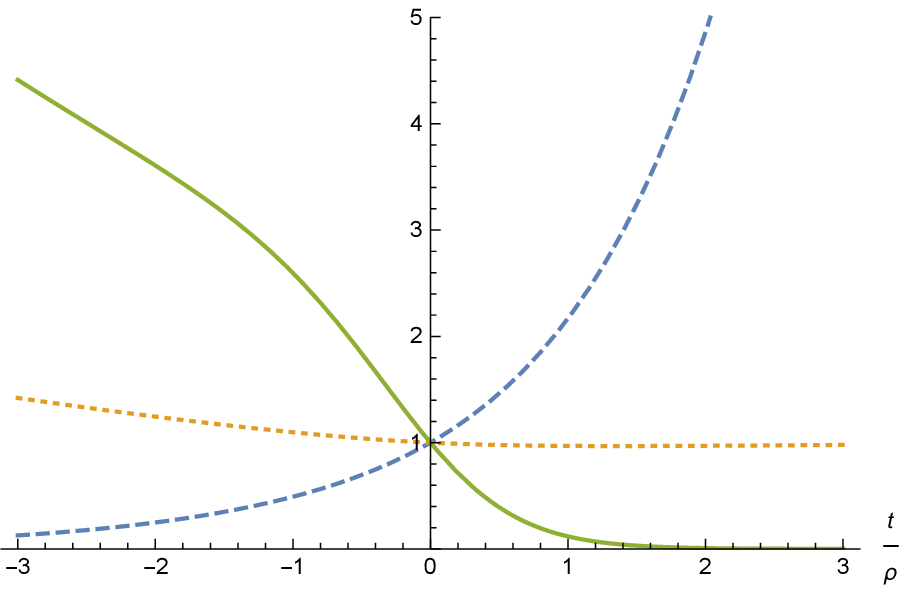} &
\includegraphics[width=60mm]{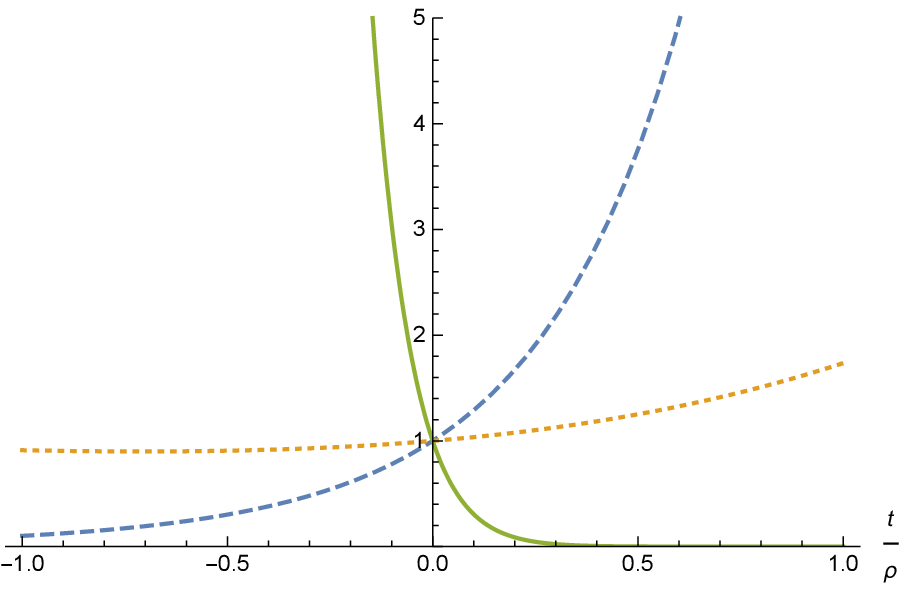} \\
\end{tabular}
\caption{\small $e^A$ (dashed), $e^C$ (dash--dotted) and $e^\phi$ (solid) for an anisotropic \emph{descending scalar} cosmology where space--time expands while the $y$ internal space contracts $(\gamma=1.6 \gamma_c,\zeta=1.2)$ (left panel), and for an anisotropic \emph{descending scalar} cosmology where all directions expand $(\gamma=1.1 \gamma_c,\zeta=1.2)$ (right panel).}
\label{fig:fig4}
\end{figure}

For $\gamma>\gamma_c$, which is a case of direct interest for String Theory, the potential is a positive exponential and the energy in eqs.~\eqref{dynamical_T} must be positive.
In this case the solution reads
\bea
ds^2 &=& - \ {\left[\Delta\,\rho\,\cosh\left(\frac{\tau}{\rho}\right)\right]^{18\,|\lambda|}}\,{e^{\,-\,\gamma\left(\phi_1\,\tau\,+\,\phi_2\right)}\ d\tau^2}
\nonumber \\ &+& e^{\,-\, \frac{\gamma\,\phi_1\,\tau}{9}}\ {\left[\Delta\,\rho\,\cosh\left(\frac{\tau}{\rho}\right)\right]^{2\,|\lambda|}} \left(e^{\,\frac{(8-p)v_1\,\tau}{9}}\,dx^2 \ + \ e^{\,\frac{-(p+1)v_1\,\tau}{9}}\,dy^2\right) \ , \nonumber \\
e^{\phi} &=&  \left[\Delta\,\rho\,\cosh\left(\frac{\tau}{\rho}\right)\right]^{\,-\,\frac{18\,\gamma}{\gamma_c^2}\,|\lambda|} \ e^{\phi_1\,\tau\,+\,\phi_2} \ ,
\eea
where $\tau \in (-\infty,+\infty)$,
with the hyperbolic parametrization of eqs.~\eqref{hamiltonian_hyperbolic}, which we display again here for the reader's benefit. We concentrate on the branch
\bea
\phi_1 &=&  -\,\frac{18|\lambda|}{\rho\,\gamma_c} \ \cosh \zeta \ , \nonumber\\
v_1 &=& \frac{12}{\rho} \ \sinh \zeta\ \sqrt{\frac{2|\lambda|}{(p+1)(8-p)}}  \ , \label{hamiltonian_hyperbolic2}
\eea
since the other, where $\phi_1 \to - \phi_1$, would simply obtain from this letting $\tau \to  - \tau$ and $\zeta \to - \zeta$.
Figs.~\ref{fig:fig3} and \ref{fig:fig4} illustrate some interesting options for this class of cosmological models.

For large values of $|\tau|$
\bea
ds^2 \!\!\!&\sim&-\ \left( \frac{\Delta\,\rho}{2}\right)^{18|\lambda|}\, e^{\,18\,|\lambda|\left[1 \ + \ \frac{\gamma}{\gamma_c}\, \sign \tau\,\cosh \zeta \right]\frac{|\tau|}{\rho}}\, e^{\,-\,\gamma\,\phi_2}\ d\tau^2  \\ &+& \!\!\!\left( \frac{\Delta\,\rho}{2}\right)^{2\,|\lambda|}\Biggl[ d\vec{x}^{\,2}\,e^{\,\frac{2\,|\lambda|\,|\tau|}{\rho}\bigl( 1 + \frac{\gamma}{\gamma_c}\, \sign \tau \cosh \zeta + \sqrt{\frac{8(8-p)}{9(p+1)|\lambda|}}\, \sign \tau\,\sinh \zeta \bigr)}\nonumber \\ &+&d\vec{y}^{\,2}\,e^{\,\frac{2\,|\lambda|\,|\tau|}{\rho}\bigl( 1 + \frac{\gamma}{\gamma_c} \sign \tau\,\cosh \zeta - \sqrt{\frac{8(p+1)}{9(8-p)|\lambda|}}\,\sign\tau\,\sinh \zeta \bigr)}\Biggr]\ , \nonumber \\
e^\phi &\sim& e^{\phi_2} \ \left(\frac{\Delta\,\rho}{2}\right)^{\,-\,\frac{18\,\gamma|\lambda|}{\gamma_c^2}} e^{\,- \, \frac{18\,|\lambda|\, |\tau|}{\rho\,\gamma_c} \left( \frac{\gamma}{\gamma_c}\, + \,\sign\tau\, \cosh \zeta \right)} \ ,
\eea
and one can see that
\begin{itemize}
\item this class of metrics has two singularities. The first, at $\tau=-\infty$, is reached within a finite amount of cosmic time, and models the Big Bang, while the second, at $\tau=+\infty$, is reached within an infinite amount of cosmic time;
\item the string coupling tends to zero as $\tau \to + \infty$ for all values of $\zeta$;
\item the string coupling tends to zero as $\tau \to - \infty$ if
$$ \frac{\gamma}{\gamma_c} \ > \ \cosh \zeta \ , $$
so that in this range the system exhibits only a \emph{climbing} behavior. If $\frac{\gamma}{\gamma_c} = \cosh \zeta$, the string coupling starts from a finite value at the initial singularity. In the isotropic case $\zeta=0$, and one recovers the result of~\cite{dks}. On the other hand, in the complementary range
 $$ \frac{\gamma}{\gamma_c} \ < \ \cosh \zeta \ , $$
 which is available in these anisotropic cosmologies, the system exhibits a purely \emph{descending} behavior;
 \item $\rho$ can be removed from these equations, up to rescalings of the $x$ and $y$ coordinates and up to a redefinition of $\phi_2$, for any value of $\gamma$. As a result, the solutions depend only on $\zeta$ and $\phi_2$. However, the potential for $X$ is now positive, and consequently the $\rho \to \infty$ limit is singular.
\end{itemize}

\section{\sc Inclusion of Form Fluxes}
\label{sec:Tn0hn0}
We can now consider the general equations of Section~\ref{sec:symmetries}, allowing for both form fluxes of the $H$--type and dilaton tadpoles. The corresponding results for the $h$ fluxes are determined by the discrete symmetry in eq.~\eqref{sym_AC}.

\subsection{\sc New variables for the general case : $H$--Fluxes}

The reduced form of the system of eqs.~\eqref{EqA_red} -- \eqref{EqB_red} is now
 \bea
 A'' &=& - \ \frac{T}{8} \ e^{\, 2\left[(p+1)A+(8-p)C + \frac{\gamma}{2}\,\phi\right]} \,+\,\frac{(7-p)}{16} \ e^{\,2\left[\beta_p\,\phi\,+\,(p+1)\,A\right]} \ H_{p+2}^2\
  \ , \nonumber \\
C''& =&  - \ \frac{T}{8} \ e^{\, 2\left[(p+1)A+(8-p)C + \frac{\gamma}{2}\,\phi\right]} \,-\, \frac{(p+1)}{16} \ e^{\,2\left[\beta_p\,\phi\,+\,(p+1)\,A\right]} \ H_{p+2}^2\ , \nonumber \\
\phi'' &=& {T\,\gamma}\ e^{\, 2\left[(p+1)A+(8-p)C + \frac{\gamma}{2}\,\phi\right]} + {\beta_p}\ e^{\,2\left[\beta_p\,\phi\,+\,(p+1)\,A\right]} \ H_{p+2}^2\ ,
\eea
and suggests to work in terms of three new variables
\bea
U &=& \phi \ + \ \big\{ \left[ (p+1)\gamma\,-\,2\,\beta_p \right] A \ + \ \left[ (7-p)\gamma\,+\,2\,\beta_p \right]C \big\}\ , \nonumber \\
X &=& (p+1)A\,+\,\frac{\gamma}{2}\,\phi + (8-p)C \ , \nonumber \\
Z &=& (p+1)A \,+ \, \beta_p\,\phi \ . \label{XYZ_2}
\eea
The last two are the combinations that enter the exponents, while the first is a combination of the original functions that is still linear in $r$, as we shall see shortly.

Inverting these relations gives
\bea
A&=& \frac{1}{\mathrm{Den}} \Bigg\{2 \beta_p X \Big[2 \beta_p+(7-p)\gamma \Big]\,-\,2\left(8-p\right) \beta_p \,U \,-\,Z \,\Big[{\gamma} [2 {\beta_p}+(7-p){\gamma} ]-2(8-p)\Big]\Bigg\} \ , \nonumber \\
C &=&  \frac{1}{\mathrm{Den}} \Bigg\{ X \Big[2 {\beta_p} [2 {\beta_p}-{\gamma} (p+1)]+2 (p+1)\Big]\nonumber \\&+&Z \Big[ {\gamma} [-2 {\beta_p}+{\gamma}( p+1)]-2(p+1)\Big]\,+\, (p+1)\, U \,(2 {\beta_p}-{\gamma})\Bigg\} \ ,  \\
\phi &=& \frac{2}{\mathrm{Den}} \Bigg\{(8 - p) (p+1) U -
 \Big[2 \beta_p + \gamma (7 - p)\Big] (p+1) X \nonumber \,+\,
 \big[16 \beta_p - \gamma (p+1)\big] Z \Bigg\} \ ,
\eea
where the denominator is
\bea
\mathrm{Den} &=& - (p+1) \Big[{\gamma} [2 {\beta_p}+(7-p){\gamma} ]-2(8-p)\Big]\,+\,2\ {\beta_p} \, \Big[18 {\beta_p}\,-\,(p+1){\gamma} \Big] \ .\label{denom}
\eea
When this expression does not vanish, letting
\bea
a&=& 4 \left( \gamma^2 \ - \ \gamma_c^2 \right) \ , \qquad 
b \ = \  8\,\beta_p\,\gamma\, \,-\,p\,-\,1 \ , \nonumber \\
c&=& 2\left[8\,\beta_p^2 \ +\ \frac{(7-p)(p+1)}{2} \right] \ ,
\eea
where, as before, $\gamma_c \ = \ \frac{3}{2}$, one can reduce the system to
\bea
U'' &=& 0 \label{kkp0_gen_gamma} \\
X'' &=& \frac{T}{8}\ a \, e^{\,2\,X}  \,+\, \frac{H_{p+2}^2}{16}\ b\, e^{\,2\,Z} \nonumber \\
Z'' &=& \frac{T}{8}\ b \, e^{\,2\,X} \,+\, \frac{H_{p+2}^2}{16}\ c\, e^{\,2\,Z} \ , \nonumber
\eea
while the corresponding expression for the Hamiltonian constraint reads
\bea
c\,(X')^2 &+& a \,(Z')^2 - 2 \,b \,X'\,Z' \nonumber \ - \ \frac{(p+1)(8 - p)}{2}\,(U')^2 \nonumber \\&+& \frac{1}{8}\,(b^2\ -\ a\,c) \left({T} \, e^{2X}\ + \ \frac{H_{p+2}^2}{2}\, e^{2Z} \right) \ =  \ 0 \ ,
\eea
and determines $(U')^2$. Notice that for $\gamma=\gamma_c$ the preceding equations take a simpler form, with $a=0$, $c=16$, and for RR forms $b=2(p-5)$. These considerations will play a role shortly.

If the denominator in eq.~\eqref{denom} vanishes, one can still reduce the system to the last two equations for $X$ and $Z$ in~\eqref{kkp0_gen_gamma}, while $\phi$ can be obtained solving
\beq
\phi'' \ = \ {T\,\gamma} \ e^{\, 2\,X} \ + \ {\beta_p}\ H_{p+2}^2\ e^{\,2\,Z}  \ .
\eeq
In this case, after obtaining $X$ and $Z$, $(\phi')^2$ can be determined by the Hamiltonian constraint, which reads
\bea
c\,(X')^2 &+& a \,(Z')^2 - 2 \,b \,X'\,Z' \nonumber \ - \ \frac{(p+1)(8-p)}{2} (\phi')^2 \nonumber \\&+& \frac{1}{8}\,(b^2\ -\ a\,c) \left({T}\, e^{2X}\ + \ \frac{H_{p+2}^2}{2}\, e^{2Z} \right) \ =  \ 0 \ .
\eea

\subsection{\sc Special Cases}

These systems are complicated, and exploring their solutions requires in general numerical techniques. Therefore, we shall now concentrate of the special cases that arise in String Theory, one of which is remarkably far simpler than one could have naively anticipated. Let us now record these special forms of the preceding expressions.
\subsubsection{\sc $USp(32)$ and $U(32)$ orientifolds with $p=1$ ``electric fluxes''}

In this case $\gamma=\frac{3}{2}$ and $\left(a,b,c\right)=\left(0,-8,16\right)$, so that the equations become
\bea
U'' &=& 0 \ , \nonumber \\
X'' &=&  - \ \frac{H_{3}^2}{2}\, e^{\,2\,Z}\ , \nonumber \\
Z'' &=& - \ {T}\ e^{\,2\,X} \ + \ H_{3}^2\, e^{\,2\,Z} \ , \nonumber \\
0&=& 2\,(X')^2\ +\ 2\,X'\,Z' \ - \ \frac{7}{8}\,(U')^2 \ + \ {T} \, e^{\, 2\,X}\  + \ \frac{H_{3}^2}{2}\, e^{\,2\,Z} \ .
\eea
The original variables are then determined as
\bea
A &=& \frac{7}{16}\, U \ -\  \frac{1}{2}\, X \ +\  \frac{1}{8}\, Z \ , \nonumber \\
C &=& - \ \frac{5}{16}\, U \ + \  \frac{1}{2}\, X \ +\  \frac{1}{8}\, Z \ , \nonumber \\
\phi &=& \frac{7}{4}\, U \ -\  {2}\, X \ -\  \frac{3}{2}\, Z \ .
\eea
\subsubsection{\sc $SO(16)\times SO(16)$ heterotic with $p=1$ ``electric'' fluxes}

In this case $\gamma=\frac{5}{2}$ and $\left(a,b,c\right)=\left(16,8,16\right)$, so that the equations are
\bea
U'' &=& 0 \ , \nonumber \\
X'' &=& {2\,T}\, e^{\,2\,X}  \ + \ \frac{H_{3}^2}{2}\ e^{\,2\,Z}\ , \nonumber \\
Z'' &=& {T} \ e^{\,2\,X} \ + \  {H_{3}^2}\, e^{\,2\,Z} \ , \label{kh0_gen_gamma52} \\
0&=& \frac{7}{24}\, (U')^2 \ - \ \frac{2}{3} \left[(X')^2 - X'\,Z' + (Z')^2\right] \ + \ {T} \, e^{\, 2\,X}\  + \ \frac{H_{3}^2}{2}\, e^{\,2\,Z}  \ . \nonumber
\eea
The original variables are then determined as
\bea
A &=& \frac{1}{48} \left(7\,U \ - \ 16\,X  \ + \ 26\,Z \right) \ ,\nonumber \\
C &=& \frac{1}{16} \left( U \  - \ 2\, Z \right) \ , \nonumber \\
\phi &=& \frac{1}{12} \left(\,-\,7\, U \ + \ 16\, X \ - \ 2\, Z \right) \ .
\eea
\subsubsection{\sc $USp(32)$ and $U(32)$ orientifolds with $p=5$ ``magnetic fluxes''} \label{sec:solvable}

In these cases $\gamma=\frac{3}{2}$ and $\left(a,b,c\right)=\left(0,0,16\right)$. 
As a result, $U$ becomes linearly dependent on $X$ and $Z$, while the non--trivial second--order equations reduce to the far simpler form
\bea
X'' &=& 0 \ , \nonumber \\
Z'' &=&  H_{7}^2\, e^{\,2\,Z} \ . \label{kh0_gen_gamma32_2}
\eea
The dilaton is then determined by
\beq
\phi'' \ = \ \frac{3}{2}\ {T} \ e^{\, 2\,X}\ + \ \frac{H_{7}^2}{2}\ e^{\,2\,Z}  \ ,
\eeq
and the corresponding Hamiltonian constraint reads
\beq
\frac{2}{3}\, X'\left( X'\,+\, Z' \,-\,2\,\phi'\right) - \ \frac{1}{2}\, (Z')^2  \ + \ {T} \, e^{\, 2\,X}\  + \ \frac{H_{7}^2}{2}\, e^{\,2\,Z} \ = \ 0 \ .
\eeq
Finally, the original variables are related to $(X,Z,\phi)$ according to
\beq
A \ = \  \frac{2\,Z \ - \ \phi}{12} \ , \qquad
C \ = \ \frac{4(X - Z) \ - \ \phi}{12} \ . \label{ACp5}
\eeq
This case is remarkably simple, and will be discussed in detail in the next section.

\subsubsection{\sc $SO(16)\times SO(16)$ heterotic with $p=5$ ``magnetic'' fluxes}

In this case $\gamma=\frac{5}{2}$ and $\left(a,b,c\right)=\left(16,-16,16\right)$, and therefore the non--trivial second--order equations are
\bea
X'' &=& {2\,T} \, e^{\,2\,X}   \ - \ {H_{7}^2}\, e^{\,2\,Z}  \ , \nonumber \\
Z'' &=& - \ {2\,T} \ e^{\,2\,X} \ + \  {H_{7}^2} \ e^{\,2\,Z}  \ , \label{kh0_gen_gamma522_2}
\eea
which imply that
\beq
X'' \ + \ Z'' \ = \ 0 \ ,
\eeq
and the dilaton equation is then
\beq
\phi'' \ = \ \frac{5}{2}\ {T} \ e^{\, 2\,X}\ - \ \frac{H_{7}^2}{2}\ e^{\,2\,Z} \ .
\eeq
The original variables are related to $(X,Z,\phi)$ according to
\bea
A &=& \frac{2\,Z \ + \ \phi}{12} \ , \nonumber \\
C &=& \frac{4(X - Z) \ - \ 7\,\phi}{12} \ ,
\eea
and the Hamiltonian constraint is
\beq
\frac{2}{3} \, (X'-2 \phi')(X'+ Z') \ - \ \frac{1}{2}\, (Z')^2  \ + \ {T} \, e^{\, 2\,X}\  + \ \frac{H_{7}^2}{2}\, e^{\,2\,Z} \ = \ 0 \ .
\eeq
The special case $X=-Z$ can be related to elliptic functions.

\subsubsection{\sc $U(32)$ orientifold with $p=3$ ``dyonic'' flux}

    This model affords this additional option, since its massless spectrum includes a self--dual five--form field strength.
    In this case the non--trivial second--order equations follow from the special system of eqs.~\eqref{eqABC_sdual},  and read
\bea
U'' &=& 0 \ , \nonumber \\
X'' &=&  - \ \frac{H_{5}^2}{8}\, e^{\,2\,Z}  \ , \nonumber \\
Z'' &=& - \ \frac{T}{2} \ e^{\,2\,X} \ + \  \frac{H_{5}^2}{2} \ e^{\,2\,Z}  \ , \label{kh0_gen_gamma522}
\eea
while the Hamiltonian constraint of eq.~\eqref{ham_sdual} becomes
\beq
-\, {5} \, (U')^2 \ + \ {8}\, (X')^2 \ + \ 4\, X'\, Z' \ + \ {T} \, e^{\, 2\,X}\  + \ \frac{H_{5}^2}{4}\, e^{\,2\,Z} \ = \ 0 \ ,
\eeq
The original variables are then determined as
\bea
A &=&  \frac{1}{4}\, Z \ , \nonumber \\
C &=& \frac{1}{4} \left( - 6\,U \ + \ 8\,X \ + \ Z\right)\  , \nonumber \\
\phi &=& 10\, U \ -\  12\, X \ -\  {3}\, Z \ .
\eea

\section{\sc Orientifolds with Tension and Flux: an Exact Solution}\label{sec:exat5b}

In terms of the variables $X$, $Z$ and $\phi$, the equations for $D=10$, $p=5$ and the ``critical'' orientifold coupling $\gamma_c=\frac{3}{2}$ reduce to the simple form discussed in Section~\ref{sec:solvable}. Consequently
\beq
X \ = \ x_1\, r \ + \ x_2 \ ,
\eeq
where $x_1$ and $x_2$ are two integration constants. 
The equations for $Z$ and $\phi$ are  then
\bea
Z'' &=& H_{7}^2\, e^{\,2\,Z} \ , \nonumber \\
\phi'' &=& \frac{3}{2}\ T\,e^{2\,x_2} \ e^{\, 2\,x_1\,r}\ + \ \frac{1}{2}\ Z'' \ . \label{eqs_simple}
\eea
The function $Z$ is determined by the first of eqs.~\eqref{eqs_simple}, and also by the Hamiltonian constraint.
The dilaton profile $\phi$ is obtained by quadratures, integrating the second of eqs.~\eqref{eqs_simple},  while $A$, $B$ and $C$ follow from eqs.~\eqref{ACp5} and the harmonic gauge condition. As before, one must treat separately the case $x_1=0$.

\subsection{\sc The Special Vacua with $x_1=0$}

Now the second of eqs.~\eqref{eqs_simple} integrates to
\beq
\phi \ = \ \frac{1}{2}\, Z \ + \ \frac{3}{4}\ T\,e^{2\,x_2}\, r^2 \ + \ \chi_1\,r \ + \ \chi_2 \ ,
\eeq
where $\chi_1$ and $\chi_2$ are two more integration constants.
The Hamiltonian constraint then becomes
\beq
(Z')^2 \ = \ H_{7}^2\, e^{\,2\,Z} \ + \ {2\,T}\, e^{\,2\,x_2} \ . \label{hamhtx10}
\eeq
Referring to Appendix~\ref{app:deq}, this energy conservation condition corresponds to a particle with positive energy subject to an inverted Newtonian potential, and therefore the solution reads
\beq
Z \ = \ -  \log\left[H_7\,\rho \, \sinh \left(\frac{r}{\rho}\right) \right]\ ,
\eeq
with
\beq
\frac{1}{\rho^2} \ = \ {2\,T} \, e^{\,2\,x_2} \ \equiv \ 2\,T_0 \ .
\eeq
Notice that these solutions are deformations, induced by the tension $T$, of those in Section~6.1 of \cite{ms_vacuum_1}. Alternatively, they are deformations, induced by the form flux parametrized by $H_7$, of those in Section~\ref{sec:susybT}. The solution for $Z$ translates into
\beq
\phi \ = \ - \ \frac{1}{2}\, \log\left[H_7\,\rho \, \sinh \left(\frac{r}{\rho}\right) \right] \ + \ \frac{3\,T_0\, r^2}{4} \ + \ \chi_1\,r \ + \ \chi_2 \ ,
\eeq
while eqs.~\eqref{ACp5} and the harmonic gauge condition lead to
\bea
A &=& - \ \frac{1}{8} \, \log\left[H_7\,\rho \, \sinh \left(\frac{r}{\rho}\right) \right]
 \ - \ \frac{T_0\, r^2}{16} \ - \ \frac{1}{12} \left(\chi_1\,r \ + \ \chi_2\right) \ , \nonumber \\
 B &=& \frac{3}{8} \, \log\left[H_7\,\rho \, \sinh \left(\frac{r}{\rho}\right) \right] \ - \ \frac{9\,T_0\, r^2}{16}\ - \ \frac{3}{4}\left( \chi_1\,r\,+\,\chi_2\right) \ + \ x_2 \ , \\
 C&=& \frac{3}{8} \, \log\left[H_7\,\rho \, \sinh \left(\frac{r}{\rho}\right) \right] \ - \  \frac{T_0\, r^2}{16} \ - \ \frac{1}{12} \,\chi_1\,r \ + \ \frac{1}{12}\left(4
  \,x_2\,-\,\chi_2\right)
   .
 \eea
 Therefore the metric is
 \bea
 ds^2 &=& e^{\ - \  \frac{T_0\, r^2}{8} \,-\,\frac{\left(\chi_1\,r+\chi_2\right)}{6}} \left( \frac{dx^2}{\left[H_7\,\rho \, \sinh \left(\frac{r}{\rho}\right) \right]^\frac{1}{4}} \ + \ {\left[H_7\,\rho \, \sinh \left(\frac{r}{\rho}\right) \right]^\frac{3}{4}} \ e^{\,\frac{2}{3}\,x_2}\ d \vec{y}^{\,2} \right) \nonumber \\
 &+& e^{\ - \  \frac{9\,T_0\, r^2}{8} \,-\,\frac{3\left(\chi_1\,r\,+\,\chi_2\right)}{2}\,+\,2\,x_2}\ {\left[H_7\,\rho \, \sinh \left(\frac{r}{\rho}\right) \right]^\frac{3}{4}} \ dr^2 \ ,  \label{metricTH}
 \eea
while the string coupling and the form field strength are
\bea
e^\phi &=& \frac{e^{\,\frac{3\,T_0\, r^2}{4} \,+\,\chi_1\,r\,+\,\chi_2}}{\left[H_7\,\rho \, \sinh \left(\frac{r}{\rho}\right) \right]^\frac{1}{2}} \ , \nonumber \\
{\cal H}_{7} &=& e^{\,\phi}\,\star\Big( H_7\, dy_1\,dy_2\,dy_3\Big) \ = \  \frac{H_7\,\epsilon_{6}\, dr}{\left[H_7\,\rho \, \sinh \left(\frac{r}{\rho}\right)\right]^2} \ . \label{stringcTH}
\eea

\subsubsection{\sc Properties of the Solutions}

Let us first note that the solutions in the preceding equations contain an inessential parameter, $x_2$, which can be eliminated rescaling the variable $r$ and the $x$ and $y$ coordinates. As a result, these vacua depend on two independent parameters, $\chi_1$ and $\chi_2$. 

A second property is that the tension grants inevitably a \emph{finite} length along the $r$ direction, and also a \emph{finite} reduced Planck mass if the $y$ coordinates parametrize a torus, independently of $\chi_1$, although their actual values depend on $\chi_1$. However, the string coupling is inevitably singular at both ends.

For large values of $r$, the preceding expressions are dominated by the Gaussian terms due to the tadpole potential, and therefore approach the flux--free solutions of Section~\ref{sec:tnohbeta}
 \bea
 ds^2 &\sim & e^{\ - \  \frac{T_0\, r^2}{8}} \left( {dx^2} \ + \ d \vec{y}^{\,2} \right) \ + \  e^{\ - \  \frac{9\,T_0\, r^2}{8}} \ dr^2 \ , \nonumber \\
 e^\phi &\sim& e^{\,\frac{3\,T_0\, r^2}{4}} \ , \nonumber \\
 {\cal H}_7 &\sim& 0 \ .
 \eea
 Letting
 \beq{}{}
 \frac{3}{4}\,T_0\,r^2 \ = \ u \ ,
 \eeq
 the dominant behavior is captured by
  \bea
 ds^2 &\sim & e^{\ - \  \frac{u}{6}} \left( {dx^2} \ + \ d \vec{y}^{\,2} \right) \ + \ \frac{1}{3\,T_0}\ e^{\ - \  \frac{3}{2}\,u} \ du^2 \ , \nonumber \\
 e^\phi &\sim& e^{\,u} \ , \nonumber \\
 {\cal H}_7 &\sim& 0 \ , \label{fluxx10asympt}
 \eea
where the numerical coefficient is front of $du^2$ is not significant, since it can be changed by a translation of $u$. This background thus coincides with eqs.~\eqref{ds2phibeta0_2}, which describe the asymptotics of the flux--free solutions of Section~\ref{sec:tnohbeta}, and actually with eqs.~\eqref{dmsol2}, which describe their isotropic form. As was shown there, this is actually the asymptotics of the tension--free and flux--free solutions of~\cite{ms_vacuum_1} reviewed in Appendix~\ref{app:kasner}.

Close to $r=0$, the Gaussian terms due to the tadpole potential become negligible compared to the other contributions, and the background approaches
 \bea
 ds^2 &=&  \frac{dx^2}{\left[H_7\,r \right]^\frac{1}{4}} \ + \ {\left[H_7\,r \right]^\frac{3}{4}} \ d \vec{y}^{\,2} \ + \  e^{\,-\,\frac{3}{2}\,\chi_2}\ {\left[H_7\,r\right]^\frac{3}{4}} \ dr^2 \ , \label{metzeroenx10}
 \eea
up to rescalings of the $x$ and $y$ coordinates, 
while the string coupling and the form field strength are dominated by
\bea
e^\phi &=& \frac{e^{\,\chi_2}}{\left[H_7\,r \right]^\frac{1}{2}} \ , \nonumber \\
{\cal H}_{7} &=& e^{\,\phi}\,\star\Big( H_7\, dy_1\,dy_2\,dy_3\Big) \ = \  \frac{H_7\,\epsilon_{6}\, dr}{\left[H_7\,r\right]^2} \ . \label{dilformzeroenx10}
\eea
This is the limiting behavior near $r=0$ of the $p=5$ fluxed background in supersymmetric strings described in~\cite{ms_vacuum_1}. Moreover, as we stressed there, when considered in the whole region $r>0$ these equations would describe a supersymmetric vacuum.

Summarizing, the solution in eqs.~\eqref{metricTH} and \eqref{stringcTH} interpolates between a supersymmetric vacuum at $r=0$ and a tension--free and flux--free isotropic vacuum with broken supersymmetry at large values of $r$, and the string coupling is infinite in both limits. 

\subsection{\sc Vacua with $x_1 \neq 0$}

In this case the solution for $X$ is a linear function of $r$,
\beq
X \ = \  x_1\,r \ + \ x_2 \ , 
\eeq
and the second of eqs.~\eqref{eqs_simple} gives
\beq
\phi \ = \ \frac{1}{2}\, Z\ + \ \frac{3\,T \, e^{\,2\,x_2}}{8\, x_1^2}\,\, e^{\,2\,x_1\,r} \ + \ \chi_1\,x_1\,r \ + \ \chi_2 \ , \label{phix1}
\eeq
where $\chi_1$ and $\chi_2$ are integration constants. The Hamiltonian constraint reduces to
\beq
(Z')^2 \ = \ H_{7}^2\, e^{\,2\,Z} \ + \ \frac{4\,x_1^2(1 \ - \ 2\, \chi_1)}{3} \ . \label{ham_complete}
\eeq
One is thus led once more to the Newtonian model of Appendix~\ref{app:deq}, with an inverted exponential potential and a total energy
\beq
E \ = \ \frac{4\,x_1^2(1 \ - \ 2\, \chi_1)}{3} \ , \label{energy_special_x1}
\eeq
which now has no definite sign. Consequently, there are three classes of solutions, which we shall discuss shortly. We write the solution for $Z$ in the form
\beq
Z \ = \ - \ \log\left[ f(r)\right] \ , \label{Zf}
\eeq
so that, using eqs.~\eqref{phix1}, \eqref{Zf}, eqs.~\eqref{ACp5} and the harmonic gauge condition
\bea
A &=& - \ \frac{1}{8} \, \log\left[f(r) \right]
 \ - \ \frac{T\,e^{2\,x_2}}{32\,x_1^2}\ e^{2\,x_1\,r} \ - \ \frac{1}{12} \left(\chi_1\,x_1\,r \ + \ \chi_2\right) \ , \nonumber \\
 B &=& \frac{3}{8} \, \log\left[f(r) \right] \ - \ \frac{9\,T\,e^{2\,x_2}}{32\,x_1^2}\ e^{2\,x_1\,r}\ - \ \frac{3}{4}\left( \chi_1\,x_1\,r\,+\,\chi_2\right) \ + \ x_1\,r \ + \ x_2 \ , \nonumber \\
 C&=& \frac{3}{8} \, \log\left[f(r) \right] \ - \  \frac{T\,e^{2\,x_2}}{32\,x_1^2} \ e^{2\,x_1\,r} \ - \ \frac{1}{12} \left(\chi_1\,x_1\,r \,+\,\chi_2\right) + \ \frac{1}{3}\left(x_1\,r\,+\,
  \,x_2\right)
   \ , \nonumber \\
   \phi &=& - \ \frac{1}{2}\, \log\left[f(r)\right] \ + \ \frac{3\,T \, e^{\,2\,x_2}}{8\, x_1^2}\,\, e^{\,2\,x_1\,r} \ + \ \chi_1\,x_1\,r \ + \ \chi_2 \ .
 \eea
Letting now
\beq
g(r) \ = \ \frac{T\,e^{2\,x_2}}{16\,x_1^2}\ e^{2\,x_1\,r} \ + \ \frac{1}{6} \left( \chi_1\,x_1\,r \,+\,\chi_2\right) \ ,
\eeq
the background takes the form
    \bea
    ds^2 &=& \ e^{\,-\,g(r)} \left( \frac{dx^2}{[f(r)]^\frac{1}{4}} \ + \  e^{\frac{2}{3}\,x_1\,r}\, [f(r)]^\frac{3}{4}\, d\vec{y}^{\,2} \right)  \ + \ e^{\,-\,9\, g(r)}\, e^{2\left(x_1\,r\,+\,x_2\right)}\,[f(r)]^\frac{3}{4}\, dr^2 \ , \nonumber \\
    e^\phi &=&  \ \frac{e^{\,6\, g(r)}}{[f(r)]^\frac{1}{2}} \ , \nonumber \\
    {\cal H}_{7} &=& H_7\, \frac{\epsilon_{6}\, dr}{\left[f(r) \right]^2} \ ,
    \eea
    where we have rescaled the $y$ coordinates to absorb contributions depending on $x_2$.
    
\subsubsection{\sc Properties of the Solutions}
    
These solutions actually depend on three parameters, $\chi_1$, $\chi_2$ and $x_2$, and a two-valued discrete one. Indeed, $\left|x_1\right|$ can be eliminated combining a rescaling of the $r$ coordinate with redefinitions of $x_2$ and $\chi_2$, together with rescalings of the $x$ and $y$ coordinates.
However, the sign of $x_1$ leads to different solutions, and in fact has important effects, since the flux introduces a restriction of $r$ to the region $0<r<\infty$. 

The function $f(r)$ depends on the energy $E$ in eq.~\eqref{energy_special_x1}.
Letting
\beq{}{}
\frac{1}{\rho} \ = \ \sqrt{\left|E\right|} \ = \ \frac{2 \left|x_1\right|}{\sqrt{3}}\ \sqrt{\left|1 \ - \ 2\, \chi_1\right|} \ ,
\eeq
one must distinguish, as usual, three cases:
\begin{itemize}
\item[1. ] If the energy $E >0$, which is the case if $\chi_1 < \frac{1}{2}$, the solution of eq.~\eqref{ham_complete} is
    \beq
    f(r) \ = \ H_7\,\rho \, \sinh \left(\frac{r}{\rho}\right)\ ,
    \eeq
    where $0< r< \infty$.
\item[2. ] If the energy $E =0$, which is the case if $\chi_1 = \frac{1}{2}$, 
    \beq
    f(r) \ = \ H_7\,r\ ,
    \eeq
    where again $0< r< \infty$.
    \item[3. ] Finally, if the energy $E < 0$,  which is the case if $\chi_1 > \frac{1}{2}$, 
    \beq
    f(r) \ = \ H_7\,\rho \, \sin \left(\frac{r}{\rho}\right)\ ,
    \eeq
    where now $0 < r < \pi\,\rho$.
\end{itemize}

The behavior near $r=0$, which is a curvature singularity where the radial variable ends, is dominated by the magnetic flux and is universal. In all cases
\beq
f(r) \ \sim \ H_7\, r \ ,
\eeq
so that
    \bea
    ds^2 &=& \ \left( \frac{dx^2}{[H_7\,r]^\frac{1}{4}} \ + \   [H_7\,r]^\frac{3}{4}\, d\vec{y}^{\,2} \right)  \ + \ e^{\,2\,x_2 \,-\,\frac{3}{2}\,\chi_2}\,[H_7\,r]^\frac{3}{4}\, dr^2 \ , \nonumber \\
    e^\phi &=&  \ \frac{e^{\,\chi_2}}{[H_7\,r]^\frac{1}{2}} \ , \nonumber \\
    {\cal H}_{7} &=& H_7\, \frac{\epsilon_{6}\, dr}{\left[H_7\,r\right]^2} \ , \label{exactp5lim0}
    \eea
up to rescalings of the $x$ and $y$ coordinates, and up to a redefinition of $\chi_2$. The dependence on $x_2$ can eliminated by a further combined redefinition of $H_7$ and $r$, after which this expression coincides with eqs.~\eqref{metzeroenx10} and \eqref{dilformzeroenx10}.
This is again the limiting behavior near $r=0$ of the $p=5$ fluxed background described in~\cite{ms_vacuum_1}, and when considered in the whole region $r>0$ eqs.~\eqref{exactp5lim0} it would describe a supersymmetric vacuum.

For $E<0$ the behavior at the other end is similarly governed by the supersymmetric solution. On the other hand, for $E \geq 0$, and thus for $\chi_1 \leq \frac{1}{2}$, the range of $r$ is unbounded and the behavior depends on the sign of $x_1$. 

\subsubsubsection{\sc Behavior for Large $r$ with $\chi_1<\frac{1}{2}$ and $x_1>0$}
Let us begin with the first case, where $E$ is positive and $x_1>0$. Letting
    \beq{}{}
    \frac{u}{6} \ = \ \frac{T\,e^{2\,x_2}}{16\,x_1^2}\ e^{2\,x_1\,r} \ ,
    \eeq
    asymptotically $g$ is dominated by
    \beq{}{}
    g(r) \ \sim \ \frac{u}{6} \ ,
    \eeq
and consequently, as $r \to +\infty$ the background approaches the isotropic behavior 
\bea{}{}
ds^2 &\sim& e^{\,-\,\frac{u}{6}} \left( dx^2 + dy^2\right) \ + \ \frac{1}{4\,x_1^2}\ e^{\,-\,\frac{3}{2}\,u}\ du^2 \ ,\nonumber \\
e^\phi &\sim& e^u \ , \qquad
{\cal H}_7 \ \sim \ 0 \ ,
\eea
and the string coupling diverges.
Up to a shift of $u$, this is the same type of isotropic behavior found for $x_1=0$ in eqs.~\eqref{fluxx10asympt}, and in fact it is again the asymptotic behavior of the Dudas-Mourad solution of~\cite{dm_vacuum}, which is also the by-now familiar isotropic tension--free and flux--free strong--coupling solution of~\cite{ms_vacuum_1}, reviewed in Appendix~\ref{app:kasner}. 

\subsubsubsection{\sc Behavior for Large $r$ with $\chi_1\leq \frac{1}{2}$ and $x_1<0$}
 On the other hand, if $x_1<0$,
    \beq{}{}
    g(r) \ \sim \  - \ \frac{1}{6}\, \chi_1\,\left|x_1\right|\,r \ ,
    \eeq
    and now $f(r)$ does play a role in the asymptotic region. Taking into account that
    \beq{}{}
    \chi_1 \ = \ \frac{1}{2}\left[1 \ - \ \frac{3}{\left(2 \rho \left|x_1\right|\right)^2}\right] \ , \label{chi1xi1}
    \eeq
  one finds  
\beq{}{}
e^{2B} \ \sim \ e^{\,-\,\frac{5\,r}{4\left|x_1\right|\rho^2}\left[ \left(\rho \left|x_1\right|  \ - \ \frac{3}{10}\right)^2 \ + \ \frac{9}{25} \right]} \ ,
\eeq
and therefore the length of the interval is always finite. Moreover, the six--dimensional effective Planck mass is always finite with an internal torus, since its asymptotic behavior is governed by
\beq{}{}
e^{2(B-A)} \ \sim \  e^{\,-\,\frac{4\,r}{3\left|x_1\right|\rho^2}\left[ \left(\rho \left|x_1\right|  \ - \ \frac{3}{8}\right)^2 \ + \ \frac{15}{64} \right]} \ ,
\eeq
which is always integrable at the right end. 
Finally, the asymptotic behavior of the string coupling is determined by
\beq{}{}
e^\phi  \ \sim \ e^{\,-\,\frac{r}{2\left|x_1\right|\rho^2}\left[ \left(\rho \left|x_1\right| \ + \ \frac{1}{2}\right)^2 \ - \ 1 \right]} \ ,
\eeq
so that there is weak coupling at the right end of the interval if $\rho \left|x_1\right| > \frac{1}{2}$. 

These solutions for $x_1<0$ were brought about by the presence of the flux, which introduces a singularity at $r=0$, so that the available region is the half-line $r>0$. This should be contrasted with the situation discussed in Section~\ref{sec:susybT}, where the sign of $x_1$ was immaterial, since in that case the range of $r$ was the whole real axis.

One can study this limiting behavior for large values of $r$ in terms of
\beq{}{}{}{}
\xi \ =\  e^{\,-\,\frac{5\,r}{8\left|x_1\right|\rho^2}\left[ \left(\rho \left|x_1\right|  \ - \ \frac{3}{10}\right)^2 \ + \ \frac{9}{25} \right]} \ ,
\eeq
so that when $\xi$ approaches zero
\bea{}{}{}{}
ds^2 & \sim & \xi^{2 \alpha_A} \, dx^2 \ + \ d \xi^2 \ + \ \xi^{2 \alpha_c} \, d{\vec y}^{\,2} \ , \nonumber \\
e^\phi &\sim & \xi^{\alpha_\phi} \ , 
\eea
where
\bea{}{}{}{}
 \alpha_A &=& \frac{1}{15} \, \frac{3 \ - \ \left(\rho|x_1| \ - \ \frac{3}{2}\right)^2}{\left(\rho \left|x_1\right|  \ - \ \frac{3}{10}\right)^2 \ + \ \frac{9}{25}} \ , \nonumber \\
\alpha_C &=& \frac{7}{15} \, \frac{\left(\rho \left|x_1\right| \ - \ \frac{9}{14}\right)^2 \,-\, \frac{15}{49}}{\left(\rho \left|x_1\right|  \ - \ \frac{3}{10}\right)^2 \ + \ \frac{9}{25}}\ , \nonumber \\
\alpha_\phi &=& \frac{4}{5}\, \frac{\left(\rho \left|x_1\right| \ + \ \frac{1}{2}\right)^2 \ - \ 1}{\left(\rho \left|x_1\right|  \ - \ \frac{3}{10}\right)^2 \ + \ \frac{9}{25}} \ .
\eea

These expressions satisfy the constraints in Appendix~\ref{app:kasner}
with
\beq
\sin\theta=\frac{3}{5}\, \frac{\left(\rho \left|x_1\right| \ + \ \frac{1}{2}\right)^2 \ - \ 1}{\left(\rho \left|x_1\right|  \ - \ \frac{3}{10}\right)^2 \ + \ \frac{9}{25}} \ ,
\eeq
so that this limiting behavior is captured once more by the tension--free and flux--free solutions in~\cite{ms_vacuum_1}, reviewed in Appendix~\ref{app:kasner}. Contrary to the $x_1\geq0$ case, the limiting behavior is now anisotropic and depends on $\chi_1$, or equivalently on the product $\rho |x_1|$, which are related to one another in eq.~\eqref{chi1xi1}, and in addition the string coupling can be weak, as we have seen.
Notice that the case $\chi_1=\frac{1}{2}$, corresponding to $\rho\rightarrow \infty$, has an interval of finite length, and in the neighborhood of its $r\rightarrow \infty$
boundary the background is captured by
\beq 
ds^2\sim d\xi^2\ +\ \xi^{-\frac{2}{5}}dx^2\ + \ \xi^{\frac{14}{15}}dy^2,\qquad e^\phi\sim \xi^{\frac{4}{5}},
\eeq 
so that the string coupling vanishes in the limit.

In brief, we have different types of solutions depending on whether $x_1$ vanishes, is positive or negative.
Within each family, the solutions also depend on  three real parameters, $\chi_1,\chi_2$ and $x_2$, and have a common behavior near the $r=0$ singularity, which is dominated by the supersymmetric solution with flux.
They also have in common a finite length of the interval.
The other properties concerning the string coupling and the behavior at the other boundary are parameter-dependent. For $x_1=0$ the background at the other end approaches the strong coupling, isotropic, flux-free and tension-free solution of~\cite{ms_vacuum_1}.
When $\chi_1 \leq \frac{1}{2}$ they are also captured by tensionless solutions: for $x_1< 0$, the background is anisotropic and the limiting behavior of the coupling can be finite, infinite or zero, while for $x_1>0$  the background is isotropic with an infinite coupling. Finally when $\chi_1>\frac{1}{2}$ the background behaves as at $r=0$, and the system is again dominated at the other end by the supersymmetric background with flux.

\subsection{\sc Cosmological Solutions}

One can obtain cosmological counterparts of the preceding solutions by an analytic continuation. This is not possible for $x_1=0$, as can be seen from eq.~\eqref{hamhtx10}, but it is possible for $x_1 \neq 0$. The Hamiltonian constraint then takes the form
\beq
(Z')^2 \ = \ - \ H_{7}^2\, e^{\,2\,Z} \ + \ \frac{4\,x_1^2(1 \ - \ 2\, \chi_1)}{3} \ , \label{ham_complete_c}
\eeq
so that now $\chi_1 < \frac{1}{2}$ and, in the conventions of Appendix~\ref{app:deq}, the potential and the energy are now positive. The solutions read
\bea
A &=& - \ \frac{1}{8} \, \log\left[f(\tau) \right]
 \ + \ \frac{T\,e^{2\,x_2}}{32\,x_1^2}\ e^{2\,x_1\,\tau} \ - \ \frac{1}{12} \left(\chi_1\,x_1\,\tau \ + \ \chi_2\right) \ , \nonumber \\
 B &=& \frac{3}{8} \, \log\left[f(\tau) \right] \ + \ \frac{9\,T\,e^{2\,x_2}}{32\,x_1^2}\ e^{2\,x_1\,\tau}\ - \ \frac{3}{4}\left( \chi_1\,x_1\,\tau\,+\,\chi_2\right) \ + \ x_1\,\tau \ + \ x_2 \ , \\
 C&=& \frac{3}{8} \, \log\left[f(\tau) \right] \ + \  \frac{T\,e^{2\,x_2}}{32\,x_1^2} \ e^{2\,x_1\,\tau} \ - \ \frac{1}{12} \left(\chi_1\,x_1\,\tau \,+\,\chi_2\right) + \ \frac{1}{3}\left(x_1\,\tau\,+\,
  \,x_2\right)
   \ , \nonumber \\
   \phi &=& - \ \frac{1}{2}\, \log\left[f(\tau)\right] \ - \ \frac{3\,T \, e^{\,2\,x_2}}{8\, x_1^2}\,\, e^{\,2\,x_1\,\tau} \ + \ \chi_1\,x_1\,\tau \ + \ \chi_2 \ , \label{cosmo_gen5}
 \eea
with
\beq
f(\tau) \ = \ H_7\,\,\rho \, \cosh \left(\frac{\tau}{\rho}\right) \ .
\eeq
where
\beq
\frac{1}{\rho} \ = \ \frac{2\,x_1}{\sqrt{3}}\, \sqrt{1 \ - \ 2\, \chi_1} \ .
\eeq
Now $- \infty < \tau < \infty$, and we take $x_1>0$ in order to have the Universe expand for increasing values of $\tau$. Actually, $x_1$ can be eliminated rescaling $\tau$, so that one can set everywhere $x_1=1$.
$\chi_1$ can then be expressed in terms of $\rho$ as
\beq
\chi_1 \ = \ \frac{1}{2}  \ - \ \frac{3}{8\,\rho^2} \ .
\eeq
\begin{figure}[ht]
\centering
\begin{tabular}{cc}
\includegraphics[width=60mm]{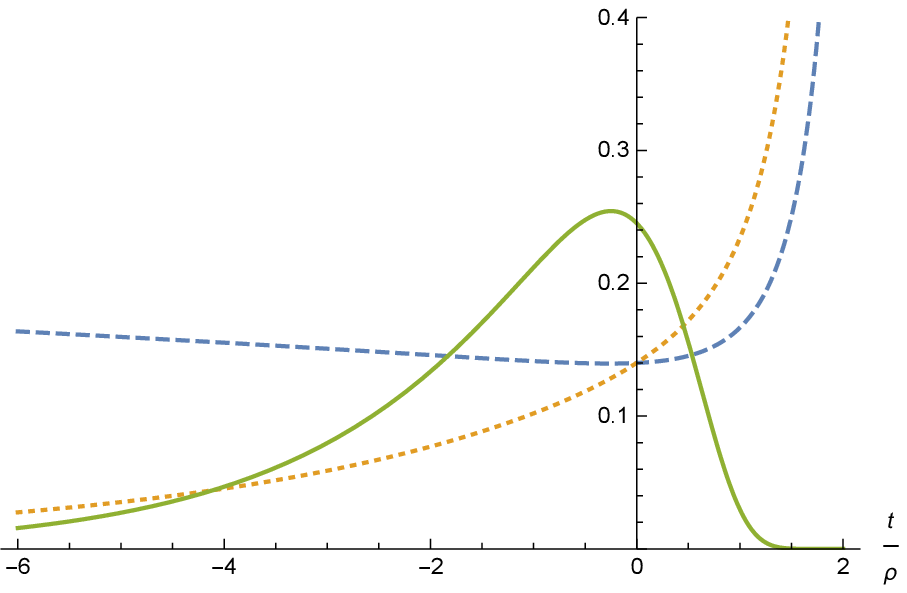} &
\includegraphics[width=60mm]{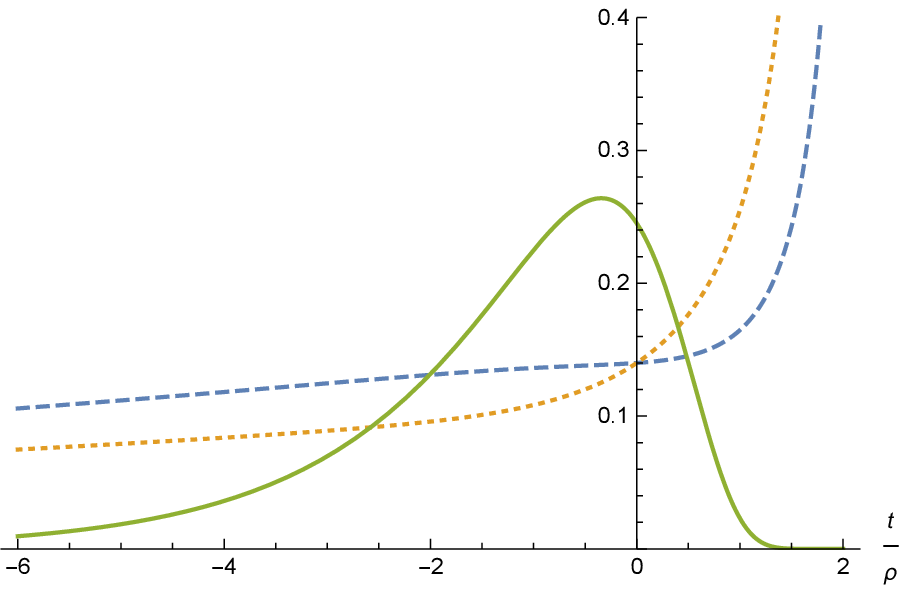} \\
\end{tabular}
\caption{\small $e^A$ (dashed), $e^C$ (dotted) and $e^\phi$ (solid) for an anisotropic four--dimensional \emph{climbing scalar} cosmology with ${\cal H}_7$ flux where space--time contracts while the $y$ internal space expands  when emerging from the initial singularity ($\rho=6$, left panel), and for an anisotropic \emph{climbing scalar} cosmology where all directions expand when emerging from the initial singularity ($\rho=1.5$, right panel).}
\label{fig:flux1}
\end{figure}
\begin{figure}[ht]
\centering
\begin{tabular}{cc}
\includegraphics[width=60mm]{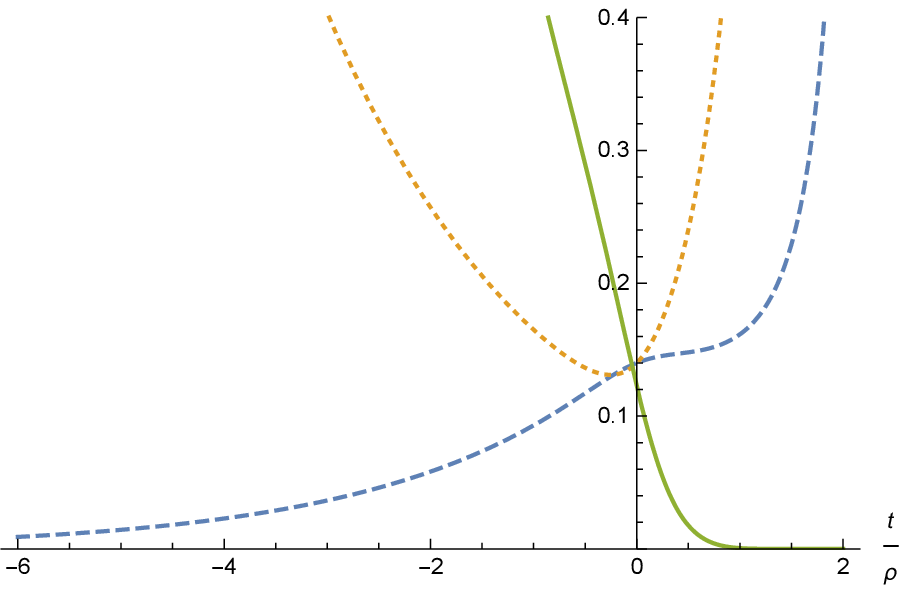} &
\includegraphics[width=60mm]{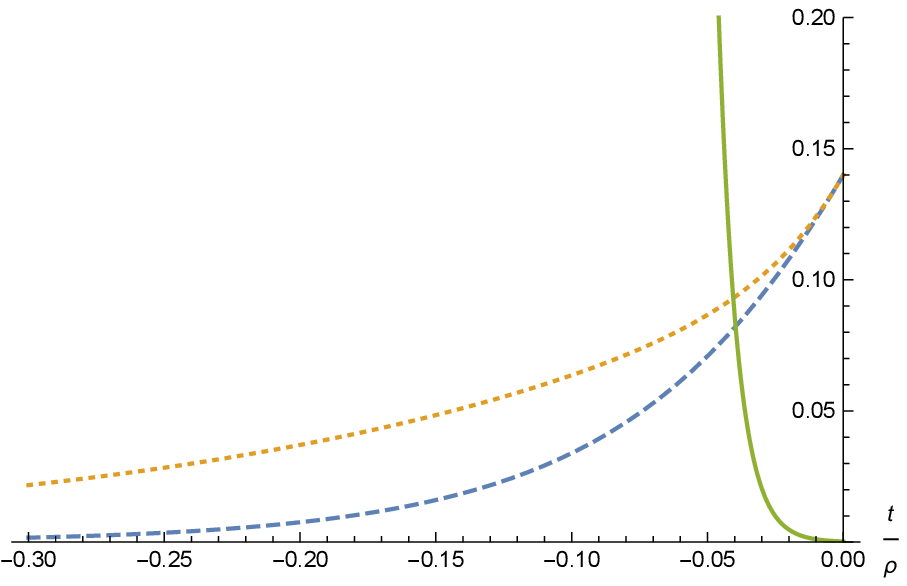} \\
\end{tabular}
\caption{\small $e^A$ (dashed), $e^C$ (dotted) and $e^\phi$ (solid) for an anisotropic four--dimensional \emph{descending scalar} cosmology with ${\cal H}_7$ flux where space--time expands while the $y$ internal space contracts when emerging from the initial singularity ($\rho=0.4$, left panel), and for an anisotropic \emph{descending scalar} cosmology where all directions expand when emerging from the initial singularity ($\rho=0.2$, right panel).}
\label{fig:flux2}
\end{figure}

\subsubsection{\sc Properties of the Solutions}

Note that at $\tau \to + \infty$, the behavior of the background is dominated by the tadpole contribution, and $B \to + \infty$ while $\phi \to \ - \ \infty$. The dominant behavior is isotropic and, in terms of the cosmic time $t \to +\infty$,
\bea{}{}
ds^2 &\sim& - dt^2 \ + \ t^\frac{2}{9}\left(dx^2 \ + \ dy^2\right) \ , \nonumber \\
e^\phi &\sim& t^{\,-\,\frac{4}{3}} \ .
\eea
On the other hand, as $\tau \to - \infty$ the behavior is dominated by the linear terms in eqs.~\eqref{cosmo_gen5}, so that
\bea
B & \sim & - \ \frac{|\tau|}{8}\left[ \left( \frac{3}{2\,\rho} \ - \ 1\right)^2 \ + \  {4} \right] \ \equiv \ - \ b \left|\tau\right| \  <   \ 0 \ , \label{Bb}
\eea
and the initial singularity lies at a finite amount of cosmic time in the past. Moreover
\beq
\phi \ \sim \ \frac{|\tau|}{2\,\rho^2} \left[\frac{1}{2} \ - \ \rho \right]\left[\frac{3}{2}\ + \ \rho \right] \ \equiv \ f \,|\tau| \ , \label{phirho}
\eeq
so that one can have \emph{climbing} behavior, and thus weak coupling, for $\rho> \frac{1}{2}$ and \emph{descending} behavior, and thus strong coupling, for $\rho < \frac{1}{2}$. There is also a third option: when $\rho = \frac{1}{2}$, the dilaton approaches a constant value at the initial singularity. 

The behavior close to the initial singularity is captured by
\bea{}{}
A &\sim&  \frac{1}{24\,\rho^2}\left[\rho \ - \ \frac{3}{2}\ - \ \sqrt{3}\right]\left[\rho \ - \ \frac{3}{2}\ + \ \sqrt{3}\right] \left|\tau\right|  \ \equiv \ a \left|\tau\right|\ , \nonumber \\
C &\sim& - \ \frac{7}{24\,\rho^2}\left[\rho \ - \ \frac{9}{14}\ - \ \frac{\sqrt{15}}{7} \right]\left[\rho \ - \ \frac{9}{14}\ + \ \frac{\sqrt{15}}{7}\right]\left|\tau\right| \ \equiv \ c \left|\tau\right| \ ,
\eea
which defines $a$ and $c$, while eqs.~\eqref{Bb} and \eqref{phirho} define $b$ and $f$, and the metric has a Kasner--like form, with
\bea{}{}
ds^2 &\sim & - \ dt^2 \ + \ t^{\,-\,\frac{2a}{b}} \, dx^2 \ + \ t^{\,-\,\frac{2c}{b}} \, d\vec{y}^2 \ , \nonumber \\
e^\phi & \sim & t^{\,-\,\frac{f}{b}} \ .
\eea
There are different options close to the initial singularity, depending of the value of $\rho$, which determines the signs of $a$, $c$ and $f$.
\begin{itemize}
\item For $\rho > \frac{3}{2} \ + \sqrt{3}$ the $x$-space contracts and the $y$-space expands, while the dilaton has a climbing behavior.
\item For $\frac{9}{14} \ + \ \frac{\sqrt{15}}{7}  < \rho < \frac{3}{2} \ + \ \sqrt{3}$ the $x$ and $y$-spaces expand, while the dilaton  has a climbing behavior.
\item For $\frac{1}{2} < \rho < \frac{9}{14} \ + \ \frac{\sqrt{15}}{7}$  the $x$-space expands, the $y$-space contracts, while the dilaton has a climbing behavior.
\item $\frac{9}{14} \ - \ \frac{\sqrt{15}}{7} < \rho < \frac{1}{2}$ the $x$-space expands, the $y$-space contracts, while the dilaton has a descending behavior.
\item For $\rho < \frac{9}{14} \ - \ \frac{\sqrt{15}}{7}$ the $x$ and $y$-spaces expand while the dilaton has a descending behavior.
\end{itemize}

Note that both in the far past and in the far future these solutions approach the results of Section~6 of~\cite{ms_vacuum_1}. The behavior is anisotropic in the far past, while it is isotropic in the far future. These different options are displayed in figs.~\ref{fig:flux1} and \ref{fig:flux2}.

\section{\sc  Conclusions}\label{sec:conclusions}
\vskip 12pt

In this paper, we have discussed in detail a class of instructive compactifications to $p+1$--dimensional Minkowski spaces in the presence of the dilaton tadpole potential of eq.~\eqref{tadpole_pot}, and also the cosmological models that can be obtained from them via analytic continuations. We have traced the behavior of the system as the exponent $\gamma$ spans the two regions $\gamma < \frac{3}{2}$ and $\gamma  \geq \frac{3}{2}$. Although only values with $\gamma \geq \frac{3}{2}$ are directly relevant to String Theory, we have pursued this scrutiny that has proved effective, in the past~\cite{dks}, in clarifying the emergence of the climbing behavior in the isotropic cosmologies of these systems. Here it unveiled the special role played, in the dynamics, by the ``critical'' value $\gamma_c=\frac{3}{2}$ that pertains to the two orientifold models of~\cite{susy95} and \cite{bsb}: it leads to the decoupling of one combination of the three functions $A$, $C$ and $\phi$ in the general case, and the behavior in the two ranges $\gamma<\gamma_c$ and $\gamma>\gamma_c$ is markedly different,  both in spatial profiles and in their cosmological counterparts. 
\begin{table}[h!]
\centering
\small{
\begin{tabular}{||c | c c c | c c c||} 
 \hline
Solution & Left(L) & Left(T) & Left($g_s$) & Right(L) & Right(T) & Right($g_s$) \\ [0.5ex] 
 \hline\hline 
$\gamma=\gamma_c$, $\beta=0$ & F & is. (0)& $\infty$ & F & is. (0) & $\infty$ \\
$\gamma = \gamma_c$, $\beta \neq 0$, $p=8$ & F & is. (0) & $0$ & F & is. (0) & $\infty$ \\
$\gamma = \gamma_c$, $\beta \neq 0$, $p < 8$ & F & anis. (0) & $A$ & F & is. (0) & $\infty$ \\ \hline
$\gamma < \gamma_c,\ p=8$ &F & is. (0) & 0 & F & is. (0) & $\infty$\\
$\gamma < \gamma_c,\ \cos\eta <\,-\,\frac{\gamma}{\gamma_c}$ & F & anis. (0) & $0$ & F & anis. (0) & $\infty$ \\
$\gamma < \gamma_c,\ \cos\eta >\,\frac{\gamma}{\gamma_c}$ & F & anis. (0) & $\infty$ & F & anis. (0) & $0$ \\
$\gamma < \gamma_c,\ \left|\cos\eta\right| <\,\frac{\gamma}{\gamma_c}$ & F & anis. (0) & $\infty$ & F & anis. (0) & $\infty$ \\ \hline
$\gamma > \gamma_c,\ p=8$, $E=0$ & F & is. ($\neq 0$) & $\infty$ & $\infty$ & is. ($\neq 0$) & 0\\
$\gamma > \gamma_c$, $E=0$, $\phi_1>0$ & F & is. ($\neq$ 0) & $\infty$ & F & anis. (0) & $\infty$ \\
$\gamma > \gamma_c$, $E=0$, $\phi_1 = 0$ & F & is. ($\neq$ 0) & $\infty$ & $\infty$ & is. ($\neq$ 0) & $0$ \\
$\gamma > \gamma_c$, $E=0$, $\phi_1 < 0$ & F & is. ($\neq$ 0) & $\infty$ & $\infty$ & anis. (0) & $0$ \\
$\gamma > \gamma_c,\ p=8$, $E>0$, $(u)$ & F & is. ($\neq 0$) & $\infty$  & F & is. (0) & 0\\
$\gamma > \gamma_c,\ p=8$, $E>0$, $(l)$ & F & is. ($\neq 0$) & $\infty$  & $\infty$ & is. (0) & 0\\
$\gamma > \gamma_c$, $E>0$, $(u)$ & F & is. ($\neq$ 0) & $\infty$ & F & anis. (0) & $A$  \\
$\gamma > \gamma_c$, $E>0$, $(l)$ & F & is. ($\neq$ 0) & $\infty$ & $\infty$ & anis. (0) & $0$  \\
$\gamma > \gamma_c$, $E<0$,  & F & is. ($\neq$ 0) & $\infty$ & $F$ & is. ($\neq$ 0) & $\infty$ 
 \\ [1ex] 
 \hline
\end{tabular}
}
\caption{A summary of the static solutions of Section~\ref{sec:susybT} with tension T and no form flux, with emphasis on their limiting behavior at their two singularities. The two groups of columns collect details on the left and right ends of the internal interval. F and $\infty$ indicate finite or infinite values for their distances from the origin $r=0$. Moreover, the labels $0,A,\infty$ indicate that the string coupling $g_s$ vanishes, is arbitrary (\emph{i.e.} can be zero, finite or infinite depending on the values of some parameters), or diverges in the limits. In addition, ``is. (0)'' indicates that the vacuum approaches asymptotically the isotropic solution of Section 5 of~\cite{ms_vacuum_1}, while ``anis. (0)" indicates that it approaches one of the corresponding anisotropic solutions, and ($\neq$0) indicates that the tension is not sub-dominant close to the singularity. Finally, $(u)$ and $(l)$ refer to the upper and lower branches of the parametrization for $\gamma> \gamma_c$.}
\label{table:tab_F}
\end{table}

We have treated the two special cases $\gamma=\frac{3}{2}$ and $\gamma=\frac{5}{2}$ in a similar fashion, as in~\cite{dm_vacuum} and then in~\cite{russo} and~\cite{dks}, since the second type of contribution is the dominant one for the heterotic $SO(16)\times SO(16)$ model of~\cite{so16so16}, although it arises from a genuine quantum effect. There are other well--known contexts where this mixing of orders plays a role in String Theory, for example the Green--Schwarz mechanism~\cite{gs} and, even more importantly in this context, the Fischler--Susskind proposal~\cite{fsuss} to deal with vacuum redefinitions in the absence of protecting mechanisms. 
The two values $\gamma=\frac{3}{2}$ and $\gamma=\frac{5}{2}$ always arise in the leading contributions to models with broken supersymmetry, so that the considerations in this paper have actually a wide range of applicability. This is true, in particular, for Scherk-Schwarz~\cite{scherkschwarz} compactifications, which afford detailed treatments in String Theory~\cite{scherkschwarz_closed}, even in the presence of open strings~\cite{scherkschwarz_open}.

When the tadpole potential~\eqref{tadpole_pot} is taken into account, the static solutions always involve internal intervals with 
singularities at one or both ends. We have seen that their lengths can be finite, as in the Dudas--Mourad vacuum of~\cite{dm_vacuum}, or infinite, depending on the value of $\gamma$ and on the values of the integration constants. 
We have presented a careful scrutiny of the different families of these exact solutions, identifying their moduli spaces and taking a close look at their asymptotics. Within the region $\gamma \leq \frac{3}{2}$, the limiting behavior is always captured by the tensionless solutions in~\cite{ms_vacuum_1}. However, within the complementary region $\gamma> \frac{3}{2}$ isotropic limiting behaviors are not always captured by the tensionless solutions in~\cite{ms_vacuum_1}. On the other hand, whenever the limiting behavior is {anisotropic}, it is captured again by the tensionless solutions in~\cite{ms_vacuum_1}. For $\gamma=\gamma_c$, the asymptotics are always governed by the tensionless solutions found in~\cite{ms_vacuum_1}. This result resonates with the existence of a non--linearly realized supersymmetry~\cite{dmnonlinear} in the $USp(32)$ Sugimoto model of~\cite{bsb}, which has this type of tadpole potential, and thus with the spontaneous character of ``brane supersymmetry breaking"~\cite{bsb}. All these properties are summarized in Table~\ref{table:tab_F}. 
\begin{table}[h!]
\centering
\small{
\begin{tabular}{||c | c ||} 
 \hline
Solution & Climbing or Descending  \\ [0.5ex] 
 \hline\hline 
$p=8, \,\gamma < \gamma_c$, $\beta \neq 0$ & (C: $\eta=\pi$), (D: $\eta=0$)  \\
$p=8\, ,\gamma \geq \gamma_c$, $\beta \neq 0$ & (C)  \\
\hline 
$p<8,\,\gamma = \gamma_c$, $\beta \neq 0$ & (C: $\theta>0$)  \\
$p<8,\,\gamma = \gamma_c$, $\beta \neq 0$ & (D: $\theta<0$)  \\
\hline 
$p<8, \ \gamma < \gamma_c$  & (D: $\cos\eta + \frac{\gamma}{\gamma_c}>0$) \\
$\qquad $ & (C: $\cos\eta + \frac{\gamma}{\gamma_c}<0$) \\ \hline
$p> 8, \ \gamma < \gamma_c$ &   (C: $\cosh\zeta < \frac{\gamma}{\gamma_c}$) \\
$\qquad $ &(D: $\cosh\zeta > \frac{\gamma}{\gamma_c}<0$) 
 \\ [1ex] 
 \hline
\end{tabular}
}
\caption{Main features of the cosmological solutions of Section~\ref{sec:susybT_c} with tension T and no flux, with emphasis on their climbing (C) or descending (D) behavior at the initial singularity.}
\label{table:tab_Fc}
\end{table}

The corresponding cosmological solutions that we have found exhibit a few novel features. They are generally anisotropic, but for $\gamma \leq \frac{3}{2}$ they approach an isotropic expansion for large values of the cosmic time. On the other hand, for $\gamma > \frac{3}{2}$ there are solutions with a Big Bang in the finite past where, for large values of the cosmic time, some dimensions continue to shrink. They can thus provide simple models of dynamical compactifications where our macroscopic four--dimensional world would have emerged from the ten--dimensional space time of String Theory via the cosmological evolution. Turning to the \emph{climbing issue}, in the anisotropic case there is still a transition point beyond which only a purely climbing behavior is possible. However, it occurs for values of $\gamma$ that lie beyond $\frac{3}{2}$, and more so the more anisotropic the expansion of the Universe is. Table~\ref{table:tab_Fc} summarizes these results for the tadpole--driven cosmologies with no flux.

\begin{table}[h!]
\centering
\small{
\begin{tabular}{||c | c c c | c c c||} 
 \hline
Solution & Left(L) & Left(T) & Left($g_s$) & Right(L) & Right(T) & Right($g_s$) \\ [0.5ex] 
 \hline\hline
$\chi_1 > \frac{1}{2}$ & F & SUSY & $\infty$ & F & SUSY & $\infty$ \\
$\chi_1 < \frac{1}{2}, x_1 \geq 0$ & F & SUSY & $\infty$ & F & is. (0) & $\infty$ \\
$\chi_1 \leq \frac{1}{2}, x_1 < 0$ & F & SUSY & $\infty$ & F & anis. (0) & $A$ 
 \\ [1ex] 
 \hline
\end{tabular}
}
\caption{The three families of exact solutions with $p=5$ flux and ``critical'' $(\gamma=\gamma_c)$ tension. As above, $A$ indicates that the limiting value of the string coupling $g_s$ at the right end can be finite, infinite or zero depending on the value of a parameter.}
\label{table:tab_FT}
\end{table}

We have also explored the general setup underlying systems that combine a symmetric form flux and the tadpole potential. These systems are more complicated and are generally not integrable, but we have identified an exactly solvable case that corresponds to the orientifold value $\gamma=\gamma_c=\frac{3}{2}$ in the presence of {an ${\cal H}_7$ flux,} the magnetic dual of a three--form field strength. This setting indicates how the tadpole-driven solutions are deformed by the flux, or alternatively how the flux-driven solutions are deformed by the tadpole. The resulting novelties can be appreciated focusing on the singularity introduced by the flux, close to which these backgrounds approach supersymmetric vacua, despite the presence of the tadpole potential. The singularity limits the range to $r \geq 0$, so that if the effective tension $T e^{\gamma\,\phi}$ builds up toward large values of $r$ one recovers the isotropic limiting behavior at strong coupling. However, if it builds up in the opposite direction, its growth is bound to terminate at $r=0$, where the interval ends, so that there are solutions where the tadpole is never dominant. Consequently, these solutions can approach an anisotropic limit for $r \to \infty$, even with small or finite asymptotic values of the string coupling. There are indeed three families of solutions, as summarized in Table~\ref{table:tab_FT}, whose limiting behavior at one end is always supersymmetric, while at the other it is tensionless, and can be isotropic, anisotropic or again supersymmetric.
We have also looked at the corresponding cosmologies, and their novelty is the option of combining contractions of some coordinates with expansions of others, which was only possible for $\gamma>\gamma_c$ without the flux.  Table~\ref{table:tab_FTC} summarizes the main properties of these cosmologies.
\begin{table}[h!]
\centering
\small{
\begin{tabular}{||c | c | c | c||} 
 \hline
Solution & $A$-space & $C$-space & Climbing or Descending \\ [0.5ex] 
 \hline\hline
$\rho > \frac{3}{2} \ + \sqrt{3}$ & contracts & expands & (C) \\
$\frac{9}{14} \ + \ \frac{\sqrt{15}}{7}  < \rho < \frac{3}{2} \ + \ \sqrt{3}$ & expands & expands & (C) \\
$\frac{1}{2} < \rho < \frac{9}{14} \ + \ \frac{\sqrt{15}}{7}$ & expands & contracts & (C) \\
$\frac{9}{14} \ - \ \frac{\sqrt{15}}{7} < \rho < \frac{1}{2}$ & expands & contracts & (D) \\
$\rho < \frac{9}{14} \ - \ \frac{\sqrt{15}}{7}$ & expands & expands & (D)
 \\ [1ex] 
 \hline
\end{tabular}
}
\caption{Main properties of the cosmological solutions with $p=5$ ${\cal H}$-flux and $\gamma=\gamma_c$.}
\label{table:tab_FTC}
\end{table}

The three ten--dimensional models of~\cite{so16so16,susy95,bsb} have brane spectra that were characterized, from the CFT perspective, in~\cite{dms_01}.  The vacua that we have constructed should prove helpful to clarify the effects of the tadpole potential on their spatial profiles,  sufficiently beyond their horizons, where curvature effects should be less important.  We hope to return to this issue soon~\cite{deformed_branes}. We also plan to investigate the role of internal spaces that are more complicated than the tori considered here, since they have the potential of providing additional interesting links to lower--dimensional Minkowski backgrounds. Setups of this type stand a chance~\cite{bms,int4d_vacuum} of bypassing the vexing stability problems~\cite{bms} encountered by $AdS$ vacua~\cite{AdStimesS} with broken supersymmetry. 

Before concluding, let us spend again a word of caution on the intrinsic limitations of the type of analysis presented here and in~\cite{ms_vacuum_1}, whose results are fully reliable for String Theory only within regions where the curvature and the string coupling are both bounded. This is typically the case far from the boundaries of the $r$-interval, where these conditions are not always fulfilled. In~\cite{ms_vacuum_1} we actually found vacuum solutions where the string coupling is bounded everywhere, and a detailed stability analysis of their spectra will be presented in~\cite{int4d_vacuum}. 
However, in none of the cases that we have analyzed is the curvature everywhere bounded. Bypassing this difficulty might be possible considering systematically higher--derivative corrections to the low--energy effective field theory, but the lowest--order corrections do not suffice to this end~\cite{conde_dud}.

\vskip 24pt
\section*{\sc Acknowledgments}

\vskip 12pt

We are grateful to E.~Dudas for stimulating discussions, and also to G.~Bogna and Y.~Tatsuta, who read an earlier version of the manuscript and made useful comments. The work of AS was supported in part by Scuola Normale, by INFN (IS GSS-Pi) and by the MIUR-PRIN contract 2017CC72MK\_003. JM is grateful to Scuola Normale Superiore for the kind hospitality extended to him while this work was in progress. AS is grateful to Universit\'e de Paris and DESY--Hamburg for the kind hospitality, and to the Alexander von Humboldt foundation for the kind and generous support, while this work was in progress.

\vskip 24pt

\vskip 24pt
\newpage
\appendix
\begin{appendices}

\section{\sc Kasner--like Solutions of Systems with $T=0$} \label{app:kasner}

In this Appendix we review briefly some results of~\cite{ms_vacuum_1} that concern static solutions in the absence of tension and flux, since this type of behavior presents itself recurrently in asymptotic regions for the systems considered in this paper. 

In the absence of tension and flux eqs.~\eqref{EqA_red} -- \eqref{Eqphi_red} reduce to
\beq
A'' \ =\  0 \ , \qquad C''\  =\  0 \ , \qquad \phi'' = 0\ ,
\eeq
and therefore the general solution in the harmonic gauge takes the form
\bea
A &=& A_1\,r \,+\, A_2 \ , \qquad B \ = \  (p+1)A+(8-p)C \ , \nonumber \\
C &=& C_1\,r \,+\, C_2  \ , \qquad \phi \ = \  \phi_1\,r \ + \ \phi_2 \ ,
\eea
where the $A_i$, $C_i$, $\phi_i$ are arbitrary constants. The constants $A_2$ and $C_2$ can be removed by rescaling all coordinates, thus bringing the solution to the form
\bea{}{}{}{}
ds^2 &=& e^{2 A_1 r}\, dx^2\ + \ e^{2\,\mu\, r}\, dr^2 \ + \ e^{2 C_1 r}\, d\vec{y}^{\,2}  \ , \nonumber \\
e^\phi &=& e^{\phi_1 r + \phi_2} \ , \label{nofluxABC}
\eea
where for simplicity we retain the same symbols, and where
\beq{}{}{}{}
\mu \ = \ (p+1)A_1+(8-p)C_1 \ . \label{mu_eq}
\eeq

The Hamiltonian constraint~\eqref{EqB_red} reduces in this case to 
\beq
\frac{\phi_1^2}{2} \ + \ (p+1) \, A_1^2 \ + \ (8-p) \, C_1^2 \ = \ \left[ (p+1)A_1 \ + \ (8-p)C_1\right]^2 \  , \label{ham_00case3}
\eeq
so that, away from the flat--space solution where $A_1$, $C_1$ and $\phi_1$ are all zero,  the parameter $\mu$ cannot vanish. Therefore, letting
\beq
\alpha_A \ = \ \frac{A_1}{\mu} \ ,  \qquad
\alpha_C \ = \ \frac{C_1}{\mu} \ , \qquad
\alpha_\phi \ = \ \frac{\phi_1}{\mu} \ ,
\eeq
the $\alpha_i$ are determined by eq.~\eqref{ham_00case3}
\beq
(p+1) \alpha_A^2 \ + \ (8-p) \alpha_C^2 \ + \ \frac{\alpha_\phi^2}{2} \ = \ 1 \ , \label{quad_constr}
\eeq
which defines an ellipsoid, while the definition of $\mu$ turns into
\beq
(p+1) \alpha_A \ + \ (8-p) \alpha_C \ = \ 1 \ , \label{linear_constr}
\eeq
which describes a plane.
The independent geometries in this class correspond to their intersections, which are the points of the ellipse
\beq{}{}{}{}
\frac{9(p+1)}{8(8-p)} \left(\alpha_A \ - \ \frac{1}{9}\right)^2 \ + \ \frac{9}{16}\, \alpha_\phi^2 \ =  \ 1 \ . \label{ellipse}
\eeq

The end result is the family of solutions
\bea
ds^2 &=& e^{2\,\alpha_A\,\mu\,r} \, dx^2 \ + \  e^{2\,\mu\,r}\,d r^2\ + \ e^{2\,\alpha_C\,\mu\,r} \, d\vec{y}^{\,2} \ , \nonumber \\
e^\phi &=& e^{\alpha_\phi\,\mu\,r} \, e^{\phi_2} \ , \label{spontaneous_r}
\eea
which comprises flat space when $\mu=0$ and, when $\mu \neq 0$, 
take the Kasner--like form
\bea
ds^2 &=& \left(\mu\xi\right)^{2\,\alpha_A} \, dx^2 \ + \ d \xi^2\ + \ \left(\mu \xi\right)^{2\,\alpha_C} \, d\vec{y}^{\,2} \ , \label{spontaneous} \nonumber \\
e^\phi &=& \left(\mu\xi\right)^{\alpha_\phi} \, e^{\phi_2} \ ,
\eea
in terms of the proper length
\beq
\xi \ = \ \frac{1}{\mu}\ e^{\mu\,r} \ , \label{rxi}
\eeq
where clearly $0 < \xi < \infty$. Using a standard parametrization for the ellipse,
the three constants can be related to an angle $\theta$ according to
\bea{}{}{}{}
\alpha_A &=& \frac{1}{9}\left[1 \ + \ 2\,\sqrt{\frac{2(8-p)}{(p+1)}} \ \cos\theta\right] \ , \nonumber \\
\alpha_C &=& \frac{1}{9}\left[1 \ - \ 2\,\sqrt{\frac{2(p+1)}{(8-p)}} \ \cos\theta\right] \ , \nonumber \\
\alpha_\phi &=& \frac{4}{3}\ \sin\theta \ \equiv \ \frac{2}{\gamma_c} \ \sin \theta \ . \label{param_theta}
\eea
The solutions are thus parametrized by $\theta$, by the constant $\phi_2$ that enters the dilaton profile and by the scale $\mu$. Here
\beq
\gamma_c \ = \ \frac{3}{2}  \label{gammac}
\eeq
is precisely the critical exponent that enters the tadpole potential of orientifold models. 

The two isotropic nine--dimensional solutions, with $\alpha_A=\alpha_C$, which are obtained for $\theta=\pm \frac{\pi}{2}$, for which
\bea
ds^2 &=& d \xi^2 \ + \ \left(\mu\,\xi\right)^\frac{2}{9} \left( dx^2 \ + \  d\vec{y}^{\,2}\right) \ , \nonumber \\
e^\phi &=& \left(\mu\,\xi\right)^{\,\pm\,\frac{4}{3}} \, e^{\phi_2} \ , \label{spontaneous_r_isotropic}
\eea
are important special cases, and play a role in several asymptotic regions considered in this paper.

These Kasner--like solutions emerge in different asymptotic regions, around $\xi=0$ or around $\xi=+\infty$. The effect of the tension will be negligible with respect to the kinetic contribution of the scalar field whenever
\beq{}{}{}{}
\frac{V(\phi(\xi))}{\left[\phi'\left(\xi\right)\right]^2} \ \to \ 0  \label{ratio}
\eeq
in the limit. This expression is proportional to
\beq{}{}{}{}
\left( \mu \xi\right)^{2\left[1 \ + \ \frac{\gamma}{\gamma_c}\, \sin\theta\right]} \ .
\eeq
As $\xi \to 0\, \left(\xi \to \infty\right)$ , the ratio vanishes provided
\beq{}{}{}{}
1 \ + \ \frac{\gamma}{\gamma_c}\, \sin\theta \ > \ 0 \  \quad \left( \ < \ 0\right)  \ .
\eeq
Hence, for $\gamma < \gamma_c$ the condition~\eqref{ratio} can only hold as $\xi \to 0$, 
so that this Kasner--like zero--tension behavior can only emerge close to boundaries at finite distance. On the other hand, for $\gamma > \gamma_c$ these zero--tension solutions can emerge in both cases, with values of $\theta$ compatible with one or the other inequality.

\section{\sc A Recurrent Newtonian System} \label{app:deq}

Here we would like to review briefly the differential equation
 \beq
 Z'' \ = \ \epsilon\,\Delta^2 \ e^{\,2\,Z} \ , \label{eqZ}
 \eeq
 where $\epsilon=\pm 1$ and $\Delta$ is real and positive, which appears in various parts of the paper. This equation has the first integral
 \beq
 (Z')^2 \ = \ E \ + \ \epsilon\, \Delta^2\, e^{2\, Z} \ , \label{conserv}
 \eeq
 and in order to proceed further one must distinguish a few cases.
 \begin{enumerate}

 \item If $\epsilon = 1$ and $E=\frac{1}{\rho^2}>0$, letting
 \beq
 x \ = \ \frac{r}{\rho}
 \eeq
 eq.~\eqref{conserv} becomes
\beq
 \left(\frac{d Z}{d x}\right)^2  \ = \ 1 \ + \ \left(\Delta\,\rho\right)^2 \, e^{2\, Z} \ ,
\eeq
and after the redefinition
\beq
Z = \ \tilde{Z} - \log(\Delta\, \rho)
\eeq
one is led to the reduced equation
\beq
 \left(\frac{d \tilde{Z}}{d x}\right)^2  \ = \ 1 \ + \ e^{2\, \tilde{Z}} \ .
\eeq
Therefore, separating variables the solutions are finally
\beq
Z \ = \ -  \log\left[\Delta\,\rho \, \sinh \left(\frac{r-r_0}{\rho}\right) \right]\ ,
\eeq
in the region $r > r_0$, or
\beq
Z \ = \ -  \log\left[\Delta\,\rho \,\sinh\left(\frac{r_ 0 - r}{\rho}\right) \right]\ ,
\eeq
in the region $r < r_0$.
\item If $\epsilon = 1$ and $E=- \frac{1}{\rho^2} < 0$, letting
\beq
 x \ = \ \frac{r}{\rho}
 \eeq
 eq.~\eqref{conserv} becomes
\beq
 \left(\frac{d Z}{d x}\right)^2  \ = \ - \,1 \ + \ \left(\Delta\,\rho\right)^2 \, e^{2\, Z} \ ,
\eeq
and after a redefinition
\beq
Z = \ \tilde{Z} - \log(\Delta\, \rho)
\eeq
one is led to the reduced equation
\beq
 \left(\frac{d \tilde{Z}}{d x}\right)^2  \ = \ - \,1 \ + \ e^{2\, \tilde{Z}} \ ,
\eeq
and therefore finally to the solution
\beq
Z \ = \ -  \log\left[\Delta\,\rho \, \cos \left(\frac{r-r_0}{\rho}\right) \right]\ ,
\eeq
which is valid for $|r-r_0|< \frac{\pi\,\rho}{2}$, and one can conveniently choose $r_0=\frac{\pi\rho}{2}$, thus recovering the solutions used in the main body of the paper. These solutions can be obtained from those of the preceding case letting
\beq
r - r_0 \ \to \ i\left( r - r_0 \ - \ \frac{\pi}{2}\right) \ , \qquad \Delta \ \to \ - \ i\, \Delta \ .
\eeq
 \item If $\epsilon=1$ and $E=0$, eq.~\eqref{conserv} reduces to
 \beq
 Z' \ = \ \pm \Delta\, e^{\,Z} \ ,
 \eeq
 which can simply integrated and yields
 \beq
 e^{-Z} \ = \ \mp \, \Delta \left(r - r_0\right) \ ,
 \eeq
 where $r_0$ is another integration constant. Here clearly the upper sign is associated to the region $r< r_0$, where the real solution reads
 \beq
 Z \ = \ - \ \log \Delta \left( r_0 - r \right) \ .
 \eeq
 while the lower one is associated to the region $r > r_0$, where the real solution reads
 \beq
 Z \ = \ - \ \log \Delta \left( r - r_0\right) \ .
 \eeq
 These types of solutions capture the limiting behavior of the preceding cases when $E$ is negligible with the respect to the exponential $e^{2 Z}$, and thus as $\rho \to \infty$.
\item Finally, If $\epsilon = -1$ and $E= \frac{1}{\rho^2} > 0$, letting
\beq
 x \ = \ \frac{r}{\rho}
 \eeq
 eq.~\eqref{conserv} becomes
\beq
 \left(\frac{d Z}{d x}\right)^2  \ = \ 1 \ - \ \left(\Delta\,\rho\right)^2 \, e^{2\, Z} \ ,
\eeq
and after a redefinition
\beq
Z = \ \tilde{Z} - \log(\Delta\, \rho)
\eeq
one is led to the reduced equation
\beq
 \left(\frac{d \tilde{Z}}{d x}\right)^2  \ = \ 1 \ - \ e^{2\, \tilde{Z}} \ ,
\eeq
and therefore finally to the solution
\beq
Z \ = \ -  \log\left[\Delta\,\rho \, \cosh \left(\frac{r-r_0}{\rho}\right) \right]\ ,
\eeq
for all real values of $r$. This can be continued to the first solution combining the two transformations
\beq
\frac{r}{\rho} \ \to \ \frac{r}{\rho} \ + \ i \frac{\pi}{2} \ , \qquad \Delta \ \to \ - \ i\, \Delta \ .
\eeq
 \end{enumerate}
 
 \section{\sc The $AdS \times S$ Vacua of~\cite{AdStimesS} in the Harmonic Gauge} \label{app:AdSxS}
In this Appendix we describe the form that the $AdS \times S$ vacua that were obtained in~\cite{AdStimesS} in the gauge $B=0$, take in the harmonic gauge used in~\cite{ms_vacuum_1} and in this paper. Their key properties are constant values for $\phi$ and $C$. The internal space is now a sphere, so that one is to start from the complete equations of ~\cite{ms_vacuum_1} with $k=0$ and $k'=1$. The equations of $C$ and $\phi$ thus reduce to the algebraic constraints
\bea
0 & =&  - \ \frac{T}{8} \ e^{\, 2\left[(8-p)C + \frac{\gamma}{2}\,\phi\right]} \ + \ \frac{(7-p)}{\alpha'}\ e^{2\left[(7-p)C\right]}\label{eqACphiF_app} \ - \ \frac{(p+1)}{16} \ e^{\,2\,\beta_p\,\phi} \ H_{p+2}^2\ , \nonumber \\
0 &=& {T\,\gamma} \ e^{\, 2\left[(8-p)C + \frac{\gamma}{2}\,\phi\right]} \ + \ {\beta_p\,}\ e^{\,2\,\beta_p\,\phi} \ H_{p+2}^2\ ,
\eea
and making use of them the Hamiltonian constraint of~\cite{ms_vacuum_1}, with $k=0$ and $k'=1$, becomes
 \begin{align}
&p(p+1)(A')^2  \, + \, {T} \, e^{\, 2\left[(p+1)A+(8-p)C + \frac{\gamma}{2}\,\phi\right]}  \nonumber \\ &- \, \frac{(7-p)(8-p)}{\alpha'}\ e^{2\left[(p+1) A+(7-p)C\right]} \,+\, \frac{1}{2}\, e^{\,2\left[\beta_p\,\phi\,+\,(p+1)\,A\right]} \ H_{p+2}^2  \,= \, 0 \ ,
 \end{align}
takes the form
 \beq
(A')^2 \label{eqBF_app} \ = \ \frac{H_{p+2}^2}{16\,(p+1)} \left[2\,\frac{\beta_p}{\gamma} \ + \ 7-p \right] \, e^{\, 2\left[(p+1)A+ \beta_p \phi\right]} \ .
 \eeq

These solutions require the two conditions
\beq
 \frac{\beta_p}{\gamma} \ < \ 0 \ , \qquad p < 7 \ .
\eeq
Different choices of $k$ would simply result in different $AdS$ slicings, as discussed in~\cite{AdStimesS}, so that here we have set for brevity $k=0$. Then the quantity
\beq
 \Delta^2 \ = \ \frac{H_{p+2}^2(p+1)}{16} \left[2\,\frac{\beta_p}{\gamma} \ + \ 7-p \right] \, e^{\, 2\,\beta_p \phi} \ > \ 0 \ , \label{Delta}
\eeq
is positive, and therefore
\beq
A \ = \ - \ \frac{1}{p+1}\ \log \left(\Delta r\right) \ ,
\eeq
according to Appendix~\ref{app:deq}, while the harmonic gauge condition $F=0$ gives
\beq
B \ = \ - \ \log \left(\Delta r\right) \ + \ (8-p)\,C \ .
\eeq

In these vacua
\bea
R^2\,g_s^{\gamma} &=& \frac{16(7-p)}{\alpha'\,T\left[ 2 \ - \ (p+1)\, \frac{\gamma}{\beta_p}\right]} \ , \nonumber \\
H_{p+2}^2 &=& \frac{\left(\alpha'\,T\right)^{p-7}}{\alpha'} \ \left(- \,\frac{\gamma}{\beta_p} \right) \left[\frac{16(7-p)}{2 - (p+1)\frac{\gamma}{\beta_p}} \right]^{8-p} \left(g_s\right)^{-\left[2\beta_p+(7-p)\gamma\right]} \label{R2H2}
\eea
where
\beq
R \ = \ e^{C} \ \quad \mathrm{and} \quad g_s \ = \ e^\phi
\eeq
are the radius of the internal sphere in units of $\sqrt{\alpha'}$ and the string coupling. As stressed in~\cite{AdStimesS}, large fluxes always translate into large sphere radii and weak string coupling.

In detail, for the $USp(32)$ and $U(32)$ orientifolds, with $p=1$, $\gamma=\frac{3}{2}$ and $\beta_p =- \frac{1}{2}$,
\beq
R^4\, g_s^3 \ \sim \ \frac{1}{T^2} \ , \qquad g_s^8 \ \sim \frac{1}{H_{3}^2\,T^6} \ ,
\eeq
while for the heterotic $SO(16) \times SO(16)$, with $p=5$, $\gamma=\frac{5}{2}$ and $\beta_p =- \frac{1}{2}$,
\beq
R^4\, g_s^5 \ \sim \ \frac{1}{T^2} \ , \qquad g_s^4 \ \sim \frac{1}{H_{7}^2\,T^2} \ ,
\eeq
and the first relation corrects a typo in~\cite{AdStimesS}.

The resulting metrics read
\beq
ds^2 \ = \ \frac{dx^2}{\left(\Delta\,r\right)^{\frac{2}{p+1}}} \ + \ \frac{dr^2}{\left(\Delta\,r\right)^2}\,R^{2(8-p)} \ + \ \alpha'\,R^2\, d\Omega^2 \ ,
\eeq
and letting
\beq
z \ = \ \frac{R^{8-p}}{\Delta}\ \log\left(\Delta\,r\right)
\eeq
they take the standard Poincar\'e form for $AdS$,
\beq
ds^2 \ = \ e^{-2\,\kappa\,z} \, dx \cdot dx \ + \ dz^2 \ + \ \alpha' R^2 \, d\Omega^2 \ ,
\eeq
with
\beq
\kappa \ = \ \frac{\Delta}{(p+1)\,R^{8-p}} \ .
\eeq
This identifies the $AdS$ radius, in string units as the sphere radius, as
\beq
R_{AdS} \ = \ (p+1) \frac{R^{8-p}}{\Delta\,\sqrt{\alpha'}} \ ,
\eeq
and making use of eqs.~\eqref{Delta} and \eqref{R2H2}, one can finally conclude that
\beq
R_{AdS} \ = \ R \sqrt{\frac{\left[2 \ - \ (p+1)\frac{\gamma}{\beta_p}\right](p+1)}{\left[-2 \ - \ (7-p)\frac{\gamma}{\beta_p}\right](7-p)}} \ .
\eeq
This gives, in particular,
\beq
R_{AdS} \ = \ \frac{R}{\sqrt{6}}
\eeq
for the orientifolds, as in~\cite{AdStimesS}, and
\beq
R_{AdS} \ = \ {R}\,\sqrt{12}
\eeq
for the heterotic $SO(16)\times SO(16)$, which corrects a misprint in~\cite{AdStimesS}. 

\end{appendices}

\end{document}